\documentclass{aa}  

\usepackage{graphicx}
\usepackage{txfonts}
\usepackage{multicol}  
\usepackage{sidecap}
\sidecaptionvpos{figure}{t}

\newcommand{\teff}{\ensuremath{T_{\rm eff}}\xspace}

\newcommand{\mearth}{\mbox{M$_\oplus$}}
\newcommand{\rearth}{\mbox{R$_\oplus$}}

\newcommand{\Kepler}{{\it Kepler}}
\newcommand{\Ktwo}{{\it K2}}
\newcommand{\TESS}{{\it TESS}}
\newcommand{\CoRoT}{{\it CoRoT}}
\newcommand{\Spitzer}{{\it Spitzer}}

\newcommand{\JWST}{{\it JWST}}

\newcommand{\ar}{\ensuremath{a/R_*}}

\renewcommand{\andname}{\ignorespaces}
\newcommand{\be}{\begin{equation}}
\newcommand{\ee}{\end{equation}}

\newcommand{\micron}{$\mu$m}

\newcommand{\cspitzer}{$c_{\textrm{AOR1}}, v_{\textrm{AOR2}}, c_{\textrm{AOR2}}, c_{\textrm{AOR3}}, c_{\textrm{AOR4}}, c_{\textrm{AOR5}}, c_{\textrm{AOR6}}$}
\newcommand{\cjoint}{$c_{\textrm{AOR1}}, v_{\textrm{AOR2}}, c_{\textrm{AOR2}}, c_{\textrm{AOR3}}, c_{\textrm{AOR4}}, c_{\textrm{AOR5}}, c_{\textrm{AOR6}}, c_{\textrm{K2C12}}, c_{\textrm{K2C19}}$}
\newcommand{\cubicgwspitzer}{$Y_{\textrm{1, AOR1}}, X_{\textrm{3, AOR2}}, X_{\textrm{1, AOR3}}, X_{\textrm{3, AOR4}}, X_{\textrm{1, AOR5}}, Y_{\textrm{3, AOR5}}, X_{\textrm{3, AOR6}}, Y_{\textrm{3, AOR6}}$}

\usepackage{color}

\usepackage{amsmath}
\usepackage[labelfont=bf]{caption}

\usepackage{lscape}

\usepackage{natbib,twoopt}
\usepackage[breaklinks=true]{hyperref} 
\bibpunct{(}{)}{;}{a}{}{,} 
\makeatletter
\newcommandtwoopt{\citeads}[3][][]{\href{http://adsabs.harvard.edu/abs/#3}%
{\def\hyper@linkstart##1##2{}%
\let\hyper@linkend\@empty\citealp[#1][#2]{#3}}}
\newcommandtwoopt{\citepads}[3][][]{\href{http://adsabs.harvard.edu/abs/#3}%
{\def\hyper@linkstart##1##2{}%
\let\hyper@linkend\@empty\citep[#1][#2]{#3}}}
\newcommandtwoopt{\citetads}[3][][]{\href{http://adsabs.harvard.edu/abs/#3}%
{\def\hyper@linkstart##1##2{}%
\let\hyper@linkend\@empty\citet[#1][#2]{#3}}}
\newcommandtwoopt{\citeyearads}[3][][]%
{\href{http://adsabs.harvard.edu/abs/#3}
{\def\hyper@linkstart##1##2{}%
\let\hyper@linkend\@empty\citeyear[#1][#2]{#3}}}
\makeatother
\hypersetup{colorlinks=true,linkcolor=blue,citecolor=blue,urlcolor=blue} 

\begin{document}

   \title{\Ktwo\ and \Spitzer\ phase curves of the rocky ultra-short-period planet K2-141\,b hint at a tenuous rock vapor atmosphere}

   
    \titlerunning{\Ktwo\ and \Spitzer\ phase curves of K2-141\,b}

   \author{S. Zieba\inst{1,2},
          \and
          M. Zilinskas\inst{2},
          \and
          L. Kreidberg\inst{1},
          \and
          T. G. Nguyen\inst{3},
          \and
          Y. Miguel\inst{4,2},
          \and
          N. B. Cowan\inst{5,6},
          \and
          R. Pierrehumbert\inst{7},
          \and
          L. Carone\inst{1},
          \and
          L. Dang\inst{6,8},
          \and
          M. Hammond\inst{9},
          \and
          T. Louden\inst{10},
          \and
          R. Lupu\inst{11},
          \and
          L. Malavolta\inst{12,13}
          \and
          K. B. Stevenson\inst{14}
          }

   \institute{Max-Planck-Institut f\"ur Astronomie, K\"onigstuhl 17, D-69117 Heidelberg, Germany\\
              \email{zieba@mpia.de}
         \and
             Leiden Observatory, Leiden University, Niels Bohrweg 2, 2333CA Leiden, The Netherlands
        \and
             Centre for Research in Earth and Space Sciences, York University, 4700 Keele St, Toronto, ON M3J 1P3, Canada
        \and             
             SRON Netherlands Institute for Space Research , Niels Bohrweg 4, 2333 CA Leiden, The Netherlands
        \and             
            Department of Earth \& Planetary Sciences, McGill University, 3450 rue University, Montréal, QC H3A 0E8, Canada
        \and             
            Department of Physics, McGill University, 3600 rue University, Montréal, QC H3A 2T8, Canada
        \and    
            Atmospheric, Oceanic \& Planetary Physics, Department of Physics, University of Oxford, Oxford OX1 3PU, UK
        \and    
            Universit\'e de Montr\'eal, Institut de Recherche sur les Exoplan\`etes, 1375 Ave.Th\'er\`ese-Lavoie-Roux, Montr\'eal, QC H2V 0B3
        \and             
            Department of Physics, Oxford University, Parks Rd, Oxford, OX1 3PJ, UK
        \and             
            Department of Physics, University of Warwick, Coventry, CV4 7AL, UK
        \and             
            Eureka Scientific Inc, Oakland, CA 94602
        \and    
            Dipartimento di Fisica e Astronomia “Galileo Galilei”, Università degli Studi di Padova, Vicolo dell’Osservatorio 3, 35122 Padova, Italy
        \and             
            INAF - Osservatorio Astronomico di Padova, Vicolo dell’Osservatorio 5, 35122, Padova, Italy
        \and             
            Johns Hopkins APL, 11100 Johns Hopkins Rd, Laurel, MD 20723, USA
             }

   \date{Received xxx; accepted xxx}

 
  \abstract
{K2-141\,b is a transiting, small $(1.5\,R_\oplus)$ ultra-short-period (USP) planet discovered by the \Kepler\ space telescope orbiting a K-dwarf host star every 6.7 hours. The planet's high surface temperature of more than 2000 K makes it an excellent target for thermal emission observations.
Here we present 65 hours of continuous photometric observations of K2-141\,b collected with \Spitzer's IRAC Channel 2 at 4.5 \micron\ spanning 10 full orbits of the planet. We measure an infrared eclipse depth of $f_p/f_* = 142.9_{-39.0}^{38.5}$ ppm and a peak to trough amplitude variation of $A = 120.6_{-43.0}^{42.3}$ ppm. The best fit model to the \Spitzer\ data shows no significant thermal hotspot offset, in contrast to the previously observed offset for the well-studied USP planet 55 Cnc\,e. 
We also jointly analyze the new \Spitzer\ observations with the photometry collected by \Kepler\ during two separate K2 campaigns. We model the planetary emission with a range of toy models that include a reflective and a thermal contribution. With a two-temperature model, we measure a dayside temperature of $T_{p,d} = 2049_{-359}^{362}$ K and a night-side temperature that is consistent with zero (\mbox{$T_{p,n} <1712$ K at $2\sigma$)}. Models with a steep dayside temperature gradient provide a better fit to the data than a uniform dayside temperature ($\Delta$BIC = 22.2). We also find evidence for a non-zero geometric albedo $A_g = 0.282_{-0.078}^{0.070}$. We also compare the data to a physically motivated, pseudo-2D rock vapor model and a 1D turbulent boundary layer model. Both models fit the data well. Notably, we find that the optical eclipse depth can be explained by thermal emission from a hot inversion layer, rather than reflected light. A thermal inversion may also be responsible for the deep optical eclipse observed for another USP, Kepler-10\,b. Finally, we significantly improve the ephemerides for K2-141\,b and c, which will facilitate further follow-up observations of this interesting system with state-of-the art observatories like \JWST.}
 

   \maketitle
%

\section{Introduction}

The field of exoplanetary science started off with the surprising discovery of planets with short orbital periods. The formation of these planets has been well-studied and still remains a puzzle to this day \citep[see e.g.,][]{Dawson2018}. Ultra-short-period (USP) planets are an extreme subgroup of this population with orbital periods shorter than one day \citep[for a review of USPs see e.g.,][]{Winn2018}. The majority of these planets have been found to be smaller than $2\,R_\oplus$ \citep{Sanchis-Ojeda2014, Lundkvist2016}. 

One early theory was that these small USP planets could be remnant bare cores of hot Jupiters that lost their envelopes due to photoevaporation, Roche overflow or other processes \citep{Jackson2013}. It was however shown that the well-known planet-metallicity correlation \citep{Fischer2005} is not observed for USP planets \citep{Winn2017}. USP planets are therefore probably not evaporated hot Jupiters, but there is still no consensus about whether they are born rocky or once had a modest hydrogen envelope
 \citep{VanEylen2018, Lopez2017}. As USPs and sub-Neptunes (planets between $2\,R_\oplus$ and $4\,R_\oplus$) orbit generally stars with similar metallicities, sub-Neptunes might be possible progenitors. Most USPs have Earth-like densities \citep{Dai2019}, but a few \citep[55 Cnc\,e;][]{Crida2018a, Crida2018b} \citep[WASP-47e;][]{Vanderburg2017} have lower densities consistent with a low iron fraction or a small volatile envelope. Direct observations of the planet's atmospheres are needed to distinguish between these scenarios.
 
\subsection{Benchmark USPs}

55 Cnc\,e \citep{Fischer2008, Dawson2010, Winn2011, Demory2011}, is one of the best-studied small USP planets and shows evidence for a thick atmosphere \citep{Demory2016a, Angelo2017}. It is one of the very few small planets ($<2\,R_\oplus$) for which thermal emission was observed in the infrared (others are: LHS 3844\,b by \citet{Kreidberg2019}, and K2-141\,b in this work). Most other USP planets have been observed in the visible light with missions like \CoRoT\ \citep{Auvergne2009}, \Kepler\ \citep{Borucki2010}, K2 \citep{Howell2014} or more recently \TESS\ \citep{Ricker2014}. Numerous observations of 55 Cnc\,e with \Spitzer\ showed some surprising results: A large hotspot offset, where the hottest region of the planet is significantly offset from the substellar point \citep{Demory2016a}. This phase curve offset could be explained by a thick atmosphere with a super-rotating jet that advects energy away from the substellar point \citep{Kite2016, Hammond2017, Angelo2017}. \citet{Demory2016b} furthermore reported varying observed dayside temperatures for 55 Cnc\,e ranging from 1300 K to 2800 K. The authors proposed that these observed changes were possibly caused by volcanic activity, leading to plumes which increase the opacity in the \Spitzer\ bandpass. \citet{Tamburo2018} reanalyzed the \Spitzer\ observations and concluded that the changing eclipse depths were best modeled by a year-to-year variability model. They also suggested that the dayside of the planet is intermittently covered with reflective grains obscuring the hot surface, possibly originating from volcanic activity or cloud variability. Despite the numerous observations of 55 Cnc\,e, its composition and structure still remains a puzzle. \citet{Dorn2019} suggested that the low observed density of 55 Cnc e (6.4 $\pm$ 0.3 g/cm$^3$) might be explained by the planet being a part of a new class of Super-Earths which formed from high-temperature condensates. Planets like this would have no core and have enhancements in Ca, Al rich minerals leading to a lower overall bulk density compared to an Earth-like (30\% Fe, 70\% MgSiO$_3$) or a pure MgSiO$_3$ composition.

Recent observations of 55 Cnc\,e with \TESS\ also showed a tentative deep optical eclipse, which could be caused by a non-zero albedo if confirmed \citep{Kipping2020}.
Other observations of 55 Cnc e in the optical by the Microvariability and Oscillations in Stars (MOST) space telescope \citep{Winn2011} showed a quasi-sinusoidal modulation of flux with the same period as the planet. The amplitude of the signal was, however, too large to be reflected light or thermal emission alone and its origin remained unclear in that study. Additional MOST observations spanning several weeks between 2011 and 2015 by \citet{Sulis2019} confirmed this optical modulation. The amplitude and phase of the signal were variable which the authors suggested might be due to some star-planet interaction or the presence of a transiting circumstellar dust torus. They also did not detect the secondary eclipse of the planet which led to an upper value for the geometic albedo of 0.47 ($2\sigma$). Furthermore, recently \citet{Morris2021} presented CHEOPS observations of the planet showing a large phase variation but they do not detect a significant secondary eclipse of the planet. The authors suggest that the origin of the signal might be from circumstellar/circumplanetary material modulating the flux of the system. This is just another example of the challenges to determine the nature of 55 Cnc\,e.

Another surprising discovery came with Kepler-10\,b \citep{Batalha2011}, the first rocky planet detected by the \Kepler\ mission. The planet showed a deep secondary eclipse that suggested an unusually high reflectivity in the \Kepler\ bandpass \citep{Batalha2011, Rouan2011, Sheets2014}. A high albedo for planets that are highly irradiated by their stars could be explained by the creation of calcium- and aluminum-rich surfaces on their dayside \citep{Leger2011, Rouan2011, Kite2016, ModirroustaGalian2021}. A subset of planets detected by \Kepler\ showed comparably high albedos in the optical wavelengths \citep{Demory2014}. Most notably, both Kepler-10\,b and Kepler-78\,b \citep{Sanchis-Ojeda2013} have albedos of 0.4-0.6 \citep{Sheets2014}. \citet{Hu2015} reanalysed the \Kepler\ data of Kepler-10\,b and did not detect any phase curve offset. They found that any model with a Bond albedo greater than 0.8 fits the visible phase curve well regardless of whether asymmetric reflection exists.

Due to the high irradiation small USP planets receive from their host stars, they are more susceptible to atmospheric loss  \citep{Lopez2017}. LHS 3844\,b \citep{Vanderspek2019}, a USP planet orbiting an M-type star, was clearly shown to lack a thick atmosphere using observations by the \Spitzer\ Space Telescope and is therefore likely a bare rock \citep{Kreidberg2019}. The 100 hour continuous phase curve of the planet showed no hotspot offset ruling out the possibility of a thick atmosphere, and any less-massive atmospheres would be unstable to erosion by stellar wind. Some planets might however retain an atmosphere by the evaporation of surface lava oceans leading to a silicate rich atmosphere \citep{Schaefer2010, Miguel2011} or might have other thick, high mean-molecular-weight atmospheres \citep{Demory2016a}. 

\subsection{The new USP K2-141\,b}

Here we present \Spitzer\ observations of the USP K2-141\,b (EPIC 246393474 b). The planet was discovered in 2018 by \citet{Malavolta2018} and independently by \citet{Barragan2018} using observations of the \Kepler\ telescope during its ``second light'' mission, K2 \citep{Howell2014}. The planet has a radius of $R_p = 1.51 \pm 0.05\,R_\oplus$ and orbits its K-dwarf host star every 0.28 days (6.7 hours). Observations of the star by the high-precision spectrograph HARPS-N measured a mass for the planet ($M_p = 5.08 \pm 0.41\,M_\oplus$). With a density of $\rho = (8.2 \pm 1.1)$ g/cm$^3$, K2-141\,b is mostly consistent with an Earth-like iron-silicate composition. The radial velocity observations furthermore confirmed another companion K2-141\,c, which is in radius and upper mass more likely to be a Neptune-like planet than a rocky planet or a HJ, orbiting further out with an orbital period of 7.7 days.

The \Kepler\ observations also revealed the optical phase curve and secondary eclipse with a depth of $23 \pm 4$ ppm. The equilibrium temperatures of K2-141\,b, Kepler-10\,b and 55 Cnc\,e are 2150K, 2170K and 1950K in case of full atmospheric heat redistribution and 2745K, 2775K and 2490K for instant reradiation, respectively\footnote{Calculated using \(T_{\text{eq}} = T_* / \sqrt{\ar} \,(1 - A_B)^{1/4}\, f^{1/4} \) while assuming Bond albedo $A_B = 0$. The heat redistribution factor, $f$, is $f = 1/4$ in case of uniform redistribution (if the planet has a thick atmosphere able to redistribute heat) and $f = 2/3$ for instant reradiation (if the planet is a bare rock) \citep{Koll2019}.}. K2-141\,b is therefore a perfect target to compare to other well studied USPs (see also Table \ref{tab:usp}). Its host star is also bright enough ($V$ = 11.5 mag, $K$ = 8.4 mag) to conduct follow-up observations of the planet's emission in the infrared as previously done with 55 Cnc\,e  ($V$ = 6.0 mag, $K$ = 4.0 mag) \citep{Demory2016a, Demory2016b}. K2-141\,b and 55 Cnc\,e are therefore the only two currently known small USPs which are accessible in both the optical and infrared which invites comparison between the two planets. One should, however, note the difference in densities for the planets: While K2-141\,b's density is consistent with an Earth-like composition (30\% Fe), is 55 Cnc e inconsistent with an Earth-like composition at over $4\sigma$.\footnote{Calculations based on the Mass-Radius tables from \citet{Zeng2019}.}

\begin{table*}[h]
\centering
\caption{Selection of USPs and their properties.}
\label{tab:usp}
\renewcommand{\arraystretch}{1.25}
\begin{tabular}{l|c|c|c|c|c|c|c|c|c|c}\hline\hline
Planet      & $P$ (days) & $R_p$ (\rearth) & $M_p$ (\mearth) & $\rho_p$ (g/cm$^3$) & $a/R_*$  & $T_{\text{eq}}^{f=1/4}$ (K) & $T_{\text{eq}}^{f=2/3}$ (K) & $T_*$ (K) & $K$ (mag) & $V$ (mag) \\\hline
\tablefootmark{(1)}K2-141\,b    & 0.28        & 1.51(5)  & 5.1(4) & 8.2(1.1) & 2.29 & 2150     & 2745          & 4599   & 8.4    & 11.5   \\
\tablefootmark{(2)}55 Cnc\,e    & 0.74        & 1.95(4)  & 8.6(4) & 6.4(0.3) & 3.52 & 1950     & 2490          & 5172   & 4.0    & 6.0    \\
\tablefootmark{(3)}Kepler-10\,b & 0.84        & 1.47(3)  & 3.3(5) & 5.5(0.9) & 3.46 & 2170     & 2775          & 5708   & 9.5    & 11.0   \\
\tablefootmark{(4)}LHS 3844\,b  & 0.46        & 1.30(2)  & ---    & ---      & 7.11 & 805      & 1030          & 3036   & 9.1    & 15.2  
\end{tabular}
\renewcommand{\arraystretch}{1.}
\tablebib{\tablefoottext{1}{\citet{Malavolta2018};} \tablefoottext{2}{\citet{Bourrier2018}, \citet{Crida2018b};}
\tablefoottext{3}{\citet{Dumusque2014};} \tablefoottext{4}{\citet{Vanderspek2019}}
}
\end{table*}
Recently, \citet{Nguyen2020} modelled the atmosphere of K2-141\,b assuming the planet has a thin rock vapor atmosphere which arises from the evaporation of the surface on the dayside. This leads to a flow that is maintained by the pressure gradient between the nightside and dayside on the planet. This flow is however not able to transport enough heat to the nightside to create a considerable thermal hotspot offset nor to heat the nightside. Previous studies of transit spectroscopy of lava planets focused on more volatile species such as Na or K \citep{Castan2011, Kite2016}. \citet{Nguyen2020} compared different atmospheric compositions expected for a rock vapor atmosphere (Na, SiO and SiO$_2$) and found that an SiO$_2$ atmosphere may be easier to observe due to the extreme geometry of this system. \citet{Nguyen2020} also noted that due to the proximity of K2-141\,b to its host star ($\ar = 2.292$), the night side (the area of the planet which never receives any incident flux) is only about a third of the planet, rather than a hemisphere. The terminator for K2-141\,b is located approximately $115^{\circ}$ from the substellar point, leading to a hemisphere-integrated night side temperature of approximately 400K for the planet\footnote{Calculations based on \citet{Kopal1954} and \citet{Nguyen2020}}. Therefore, the regions probed during a transit range from $\sim$65$^{\circ}$ to $\sim$115$^{\circ}$ from the substellar point. If the planet is further away from the star the region probed during a transit is approximately $90^{\circ}$ from the substellar point. This effect is so small, however, that the flux emitted from the night side would not be detectable within the measurement precision of our data. We therefore adopt a night hemisphere in this paper.

The paper is structured as follows: Section \ref{sec:data} describes the data reduction of the K2 and \Spitzer\ observations used in this paper. Section \ref{sec:analysis} discusses the different models which were fit to the K2 and \Spitzer\ data to extract information on the reflective and thermal emission coming from the planet. Section \ref{sec:composition} compares the observations to two different atmospheric models: a pseudo-2D rock vapor model and a 1D Turbulent Boundary Layer model, the latter being recently published in \citet{Nguyen2020} and further improved in \citet{Nguyen2022}. In Section \ref{sec:discussion} we discuss our findings and summarize our conclusions in Section \ref{sec:summary}.

\section{Observations and Data Reduction} \label{sec:data}

\begin{table*}[]
\captionsetup{justification=centering}
\centering
\caption{Observations with \Kepler\ and \Spitzer\ used in this work.}
\label{tab:dates}
\renewcommand{\arraystretch}{1.2}
\begin{tabular}{c|c|c|c|c}\hline\hline
Observation & AORKEY & Start Date & End Date & Exp. time (s) \\\hline
K2 C12       & --- & 2016-12-21 22:41:48 & 2017-03-04 12:56:44 & 1800 \\
K2 C19       & --- & 2018-09-08 02:48:49 & 2018-09-15 03:00:18 & 60 \\
\Spitzer\ AOR1 & 66695168 & 2018-10-09 01:52:12 & 2018-10-09 13:41:46 & 2 \\
\Spitzer\ AOR2 & 66694912 & 2018-10-09 13:47:34 & 2018-10-10 01:37:08 & 2 \\
\Spitzer\ AOR3 & 66694656 & 2018-10-10 01:42:56 & 2018-10-10 13:32:30 & 2 \\
\Spitzer\ AOR4 & 66694400 & 2018-10-10 13:38:18 & 2018-10-11 01:27:52 & 2 \\
\Spitzer\ AOR5 & 66694144 & 2018-10-11 01:33:40 & 2018-10-11 13:23:14 & 2 \\
\Spitzer\ AOR6 & 66693632 & 2018-10-11 13:29:02 & 2018-10-11 18:28:51 & 2
\end{tabular}
\renewcommand{\arraystretch}{1.2}
\end{table*}

\subsection{\Spitzer\ photometry}

We observed the K2-141 system with the \Spitzer\ InfraRed Array Camera \citep[IRAC;][]{Fazio2004} for about 65 hours between October 09 and October 11, 2018 \citep[Program 14135,][]{Kreidberg2018}. We used Channel 2 on IRAC (equivalent to a photometric bandpass of 4 -- 5 \micron) with a two-second exposure time. The observations began with a 30-minute burn-in to allow for the telescope to thermally settle. Following this procedure, we placed the target on the ``sweet spot'', a pixel on the detector which is known to have a minimal intra-pixel sensitivity variation.

We split the observations into six sequential datasets (AORs, Astronomical Observation Requests) which we downloaded from the \Spitzer\ Heritage Archive\footnote{\url{https://sha.ipac.caltech.edu/applications/Spitzer/SHA/}} (see Table \ref{tab:dates} for the start and end time of each individual AOR). We reduced the Basic Calibrated Data (BCD, provided by the \Spitzer\ Science Center) with the Photometry for Orbits, Eclipses, and Transits (\texttt{POET}) pipeline (which is available open-source on GitHub\footnote{\url{https://github.com/kevin218/POET}}) developed by \citet{Stevenson2012} and \citet{Cubillos2013}. It performs centroiding on each frame by fitting a 2D Gaussian profile to the stellar image \citep{Lust2014} in each \Spitzer\ exposure after upsampling by a factor of five in each spatial direction \citep{Harrington2007}. The target remains centered near the sweet spot for the entire AOR in each observation, with the majority of the exposures being well within of 0.1 pixels from the sweet spot (see plots in Section \ref{sec:pointing}). \texttt{POET} then identifies and flags bad pixels using an iterative sigma-clipping procedure along the time axis and then sums the flux in each fixed aperture. We have chosen a grid of apertures with radii ranging from two to four pixels in 0.25 pixel steps for every AOR and used the aperture which minimizes the residual noise in each of the extracted light curves (a list of the apertures can be found in Table \ref{tab:SpitzerAORs}). For the median background flux estimation, we used an annulus with an inner radius of 7 pixels and outer radius of 15 pixels.

The dominant systematics for the 4.5 \micron\ \Spitzer\ channel are intrapixel sensitivity variations \citep{Charbonneau2005, May2020}. We therefore fitted for them by using the BiLinearlyInterpolated Subpixel Sensitivity (BLISS) map technique introduced by \citet{Stevenson2012} (see plots in Section \ref{sec:bliss} to see the determined intrapixel sensitivity variations across the detector). We determined the pixel bin size used for the map for every AOR and listed it in Table \ref{tab:SpitzerAORs}.

\begin{table}[]
\centering
\caption{Parameters used for the data reduction of every \Spitzer\ AOR determined by minimizing the Bayesian Information Criterion (BIC).}
\renewcommand{\arraystretch}{1.2}
\begin{tabular}{l|l|l|l|l}\hline\hline
AOR & aper. size & bin size & ramp model & PRF-FWHM \\ \hline
1 & 3.00 & 0.013 & constant & $Y_1$ \\
2 & 3.00 & 0.013 & linear & $X_3$ \\
3 & 3.00 & 0.015 & constant & $X_1$ \\
4 & 3.00 & 0.013 & constant & $X_3$ \\
5 & 3.25 & 0.011 & constant & $X_1$, $Y_3$ \\
6 & 3.00 & 0.012 & constant & $X_3$, $Y_3$
\label{tab:SpitzerAORs}
\end{tabular}
\renewcommand{\arraystretch}{1.}
\tablefoot{
aper. size: aperture size given in pixels; bin size: the optimal resolution for BLISS mapping; ramp model: the optimal ramp model $R(t)$ for the AOR (see Equation \ref{eq:flux}); PRF-FWHM: the complexity of the PRF fit $G(X,Y,t)$. The latter is described by the dimension ($X$ or $Y$) and the degree of the fit (1 for linear, 2 for quadratic and 3 for cubic).
}
\end{table}

We visually inspected the data and removed three short segments of data (two in AOR3 and one in AOR5) making up approximately 5$\%$ of the observations that showed strong correlated noise in the residuals. After visually inspecting the individual frames during the discarded segment in AOR5 it was able to attribute this event to a strong cosmic ray hit on the detector (see Fig. \ref{fig:frames}). The other segment in AOR3 showed no noticeable trends in PRF width or other parameters (see Fig. \ref{fig:systematics}). Similar outliers were also observed in previous observations using \Spitzer's IRAC Channel 2 \citep[e.g.,][]{Kreidberg2019, Challener2020}. For the majority of the observations, the target fell well within 0.5 pixels around the sweet spot. We removed an additional 0.3\% of exposures in the third AOR because these exposures had a centroid position shifted by a whole pixel in the x-dimension. 

\subsection{\Ktwo\ photometry}

\subsubsection{Campaign 12}
K2’s campaign 12 (C12) was carried out by the \Kepler\ space telescope from December 15, 2016, to March 04, 2017. The photometric data were taken in the long cadence sampling rate ($\sim$30 minutes) over a timespan of about 79 days. The data extraction and reduction is described in \citet{Malavolta2018} and was reused for this analysis.
\subsubsection{Campaign 19}
K2's campaign 19 (C19) was carried out from August 30, 2018 to September 26, 2018 and marked the last observations taken by the \Kepler\ space telescope before running out of fuel. C19's field-of-view overlapped with C12 leading to a reobservation of K2-141\footnote{The star was proposed to be observed in short cadence mode in K2 Campaign 19 from the following K2 General Observer programs: GO19027, A. Vanderburg; GO19065, C. Dressing}. The fuel shortage on the spacecraft, however, led to a shorter campaign. These last data taken by \Kepler\ span only about a month and suffer from erratic pointing at the beginning and end of the campaign. We removed approximately 54\% of the data leaving us with approximately 7 days of observations with a photometric precision comparable to other K2 Campaigns. Approximately 8.5 days at the beginning of the observations were removed because the boresight of the telescope was off-nominal leading to K2-141 being completely out of the pixel stamp. We also removed the final 11 days of the observations where the boresight and roll of \Kepler\ fluctuated erratically\footnote{see the Data Release Notes for K2 Campaign 19 for further information: \url{https://archive.stsci.edu/missions/k2/doc/drn/KSCI-19145-002_K2-DRN29_C19.pdf}}. In contrast to C12, the observations of K2-141 in C19 were taken in the short cadence mode with a sampling rate of about one minute. Due to this higher temporal resolution, there is no need to oversample the C19 data.

The photometry was accessed with the python package \texttt{lightkurve} \citep{Lightkurve2018}, which retrieves the data from the MAST archive\footnote{\url{https://archive.stsci.edu/k2/data_search/search.php}}. We downloaded the Simple Aperture Photometry (SAP) data and removed every measurement with a non-zero ``quality'' flag, which indicates events like thruster firings or cosmic ray hits. 

We used the Self-Flat-Fielding (SFF) procedure described in \citet{Vanderburg2014} and \citet{Vanderburg2016}, which is implemented in \texttt{lightkurve}, to correct for the systematic flux variations of the K2 data caused by thruster firings every six hours. In addition to the six hour back and forth movement throughout the K2 mission, there was also transverse spacecraft drift on timescales longer than 10 days \citep{Vanderburg2016}. We therefore subdivided the 8 days of data into two ``windows'' and performed the SFF independently in each of them. To remove outliers, we fitted a cubic spline to the data and performed an iterative sigma clipping with respect to the median to mask outliers at 5$\sigma$ below and 3$\sigma$ above the light curve, which removed 0.3\% of the remaining C19 data. We then normalized the data by dividing by the median of the flux.

To remove the stellar variability \citep[$P$$_{rot}$$~\sim$ 14 days,][]{Malavolta2018}, we fitted a linear function of time to the out-of-transit data for each orbit (eclipse to eclipse) while masking the transits following \citet{Sanchis-Ojeda2013}. For each orbit, the linear function was then subtracted from the data and unity was added. The observations contain 25 full phases (eclipse to eclipse) of the planet around the star. The C19 observations also contain one transit of K2-141\,c which we removed from our analysis, but use in Appendix \ref{sec:K2-141c_ephemeris} to improve the ephemeris of K2-141\,c. 

\begin{figure}
\centering
\includegraphics[width=0.5\textwidth]{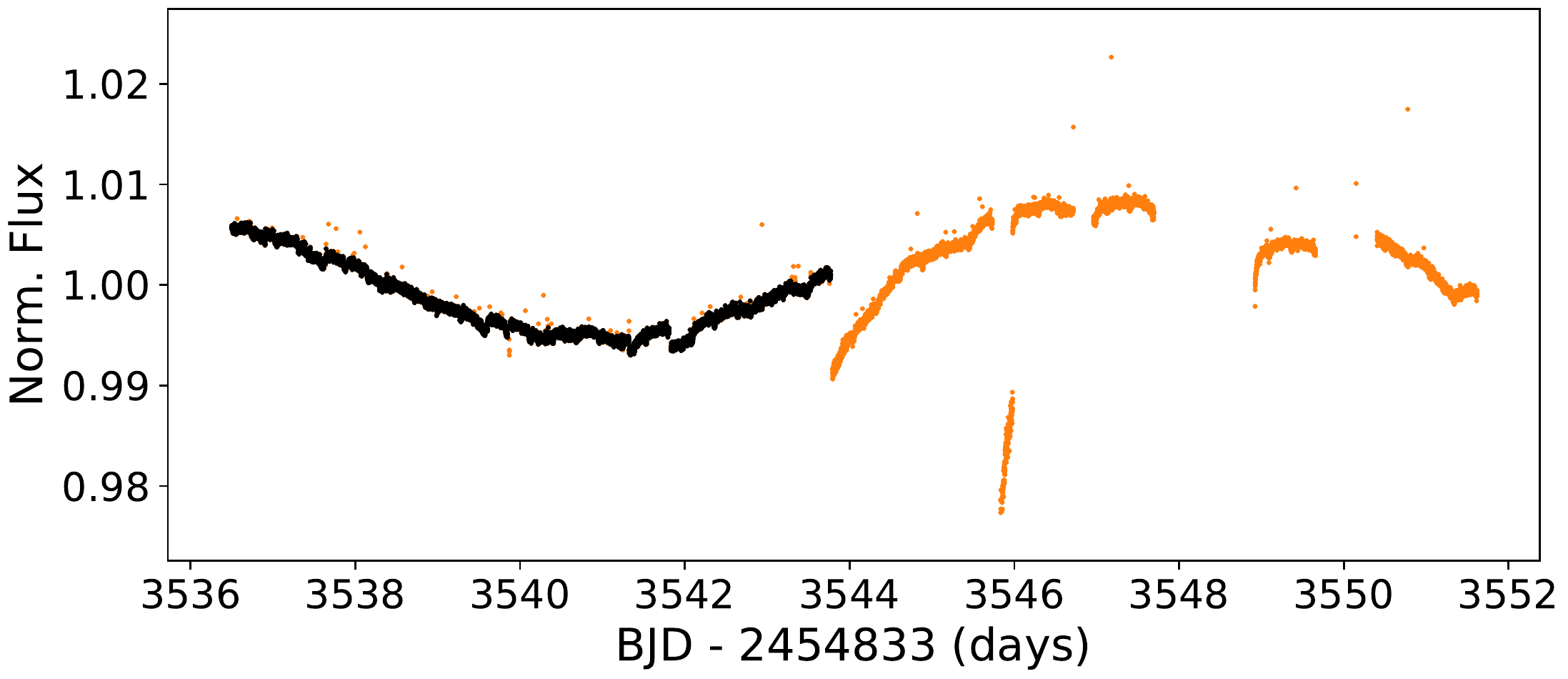}
\includegraphics[width=0.5\textwidth]{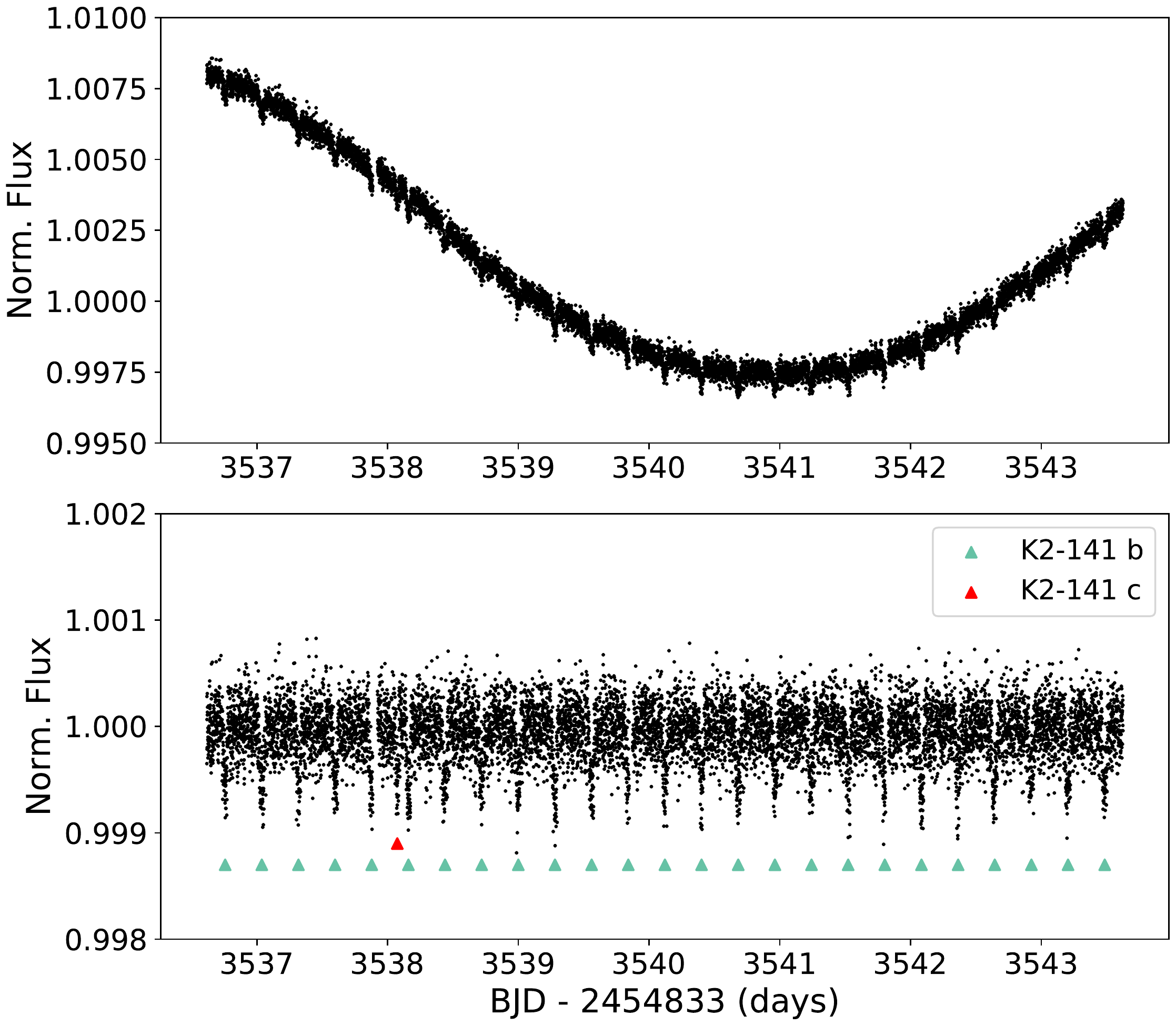}
\caption{The K2 Campaign 19 data of K2-141. \textit{Upper panel:} The full K2 Campaign 19 observations. We removed outliers and data suffering from decreased photometric precision due to \Kepler\ running out of fuel (marked in orange). \textit{Middle panel:} K2 C19 data with instrumental systematics removed using the SFF algorithm. \textit{Lower panel:} The K2 C19 observations after removal of the stellar variability. The 25 transits by K2-141\,b have been marked in green. An additional transit by K2-141\,c is denoted in red.}
\label{fig:C19}
\end{figure}

\section{Light curve fits} \label{sec:analysis}

We considered fits to the \Spitzer\ data alone (see Section \ref{sec:fitSpitzer}), and to the joint \Spitzer\ and \Kepler\ dataset (see Section \ref{sec:fitJoint}). For every model described in the following section we performed a differential evolution Markov chain Monte Carlo (MCMC) \citep{Braak2006} analysis to estimate parameter uncertainties. We rescaled the uncertainties for every data point by a constant factor so that the reduced chi-squared is unity and we get realistic uncertainties for the fitted parameters. The chi-squared values before rescaling based on the final best fitting model in our analysis can be found in Table \ref{tab:rms_all_AORS}. We ran the MCMC until all free parameters of four initialized chains converged to within 1\% of unity according to the Gelman Rubin statistic \citep{Gelman1992}. Each chain consisted out of 10000 steps and we discarded the first 50\% of the MCMCs as burn-in. This leaves us with a total of 20000 steps for each run. We include plots of the resulting posterior distributions in the Appendix (\ref{fig:corner1} to \ref{fig:corner6}). 

The transit model implemented in \texttt{BATMAN} \citep{Kreidberg2015} which was used in every fit, consists of the time of central transit $t_0$, the radius of planet in units of stellar radii $R_p/R_*$, the orbital period $P$, semi-major axis in units of stellar radii \ar\ and the cosine of the inclination $\cos i$. We fixed the eccentricity $ecc$ and the argument of periastron $\omega$ to zero. This is justified due to the very short circularization time scale of these USPs. Following equation 3 from \citet{Adams2006}, the circularization time scale for K2-141\,b is only about $\tau_{circ} = 3.1$ Myrs assuming a tidal quality factor of Q$_P = 10^6$. We used the \texttt{ExoCTK} limb darkening calculator \citep{Batalha2017} and the stellar parameters given in \citet{Malavolta2018} to determine and fix the linear and quadratic limb-darkening coefficients $u_1$ and $u_2$. They are $u_1=0.105$ and $u_2=0.119$ in the \Spitzer\ IRAC Channel 2 bandpass and $u_1=0.666$ and $u_2=0.062$ in the \Kepler\ bandpass.

\subsection{\Spitzer\ only fit}\label{sec:fitSpitzer}

\begin{figure}
\centering
\includegraphics[width=0.5\textwidth]{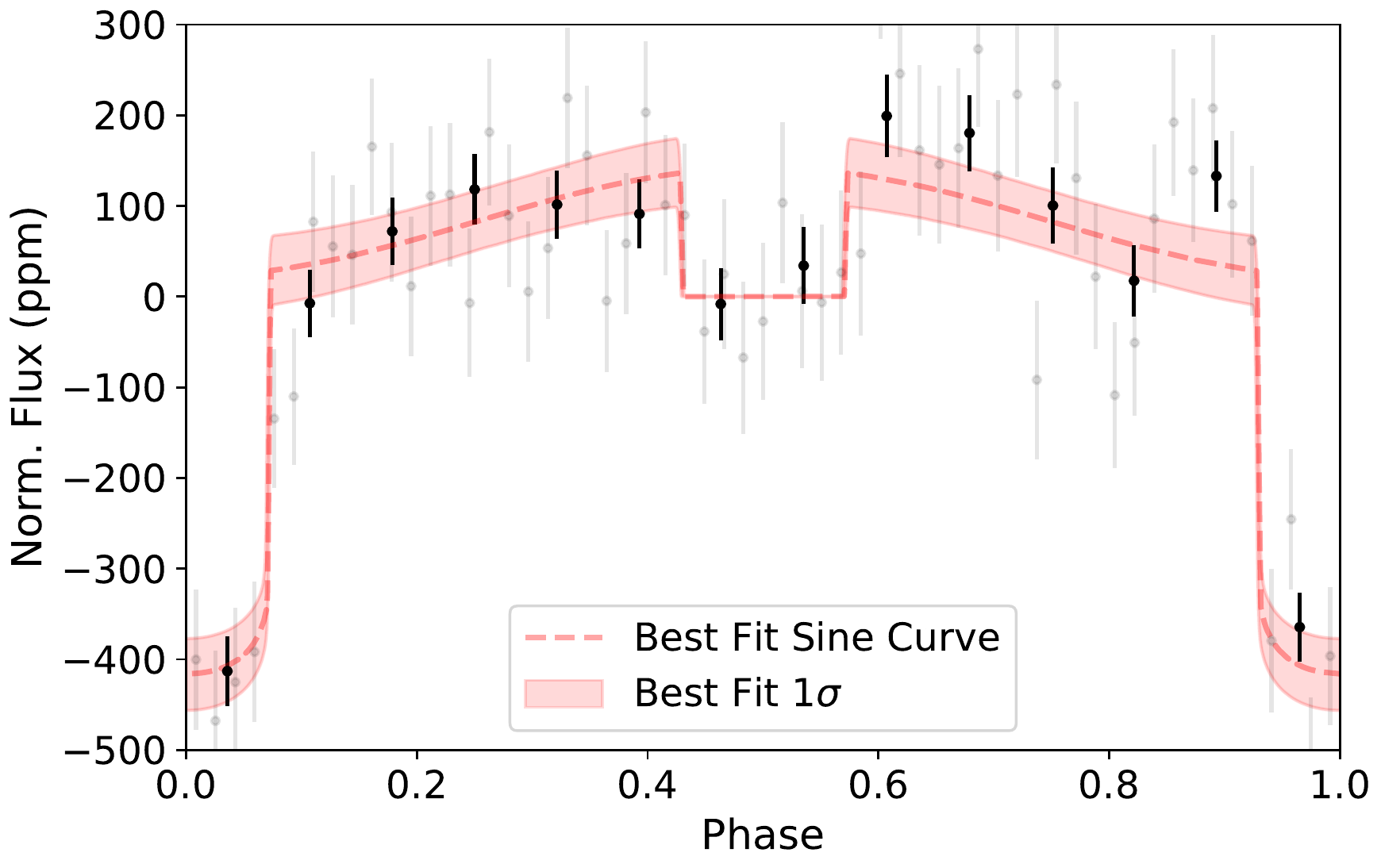}
\caption{Phase curve of K2-141\,b after phase folding the complete \Spitzer\ observations. The eclipse is at phase = 0.5. The best fitting model shown in red describes the observed thermal emission from the planet as a sinusoidal function. Each bin in black (grey) consists out of 7200 (1800) \Spitzer\ 2-second exposures. The pairs plot for this model can be found in the Appendix, Section \ref{fig:corner1}.}
\label{fig:spitzer_phasefold}
\end{figure}

We model the full \Spitzer\ light curve, $F(x,y,X,Y,t)$, as:

\begin{equation}
\label{eq:flux}
    F(x,y,X,Y,t) = F_s\,R(t)\,M(x,y)\,T(t)\,E(t)\,P(t)\,G(X,Y),
\end{equation}

\noindent where $F_s$ is the constant out-of-transit flux (i.e., the stellar flux), $R(t)$ is the ramp model, $M(x,y)$ is the BLISS map with $(x,y)$ describing the position of the star on the detector, $T(t)$ is the \citet{Mandel2002} transit model implemented in \texttt{BATMAN} \citep{Kreidberg2015}, $E(t)$ is the eclipse model implemented in \texttt{BATMAN} and $P(t)$ the phase variation in \texttt{SPIDERMAN} \citep{Louden2018} or \texttt{POET} \citet{Cubillos2013} (depending on which specific model was used). $G(X,Y)$ is a term fitting for variations in the pixel response function (PRF) using a 2D cubic \citep[i.e., PRF-FWHM deterending, see e.g., ][]{Lanotte2014, Mendonca2018, May2020} and has the following form:

\begin{equation}
\begin{split}
    G(&X,Y) = \\
    &= X_1\,s_x + X_2\,s_x^2 + X_3\,s_x^3 + Y_1\,s_y + Y_2\,s_y^2 + Y_3\,s_y^3 + 1,
\end{split}
\end{equation}

\noindent where $X_1$ ($Y_1$), $X_2$ ($Y_2$), and $X_3$ ($Y_3$) are the linear, quadratic and cubic coefficients in the $X$ ($Y$) dimension, respectively, and $s_x$ and $s_y$ the Gaussian widths of the pixel response function in the $X$ ($Y$) dimension, respectively. The optimal resolution for BLISS mapping, the ramp model $R(t)$ (either a constant or linear ramp) and the order of the PRF fit $G(X,Y)$ were independently determined by minimizing the Bayesian Information Criterion (BIC) for every AOR and are listed in Table \ref{tab:SpitzerAORs}. 

Equation \ref{eq:flux} describes the general model which was fitted to the \Spitzer\ data. For this model, the orbital period $P$ was fixed to the literature value reported in \citet{Malavolta2018}. For parameters which are more precisely determined by the K2 data, namely $t_0$, $R_p/R_*$, \ar\ and $\cos i$, we used Gaussian priors in our analysis based on the literature values. 

We fit three different phase variation models $P(t)$ to the \Spitzer\ data:
\begin{enumerate}
    \item a sinusoid with amplitude $A$ multiplied by an eclipse model with eclipse depth $f_p/f_*$ and including a phase offset $\phi$ of the hotspot
    \item same as 1. but without a phase offset
    \item a two temperature model for the planet with a constant temperature on the dayside, $T_{p,d}$, and on the nightside, $T_{p,n}$.
\end{enumerate}
A list of the free parameters for every model is listed in the Appendix (see Table \ref{tab:Spitzer_models}). 

A commonly approach for model selection in the exoplanet literature is using the Bayesian Information Criterion \citep[BIC,][]{Schwarz1978,Kass1995,Liddle2007}. It approximates the evidence $E$ and has the following form:

\begin{equation}
\begin{split}
\ln E \approx -\frac{1}{2} \rm{BIC} &= \ln \mathcal{L}_{max} - \frac{1}{2} k \ln N\\
\implies \rm{BIC} &= - 2 \ln \mathcal{L}_{max} + k \ln N
\end{split}
\end{equation}

\noindent
where $\mathcal{L}_{max}$ is the maximum likelihood of the model, $k$ the number of free parameters of the model and $N$ the number of data points. The BIC therefore penalizes models with more free parameters and the best-fitting model is the one with the lowest BIC (i.e., the largest evidence). We compare models by calculating $\Delta$BIC:

\begin{equation}
\Delta \rm{BIC}_i = \rm{BIC}_i - \rm{BIC_{min}},
\end{equation}

\noindent
with $\rm{BIC_{min}}$ being the smallest BIC of the set of models being compared. By taking the difference of $\rm{BIC}_i$ and $\rm{BIC_{min}}$, several constants cancel out and we are left with: \( \rm{BIC} = \chi^2 + k\,\ln N  \), which now includes the $\chi^2$ value of the model. When comparing two models, \citet{Kass1995} lists a $\Delta$BIC > 3.2, $\Delta$BIC > 10, $\Delta$BIC > 100 as being a substantial, strong, decisive evidence for the model with the lower BIC, respectively. 

The symmetric sinusoidal model with no hotspot offset is statistically substantially preferred with $\Delta$BIC $>$ 8.8 (see Table \ref{tab:Spitzer values}) compared to a sinusoid with an offset. We therefore find the data are consistent with the peak brightness occurring at the substellar point, in contrast to the prominent USP planet 55 Cnc\,e which has an eastward offset of $41^{\circ} \pm 12^{\circ}$ \citep{Demory2016a}. When we include an offset as a free parameter, we obtain $\phi = - 34.5_{-14.6}^{15.3}$ (the negative sign denotes an offset westwards from the substellar point). We measure an eclipse depth in the \Spitzer\ bandpass $f_p/f_* = 142.9_{-39.0}^{38.5}$ ppm and an amplitude variation $A = 120.6_{-43.0}^{42.3}$ ppm. Note, that the reflected light contribution at 4.5 \micron\ has not been subtracted from the brightness temperature computation. Table \ref{tab:Spitzer values} lists all best fit parameters and their uncertainties. A comparison of the BIC between the models showed that the sinusoidal model with no hotspot offset fits best to the \Spitzer\ data. We show the best fitting model with the \Spitzer\ observations in Fig. \ref{fig:spitzer_phasefold}.

\begin{table*}[]
\centering
\caption{All models fitted to the \Spitzer\ data alone.}

\renewcommand{\arraystretch}{1.6}

\begin{tabular}{ccc|ccc}\hline\hline
 &  & & \multicolumn{3}{c}{Model Name} \\
Parameter   & Units                              & Prior                         & Sinusoidal ($\phi$ = 0) & Sinusoidal ($\phi$ free)& Two Temp. \\ \hline
$t_0$       & BJD$_{\text{TDB}}$ - 2457744.0 d   & $\mathcal{N}$(0.07160,0.00022)            & $0.07191_{-0.00019}^{0.00019}$& $0.07189_{-0.00021}^{0.00020}$& $0.07191_{-0.00020}^{0.00019}$\\
$R_p/R_*$   & ---                                & $\mathcal{N}$(0.02037,0.00046)            & $0.02038_{-0.00042}^{0.00041}$& $0.02041_{-0.00041}^{0.00041}$& $0.02035_{-0.00040}^{0.00039}$ \\
$a/R_*$     & ---                                & $\mathcal{N}$(2.292,0.056)                & $2.278_{-0.038}^{0.040}$      & $2.277_{-0.040}^{0.041}$      & $2.278_{-0.040}^{0.039}$ \\
$\cos i$    & ---                                & $\mathcal{N}$(0.064532,0.064)             & $0.068_{-0.039}^{0.042}$      & $0.066_{-0.038}^{0.045}$      & $0.070_{-0.040}^{0.044}$ \\
$A$         & ppm                                & $\mathcal{U}$(0,400)                      & $120.6_{-43.0}^{42.3}$        & $142.2_{-43.1}^{42.7}$        & --- \\
$f_p/f_*$   & ppm                                & $\mathcal{U}$(0,400)                      & $142.9_{-39.0}^{38.5}$        & $144.7_{-39.0}^{39.4}$        & --- \\
$\phi$      & degrees                            & $\mathcal{U}$($-180^{\circ},180^{\circ}$) & ---                           & $-34.5_{-14.6}^{15.3}$         & --- \\
$T_*$       & K                                  & $\mathcal{N}$(4599,79)                    & ---                           & ---                           & $4602_{-78}^{76}$ \\
$T_{p,n}$   & K                                  & $\mathcal{U}$(0,4599)                     & ---                           & ---                           & $909_{-560}^{497}$ \\
$T_{p,d}$   & K                                  & $\mathcal{U}$(0,4599)                     & ---                           & ---                           & $2233_{-335}^{330}$ \\\hline
\multicolumn{3}{c}{$\Delta$BIC}                                                  & 0                             & 8.8                           & 9.6

\end{tabular}

\renewcommand{\arraystretch}{1}
\tablefoot{
The Gaussian priors on $t_0, R_p/R_*, \ar\, \cos i$ and $T_*$ are based on the values reported in \citet{Malavolta2018}. The orbital period $P$ was fixed in these runs to $P = 0.2803244$ days \citep{Malavolta2018}.
$\mathcal{N}$ and $\mathcal{U}$ denote a Gaussian and uniform prior, respectively.\\
}
\label{tab:Spitzer values}
\end{table*}

\subsubsection{Goodness of \Spitzer\ only fit}
\label{sec:goodness_spitzer}

\begin{figure}
\centering
\includegraphics[width=0.5\textwidth]{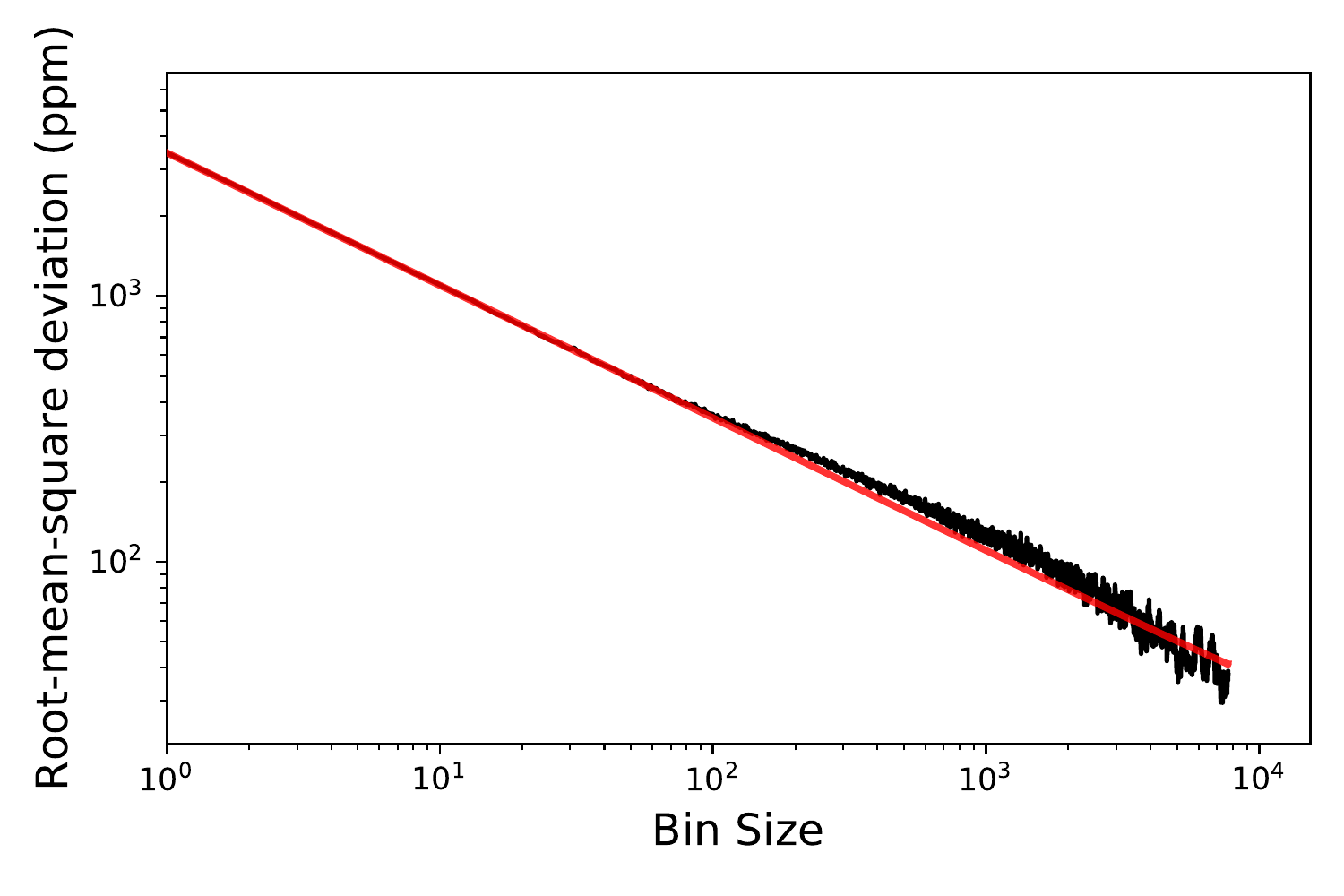}
\caption{Allan deviation plot of the \Spitzer\ data: root-mean-square (rms) of the fit residuals of the \Spitzer\ data using the sinusoidal model without a hotspot offset (black curve) as a function of the number of data points per bin. A bin size of one depicts no binning at all. The red line shows the expected rms for Gaussian noise following the inverse square root law.}
\label{fig:rms}
\end{figure}

We tested for the presence of red noise by comparing the root-mean-square (rms) of the binned residuals of the light curve with the predictions from white noise. If the data are uncorrelated (white) in time, the rms of the residuals is expected to decrease with $\sqrt N$, where $N$ is the number of data in a bin. A bin size of one, therefore, denotes no binning at all. We combined all 6 datasets and subtracted the best fitting model. Figure \ref{fig:rms} shows that the residuals of the \Spitzer\ data agree well with the expectations from uncorrelated noise. The same figures for each individual \Spitzer\ AOR can be found in Appendix \ref{sec:all_allan_spitzer}.

\subsection{Joint \Spitzer\ and \Kepler\ fit}\label{sec:fitJoint}

\begin{figure}
\centering
\includegraphics[width=0.5\textwidth]{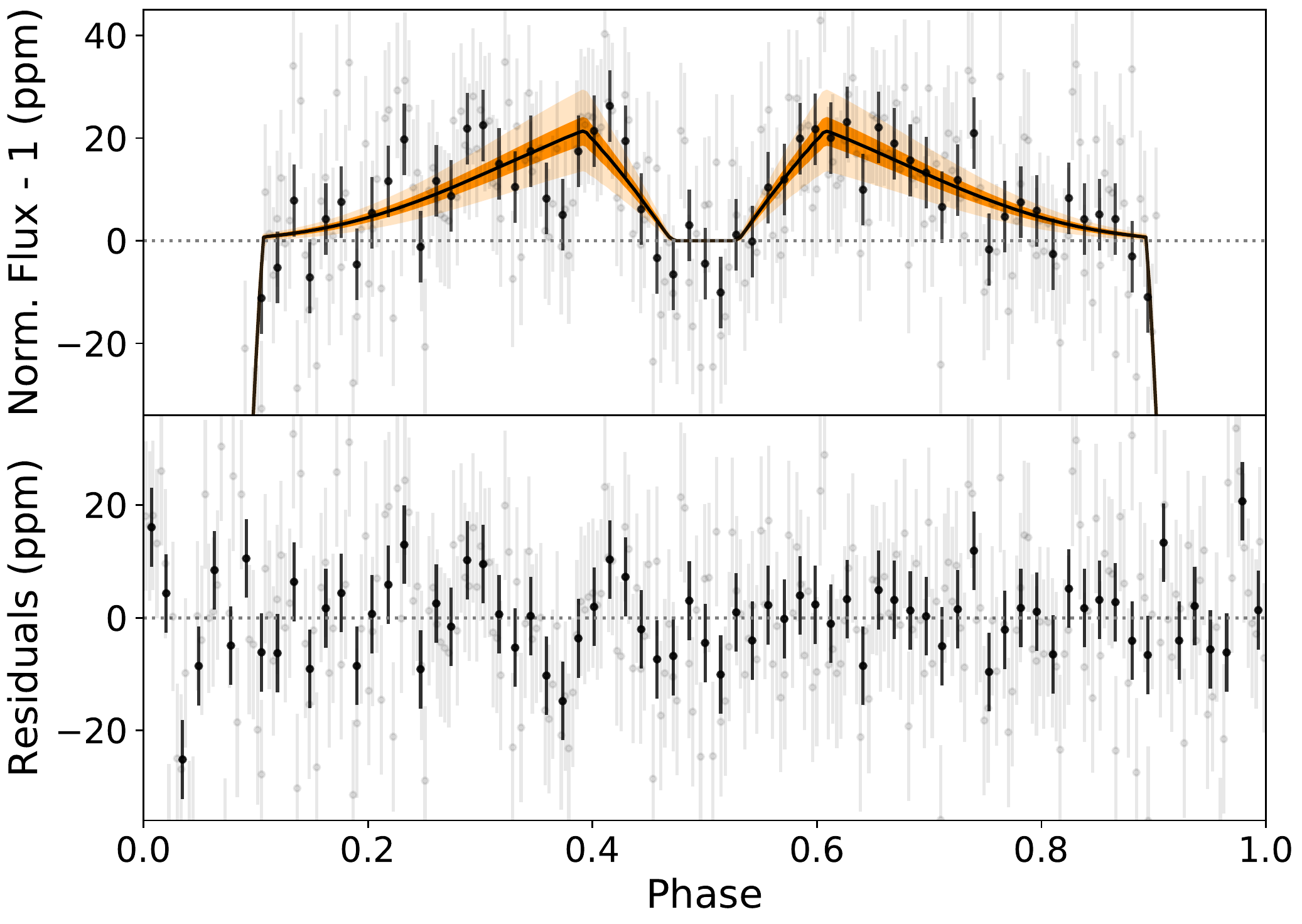}
\caption{Phase curve of K2-141\,b as seen by \Kepler\ during Campaign 12 and 19. \textit{Upper panel:} The solid black line shows the phase curve based on the values in our best fit model (toy model without heat redistribution for the thermal emission and a uniform Lambertian reflector for the reflected light contribution). The dark orange and light orange areas around the best fitting model are the 1$\sigma$ and 3$\sigma$ uncertainties, respectively. We binned the Campaign 19 data that was collected in short cadence mode ($\sim$ $1$ minute) into bins of 29.4 minutes to have the same temporal resolution as the Campaign 12 data. We then phase folded the \Kepler\ observations and binned the data for clarity. Each bin in black (grey) consists out of 180 (45) \Kepler\ exposures. \textit{Lower panel:} The residuals of the \Kepler\ observations to the best fitting model.}
\label{fig:kepler_phasefold}
\end{figure}

The phase variation $P(t)$ for the joint fit consists of a thermal $P_\textrm{therm}(t)$ and a reflective $P_\textrm{refl}(t)$ contribution, with $P(t) = P_\textrm{therm}(t) + P_\textrm{refl}(t)$. For the reflected light component $P_\textrm{refl}(t)$, we assumed a uniform Lambertian reflector \citep{Seager2010}:

\begin{equation}
    P_\textrm{refl}(t)=A_{\mathrm{g}} \frac{\sin z(t)+(\pi-z(t)) \cos z(t)}{\pi},
\end{equation}

\noindent where $A_g$ is the geometric albedo and the orbital phase z(t) is described by:

\begin{equation}
    z(t)=\arccos \left(-\sin i \cos \left(2 \pi \frac{t-t_0}{P}\right)\right).
\end{equation}

We fitted three thermal emission models to the combined \Spitzer\ and K2 dataset:

\begin{enumerate}
    \item A toy model described in \citet{Kreidberg2016} with the planet's heat redistribution $F$ as a free parameter
    \item same as 1. but with the redistribution fixed to zero leading to a nightside temperature of 0 K and 
    \item a two temperature model with a constant temperature on the dayside, $T_{p,d}$, and on the nightside, $T_{p,n}$.
\end{enumerate}

A list of the free parameters for every model is listed in the Appendix (see Table \ref{tab:joint_models}).  The toy heat redistribution model described in \citet{Kreidberg2016} expresses the temperature of the planet $T(z)$ as a function of the zenith angle $z$ using the following form:

\begin{equation}\label{eq:toymodelKreidberg}
\sigma T(z)^4 =
\begin{cases}
S \, (1 - A_B) \, F/2, & \rm nightside \\
S \, (1 - A_B) \, (F/2 + (1 - 2F)\cos{z}), & \rm dayside 
\end{cases}
\end{equation}
where $\sigma$ is the Stefan-Boltzmann constant, $S = \sigma \frac{\teff^4}{(\ar)^2}$ the insolation, $A_B$ the Bond albedo, $z$ the zenith angle and $0 < F < 0.5$ the heat redistribution parameter. For $F = 0$ no heat is being distributed and the nightside has a temperature of 0K. If $F = 0.5$, half of the energy received by the dayside is being transported to the nightside and the whole planet is isothermal.

With the full \Spitzer\ and \Kepler\ dataset, we now also fit for the orbital period $P$. We use Gaussian priors for $T_*$ and $a/R_*$  based on values reported in \citet{Malavolta2018}: The prior for the stellar temperature is $T_* = (4599 \pm 79)$ K and for semi-major axis in units of the stellar radius we use $a/R_* = 2.36 \pm 0.06$, which we derive from the stellar density $\rho_* = (2.244 \pm 0.161) \rho_o $ following $a/R_* ~\propto ~(\rho_* P^2)^{1/3}$. 

The K2 data in Campaign 12 was collected in the long cadence mode with a sampling rate of approximately 30 minutes. We oversample the data by a factor of 11 as in \citet{Malavolta2018} to account for the long exposure time. The data from Campaign 19 has a shorter sampling rate of about a minute and we therefore do not oversample this data set.

The insolation parameter $S$ for the toy model was calculated at every step in a self-consistent way, assuming $S ~ \propto ~ \teff^4 / (\ar)^2$. We fit for the stellar temperature in the toy model,  to take into account its uncertainty in the calculation of the insolation. In every step of the MCMC, we calculate a Kurucz model \citep{Kurucz1993} for the host star using the priors on the stellar temperature and stellar properties from \citet{Malavolta2018}.

We tested using separate geometric albedos for the \Kepler\ and the \Spitzer\ dataset ($A_{g, K2}$ and $A_{g, \Spitzer}$), but obtained a  uniform posterior distribution for $A_{g, \Spitzer}$, indicating that the \Spitzer\ data are not able to constrain the albedo at 4.5 $\mu$m (where thermal emission dominates). We therefore used a wavelength-independent geometric albedo $A_g$ in all subsequent fits. 

The toy model includes the Bond albedo as a parameter to regulate the radiation balance of the planet. Since we assume Lambertian reflection in our analysis, the Bond albedo $A_B$ and the geometric albedo $A_g$ are related by: $A_B$ = 3/2 $A_g$. While Lambertian reflectance is not an accurate model for the rocky bodies in the Solar System \citep{Mayorga2016}, this simplifying assumption is appropriate given the precision of our data and the unknown surface properties of K2-141\,b.

\begin{table*}[]
\centering
\caption{All models fitted to the joint \Spitzer\ and K2 dataset.}

\renewcommand{\arraystretch}{1.6}
\makebox[\linewidth]{
\begin{tabular}{ccc|ccc}\hline\hline
 &  &  & \multicolumn{3}{c}{Model Name} \\
Parameter & Unit                              & Prior                 & \tablefootmark{(1)}Toy Model ($F$=0)         & Toy Model ($F$ free)              & Two Temp. Model \\ \hline
$P$       & $P$ - 0.2803 d                    & $\mathcal{U}$($1.690\mathrm{e}{-5}, 3.190\mathrm{e}{-5}$) & $2.4956\mathrm{e}{-5}_{-0.0065\mathrm{e}{-5}}^{0.0067\mathrm{e}{-5}}$  & $2.4957\mathrm{e}{-5}_{-0.0069\mathrm{e}{-5}}^{0.0068\mathrm{e}{-5}}$  & $2.4955\mathrm{e}{-5}_{-0.0066\mathrm{e}{-5}}^{0.0068\mathrm{e}{-5}}$ \\
$t_0$     & BJD$_{\text{TDB}}$ - 2457744.0 d  & $\mathcal{U}$(0.07094, 0.07226) & $0.071508_{-0.000103}^{0.000103}$ & $0.071505_{-0.000103}^{0.000106}$ & $0.071505_{-0.000101}^{0.000108}$ \\
$R_p/R_*$ &  ---                              & $\mathcal{U}$(0.01807, 0.02267) & $0.02061_{-0.00013}^{0.00020}$    & $0.02065_{-0.00015}^{0.00020}$    & $0.02064_{-0.00015}^{0.00021}$ \\
$a/R_*$   & ---                               & $\mathcal{N}$(2.36, 0.06)         & $2.365_{-0.052}^{0.032}$          & $2.354_{-0.050}^{0.037}$          & $2.356_{-0.055}^{0.037}$ \\
$\cos i$  & ---                               & $\mathcal{U}$(0, 0.36975)       & $0.083_{-0.047}^{0.048}$          & $0.095_{-0.048}^{0.044}$          & $0.093_{0.052}^{0.047}$ \\
$F$       & ---                               & $\mathcal{U}$(0, 0.5)             & 0 (fixed)                         & $0.156_{-0.098}^{0.141}$          & --- \\
$T_*$     & K                                 & $\mathcal{N}$(4599, 79)           & $4593_{-81}^{80}$                 & $4603_{-79}^{80}$                 & $4604_{81}^{77}$ \\
$T_{p,n}$ & K                                 & $\mathcal{U}$(0, 4599)            & ---                               & ---                               & \begin{tabular}[c]{@{}c@{}}$956_{-556}^{489}$  \\(<1712K $2\sigma$, <2085K $3\sigma$)\end{tabular}\\
$T_{p,d}$ & K                                 & $\mathcal{U}$(0, 4599)            & ---                               & ---                               & \begin{tabular}[c]{@{}c@{}}$2049_{-359}^{362}$ \\(<2635 $2\sigma$, <2857K $3\sigma$)\end{tabular}\\
$A_g$     & ---                               & $\mathcal{U}$(0, 1)                & $0.282_{-0.078}^{0.070}$          & $0.298_{-0.068}^{0.062}$          & $0.308_{-0.071}^{0.057}$ \\\hline
\multicolumn{3}{c}{$\Delta$BIC}                                       & 0                                 & 12.0                              & 22.2

\end{tabular}
}
\renewcommand{\arraystretch}{1}

\tablefoot{
The uniform priors on $P, t_0, R_p/R_*, \cos i$ are based on the $5\sigma$ confidence interval of these parameters reported in \citet{Malavolta2018}. The Gaussian prior for $T_*$ and $a/R_*$ are also from \citet{Malavolta2018}. We derived the Gaussian prior for $a/R_*$ from the stellar density $\rho_* = (2.244 \pm 0.161) \rho_o $ following $a/R_* ~\propto ~(\rho_* P^2)^{1/3}$. $\mathcal{N}$ and $\mathcal{U}$ denote a Gaussian and uniform prior, respectively.\\
\tablefoottext{1}{The Toy model without any redistribution ($F=0$) provides the best fit to our data. We therefore recommend using the planetary parameters ($P, t_0, R_p/R_*, a/R_*, \cos i$) used in this column.}
}

\label{tab:Joint values}
\end{table*}

\begin{table}[]
\centering
\caption{Derived parameters for K2-141\,b from the best fitting model (Toy Model with $F$=0) presented in Table \ref{tab:Joint values}.}
\label{tab:derived_params}
\renewcommand{\arraystretch}{1.6}
\begin{tabular}{l|l|l}\hline\hline
Parameter & Unit & Value \\\hline
$i$                   & $^{\circ}$ & 85.2 $\pm$ 2.7 \\
$R_p$                 &\rearth     & 1.53 $\pm$ 0.04 \\
$\rho_p$              & g/cm$^3$   & 7.82 $\pm$ 0.90 \\
$f_p/f_{*,~\text{K2}}$ & ppm        & $26.4_{-2.5}^{3.5}$ \\
$T_{14}$              & hours      & $0.939_{-0.004}^{0.005}$ \\
$(R_p/R_*)^2$         & ppm        & $424.8_{-5.2}^{8.1}$
\end{tabular}
\renewcommand{\arraystretch}{1}
\tablefoot{
The calculations use the measured stellar radius and planetary mass reported in \citet{Malavolta2018}.
}

\end{table}

\begin{figure}
\centering
\includegraphics[width=0.5\textwidth]{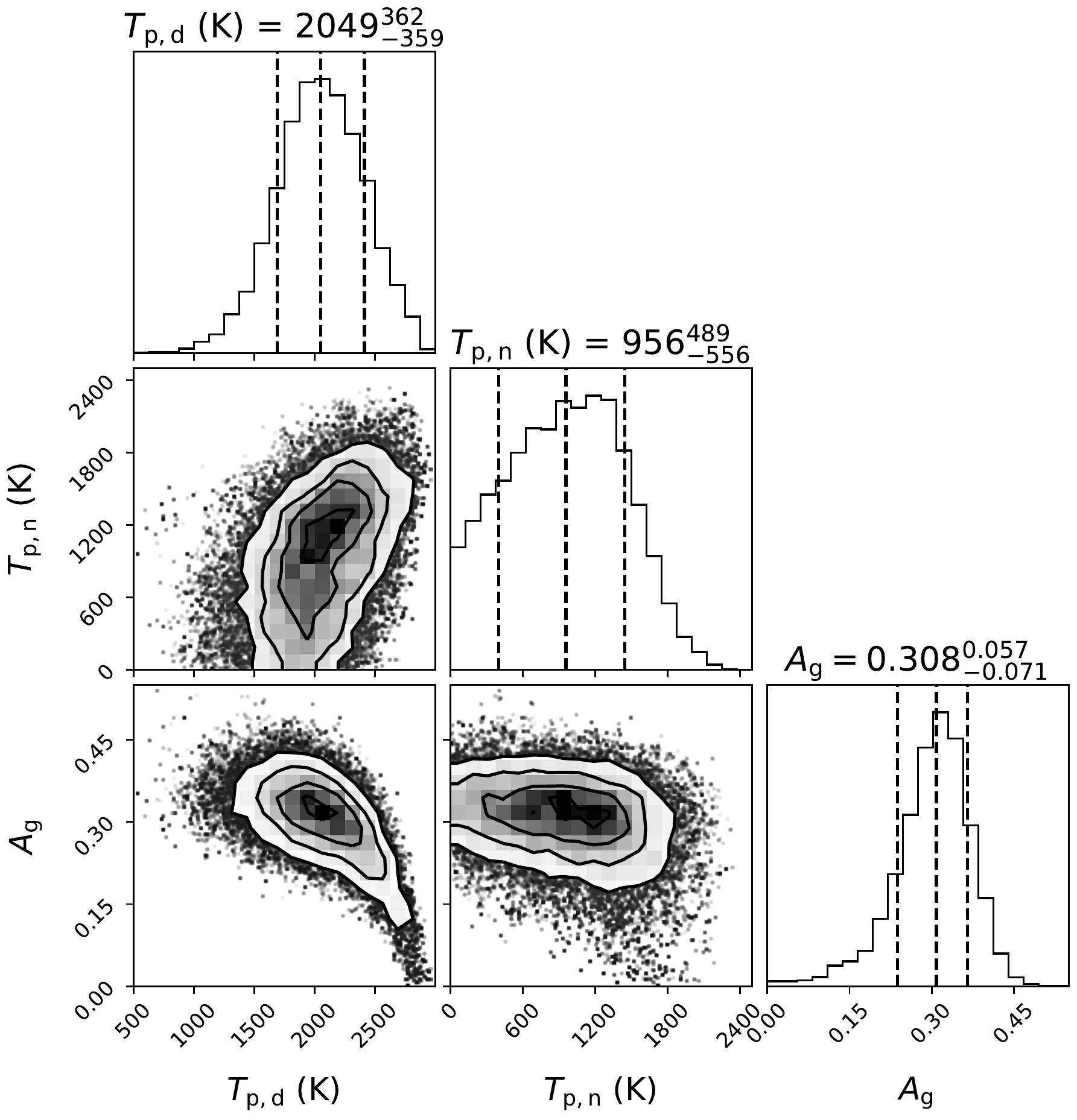}
\caption{Corner plot of the best fitting model to the joint K2 and \Spitzer\ dataset for the dayside temperature $T_{p,d}$, the nightside temperature $T_{p,n}$ and the geometric albedo $A_g$.}
\label{fig:small_corner}
\end{figure}

\subsubsection{Goodness of Joint \Spitzer\ and \Kepler\ fit}

\begin{figure*}
\centering
\includegraphics[width=0.46\textwidth]{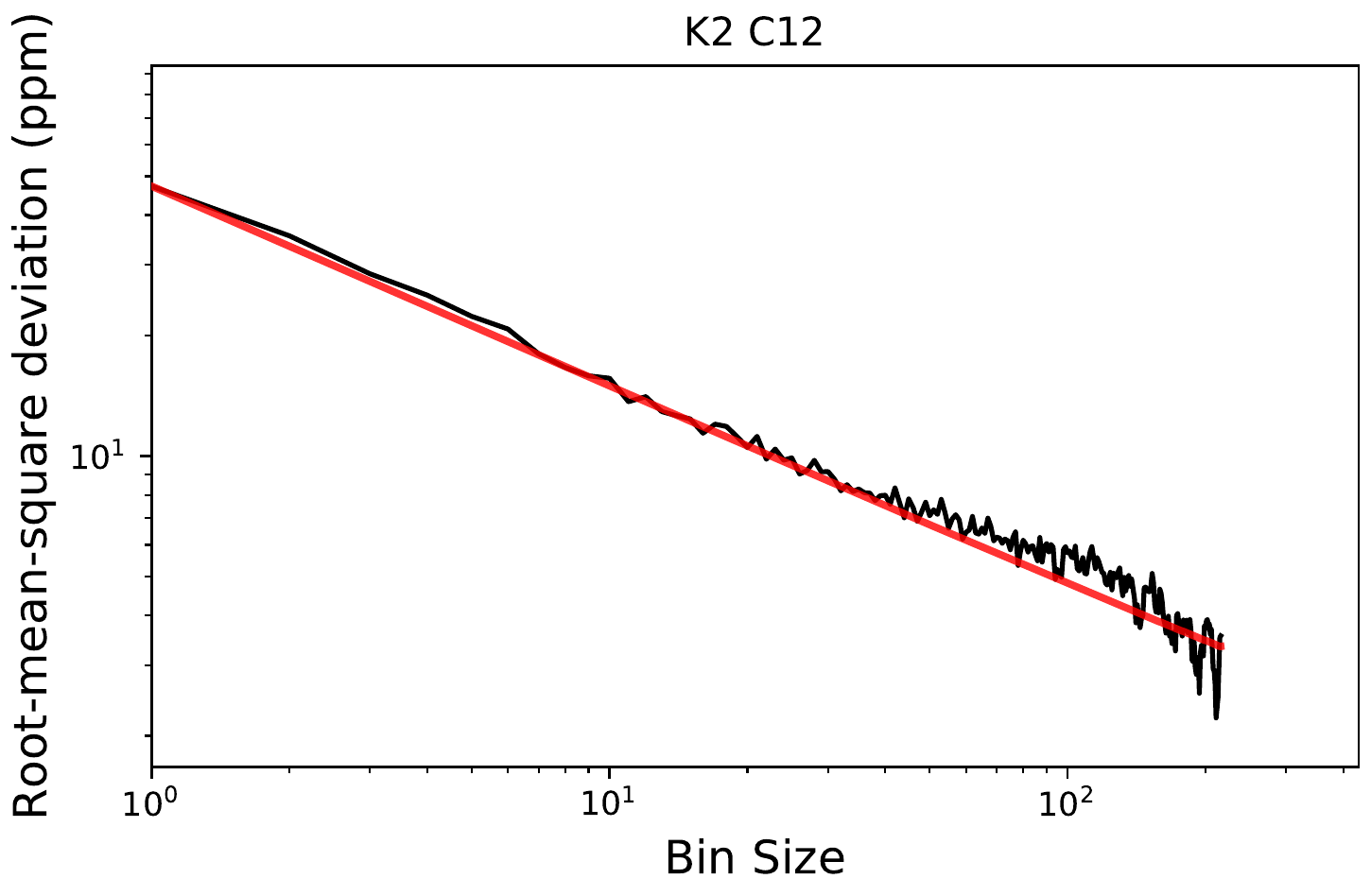}
\includegraphics[width=0.46\textwidth]{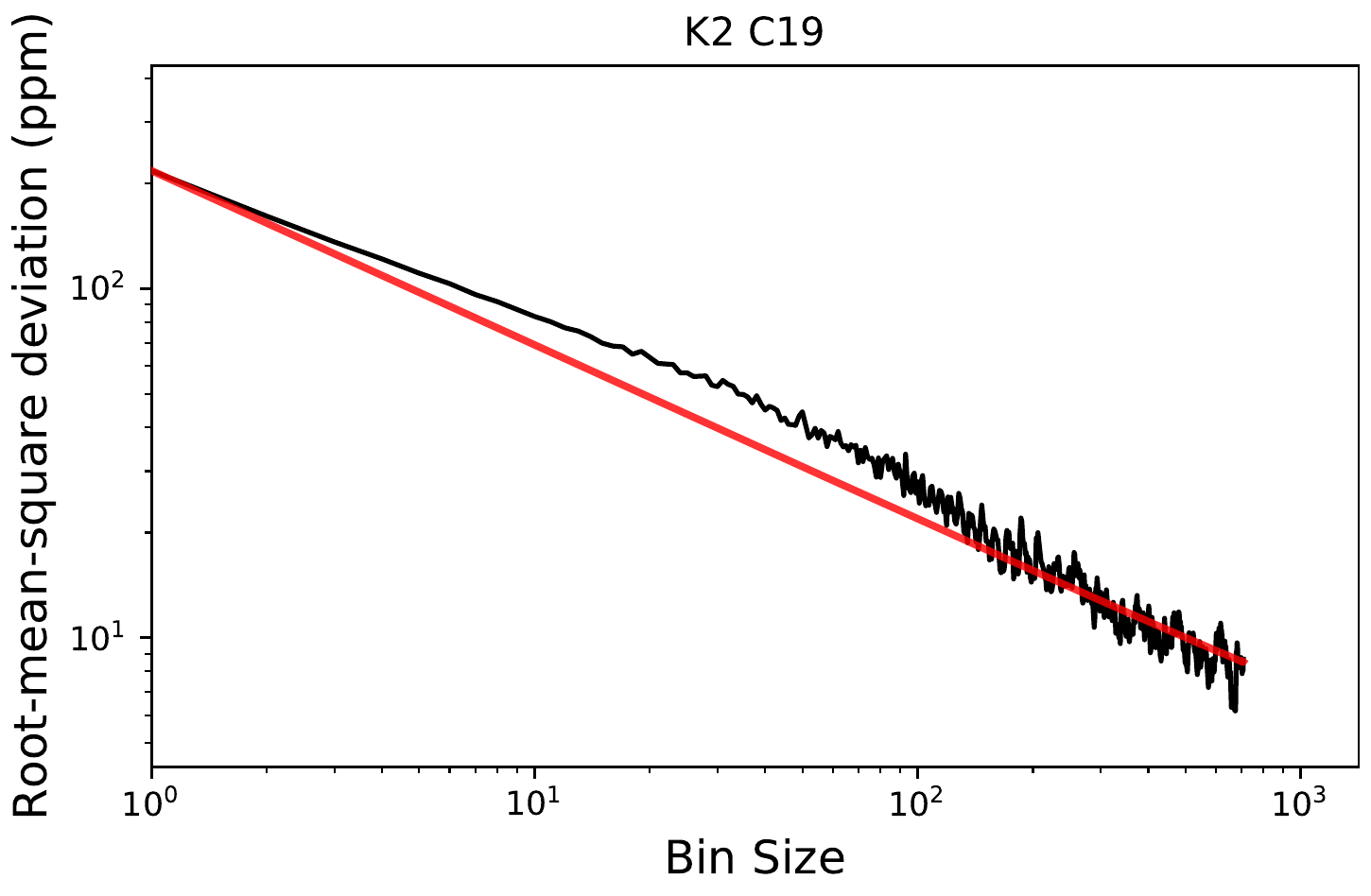}
\caption{Allan deviation plot for K2 C12 in the \textit{left panel} and K2 C19 in the \textit{right panel}.}
\label{fig:rms_kepler}
\end{figure*}

As in Section \ref{sec:goodness_spitzer} we took the observations and subtracted the best fitting model to compare the rms of the fit residuals with the expected rms for Gaussian noise. The \Spitzer\ data agrees again well with the expectations from uncorrelated noise and looks very similar to Figure \ref{fig:rms}. Figure \ref{fig:rms_kepler} shows the Allan deviation plots for the two K2 Campaigns. 

\subsection{Improved Ephemerides}

Having a precise ephemeris is crucial to schedule follow-up observations of a planet. K2-141\,b is an exciting target to be observed with observatories like \JWST. In fact, Cycle 1 of \JWST\ includes two programs to observe the planet \citep{Dang2021, Espinoza2021}. We improved the orbital period $P$ and the transit time $t_0$ significantly in our joint analysis using the three different datasets: the long cadence K2 C12 observations used in the discovery papers \citep{Malavolta2018, Barragan2018}, new short cadence observations during Campaign 19 of K2 and new \Spitzer\ observations. The updated parameters for $P$ and $t_0$ are listed in Table \ref{tab:K2-141b_ephemeris}. They are based on our joint fit using the toy model \citep{Kreidberg2016} with the heat redistribution $F$ set to zero as the resulting fit agrees best with our data. With the additional data, the $3\sigma$ uncertainty on the predicted transit time in 2024 decreases from about an hour to just 2.7 minutes. We also used the one transit of K2-141\,c observed in K2 C19 (see Figure \ref{fig:C19}) to improve the ephemeris of the planet. Future observers can use the updated $P$ and $t_0$ of K2-141\,c to avoid scheduling conflicts with planet b. The analysis for K2-141\,c can be found in Appendix \ref{sec:K2-141c_ephemeris}.

\begin{table}[]
\centering
\caption{Updated ephemeris for K2-141\,b and the $3\sigma$ uncertainty on the predicted transit time in 2022 and 2024.}
\label{tab:K2-141b_ephemeris}
\renewcommand{\arraystretch}{1.6}
\begin{tabular}{l|l|l}\hline\hline
K2-141\,b              & \tablefootmark{(1)}Discovery        & Updated                               \\\hline
$P$ (d)               & $0.2803244 \pm 0.0000015$  & $0.280324956_{-6.5e-08}^{6.7e-08}$ \\
\tablefootmark{(2)}$t_0$    & $7744.07160 \pm 0.00022$   & $7744.071508_{-0.000103}^{0.000103}$  \\
3$\sigma_{\rm{2022}}$ & 42 minutes                 & 1.9 minutes                            \\
3$\sigma_{\rm{2024}}$ & 59 minutes                 & 2.7 minutes                          
\end{tabular}
\renewcommand{\arraystretch}{1}
\tablefoot{
The same analysis for K2-141\,c can be found in the Appendix (see Table \ref{tab:K2-141c_ephemeris}).\\
\tablefoottext{1}{Based on \citet{Malavolta2018}}\\
\tablefoottext{2}{Expressed as BJD$_{\rm{TDB}}$ - 2450000.0 d}
}

\end{table}

\subsection{Results}

We performed three different fits for the \Spitzer\ data and for the joint dataset. We measured the eclipse depth in the \Spitzer\ bandpass $f_p/f_* = 142.9_{-39.0}^{38.5}$ ppm and an amplitude variation (peak to trough) $A = 120.6_{-43.0}^{42.3}$ ppm. The best fit is a two temperature model for the planet without a hotspot offset $\phi$. When we, however, let $\phi$ vary, we find a value of $\phi = - 34.5_{-14.6}^{15.3}$, which is at a 3.9$\sigma$ level strongly inconsistent with the value obtained for 55 Cancri e of $41^{\circ} \pm 12^{\circ}$ \citep{Demory2016a}. For the joint analysis (\Spitzer\ observations and the two K2 Campaigns) we find that a toy heat redistribution model from \citet{Kreidberg2016} without heat redistribution is most preferred. We measure a geometric albedo of $A_g = 0.282_{-0.078}^{0.070}$, a dayside temperature of $T_{p,d} = 2049_{-359}^{362}$K and a nightside temperature of $T_{p,n} = 956_{-556}^{489}$K (<1712K at $2\sigma$). We found an eclipse depth in the \Kepler\ bandpass of $f_p/f_* = 26.4_{-2.5}^{3.5}$ ppm which is consistent with the value reported in the discovery paper \citep[23 $\pm$ 4 ppm,][]{Malavolta2018}. We therefore robustly detect emission coming from the dayside of K2-141 b in the optical light. As a comparison, 55 Cnc e's secondary eclipse detection was only seen in the TESS observations at a significance of $3\sigma$ \citep{Kipping2020}. We show the best fitting model to the joint data set with the \Kepler\ observations in Fig. \ref{fig:kepler_phasefold}.

\section{Atmospheric Constraints} \label{sec:composition}

In addition to the toy models presented in Section 3, here  we compare the data to physically motivated models. K2-141\,b is expected to have a molten surface with a thin rock vapor atmosphere. To model the atmosphere, we used two different approaches: (1) a pseudo 2-D model that includes radiative transfer for plausible chemical species, and (2) 1D Turbulent Boundary Layer model that includes mass transfer between the planet's surface and the atmosphere.

\subsection{Pseudo-2D rock vapor model} \label{subsec:model1}

We calculated pseudo-2D models for the atmosphere by dividing the planet into concentric rings in 10 degree radial increments starting at the substellar point and finishing at a zenith angle of $80^{\circ}$ (for angles $>80^{\circ}$ the outgassed atmosphere becomes too tenuous resulting in numerical instabilities). 

This modeling approach is accurate in the limit that each column of atmosphere equilibrates locally with the magma ocean, without any influence of heat or mass transport from neighboring columns. For each increment, we calculated the outgassed chemistry and temperature-pressure structure of a gas-melt equilibrium atmosphere. Our outgassed elemental budget and atmospheric pressure are determined by the results of the melt-gas equilibrium code \texttt{MAGMA} \citep{Fegley1987,Schaefer2004}. This is done for a volatile-free komatiite\footnote{Komatiites are magnesium-rich, ultramafic lavas which formed on Earth during the Archaean (3.8 -- 2.5
billion years ago) when the Earth had higher surface temperatures \citep{McEwen1998, Schaefer2004}.} composition with no fractional vaporization (removal of vapour from the atmosphere) \citep{Schaefer2004,Miguel2011}. The outgassed chemistry and pressure are consistently adjusted for a surface temperature computed using radiative-transfer models, which are described below. Note that possible melt compositions for exoplanets are currently not known. Our choice of komatiite is based on early Earth \citep{Miguel2011}. Different melt compositions or evaporated atmospheres may result in chemistry and thermal structure changes (Zilinskas et al. in prep). 

Equilibrium gas chemistry in the atmosphere is computed using a thermochemical equilibrium model \texttt{FastChem}\footnote{\url{https://github.com/exoclime/FastChem}} \citep{Stock2018}. The chemistry considered includes over 30 different species for elements: O, Na, Si, Fe, Mg, K, Ti, Al, Ca and does not include ions. We do not consider the possible temporal evolution of chemistry through disequilibrium processes such as photochemistry or atmospheric mixing.

\begin{figure}
\centering
\includegraphics[width=0.5\textwidth]{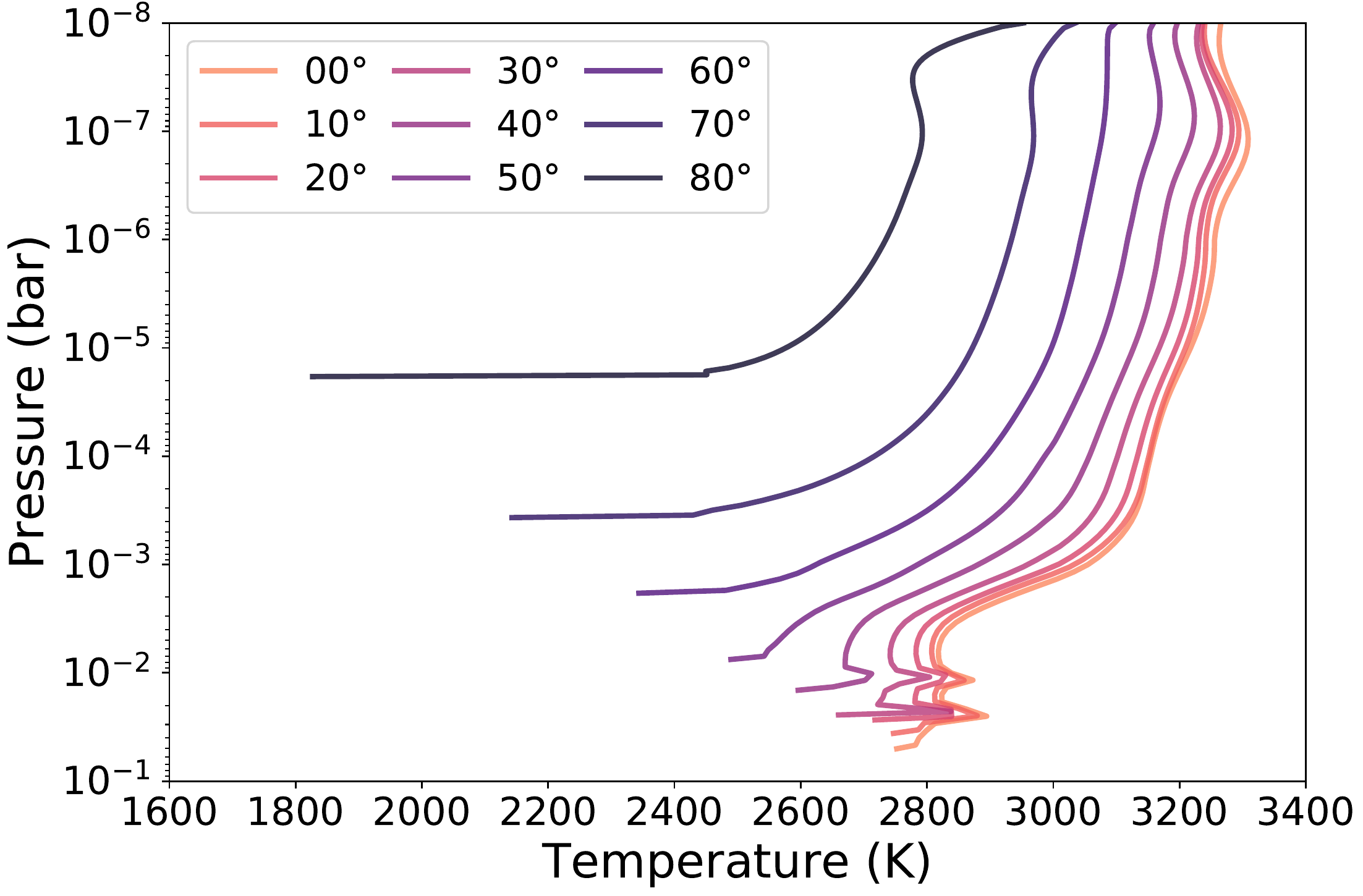}
\caption{T--P profile of K2-141\,b based on the pseudo-2D rock vapor model as described in Section \ref{subsec:model1} at different angles from the substellar point.}
\label{fig:tp_profile}
\end{figure}

The temperature profile of the atmosphere is modelled in a radiative-convective equilibrium using a radiative transfer code \texttt{HELIOS}\footnote{\url{https://github.com/exoclime/HELIOS}} \citep{Malik_2017,Malik_2019}. As absorbers we include Na and SiO, for which we use a sampling wavelength resolution of $\lambda/\Delta\lambda = 1000$ and a range of 0.06 -- 200 $\mu$m. Na opacity is computed using Vienna Atomic Line Database (VALD3) line list \citep{Ryab_2015}. We use the Voigt profile approximation for all, but the 0.6 $\mu$m doublet. The doublet is instead fitted using unified line-shape theory of \citet{Rossi_1985} and \citet{Allard_2007a,Allard_2007b}. The opacity of SiO is constructed using the EBJT \citep{Barton_2013} line list for ground state transitions and the Kurucz \citep{Kurucz_1992} line list for shortwave bands. For simplicity we assume null surface albedo and blackbody stellar irradiation, which may slightly overestimate incident shortwave flux. As with chemistry, temperature profiles, including the surface temperature, are consistently adjusted depending on the outgassed material. The temperature-pressure profiles of K2-141\,b at different zenith angles for this model are shown in Figure \ref{fig:tp_profile}. All zenith angles show a thermal inversion due to short-wavelength Na absorption, with a sharp increase in temperature starting at a few millibar. The amount of heating is sensitive to the UV spectrum of the star, which is unknown; however, in general thermal inversions should be expected in rock vapor atmospheres \citep{Zilinskas2021}. Future UV characterization of K2-141 would refine the theoretical predictions of the temperature structure.

To simulate emission spectra for each radial segment we use the radiative-transfer code \texttt{petitRADTRANS}\footnote{\url{https://gitlab.com/mauricemolli/petitRADTRANS}} \citep{Molliere_2019} with the same wavelength resolution and opacities as for the T-P profile calculation. We sum the fluxes weighted by the area of each concentric ring to calculate the total flux from the planet. Finally, we divide the planet flux by a PHOENIX stellar spectrum \citep{Husser_2013} to determine the planet-to-star contrast.

\begin{figure*}
\centering
\includegraphics[width=0.49\textwidth]{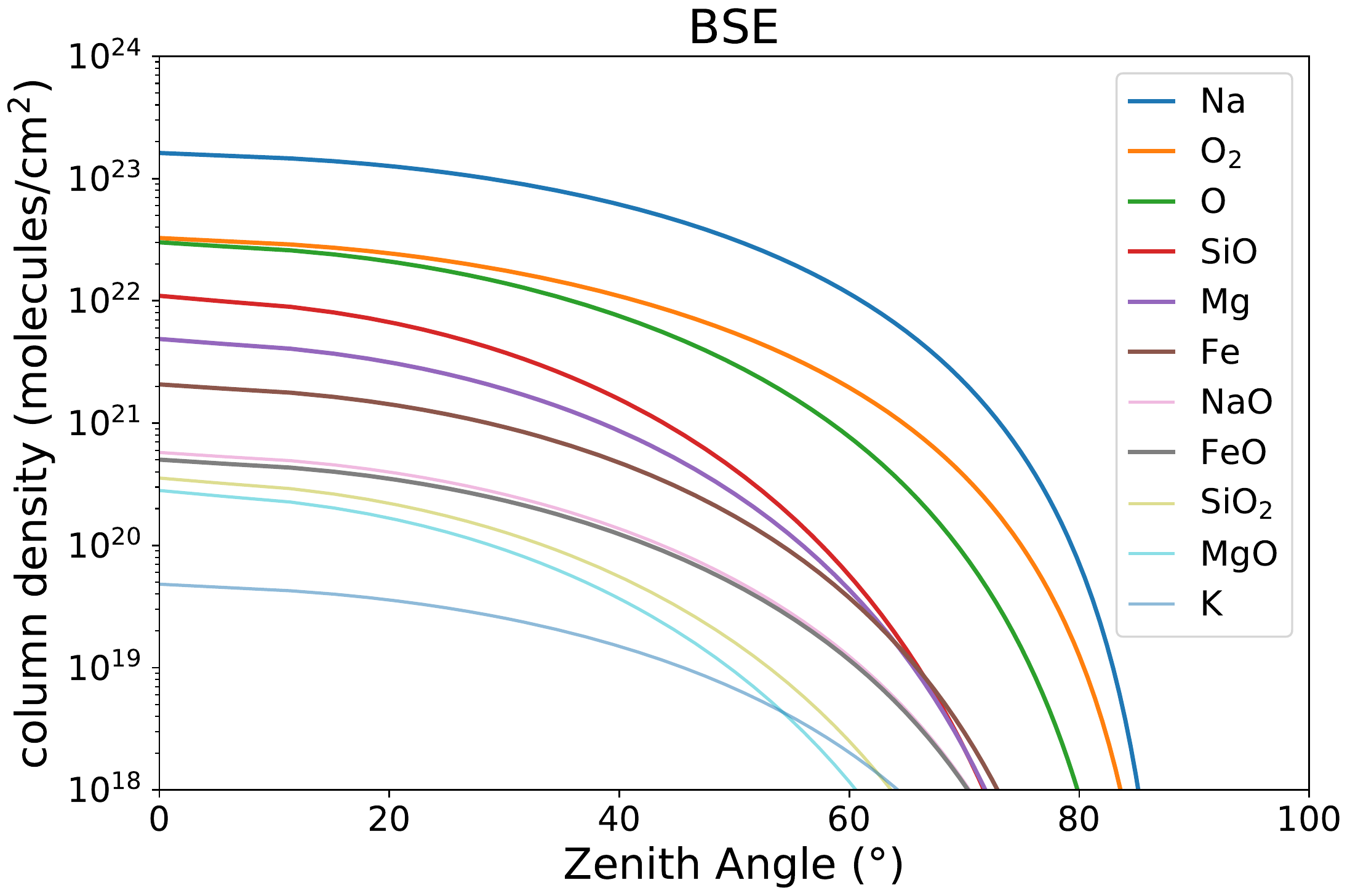}
\includegraphics[width=0.49\textwidth]{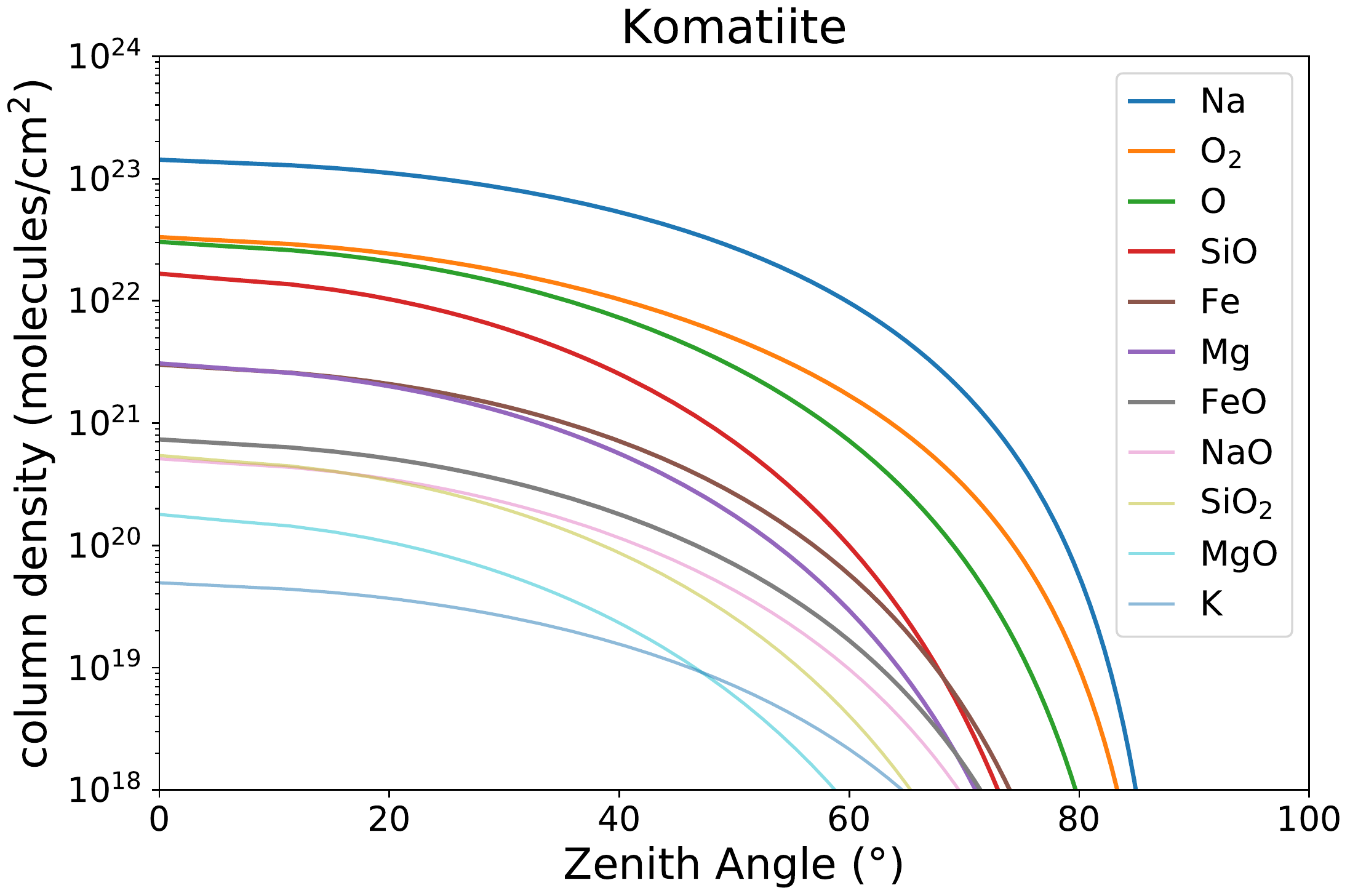}
 \caption{Column density of various species as a function of the zenith angle of the planet for a Bulk-Silicate-Earth composition (\textit{left panel}) and a komitiite composition (\textit{right panel}) based on calculations in \citet{Miguel2011}. The temperature as a function of zenith angle assumed for this plot is based on the our best fit model presented in \citet{Kreidberg2016} without heat redistribution ($F$ = 0). It assumes that the temperature at the substellar point is $T_{\text{substellar}} = (\frac{S (1-A_B)}{\sigma})^{1/4} = \frac{\teff}{\sqrt{\ar}} (1-A_B)^{1/4}$ and $T_{\text{terminator}} = 0$ K at a zenith angle of $z = 90^{\circ}$, with $S$ being the insolation, $A_B$ the Bond albedo and $\sigma$ the Stefan–Boltzmann constant. The species in the legend are sorted by the column density at the substellar point in descending order.}
\label{fig:columndensity}
\end{figure*}

Equilibrium gas chemistry in the atmosphere leads to a decreasing surface pressure with zenith angle. Figure \ref{fig:columndensity} shows the column density as a function of zenith angle for different species expected at the temperatures of K2-141\,b based on calculations in \citet{Miguel2011}. We show the densities for a Bulk-Silicate-Earth composition and a komatiite composition which show similar results with Na being the most abundant in both of them.

\subsection{1D Turbulent Boundary Layer model}

In addition to the pseudo-2D model described above, we also computed an atmospheric circulation model following \cite{Nguyen2020} and \cite{Castan2011}. The model calculates the steady-state flow induced by constant evaporation on the dayside and condensation on the nightside. Being tidally locked, we can impose symmetry across the substellar-antistellar axis by neglecting Coriolis forces.  By assuming a turbulent boundary layer (TBL), we can marginalize over the vertical dimension and further reduce the problem to 1D: distance from the sub-stellar point.

This model assumes a boundary layer that is: hydrostatically-bound and behaves like a continuous fluid (atmosphere does not escape K2-141\,b and we can apply fluid dynamic equations), turbulent (for vertically-constant wind speeds), and optically thin (no radiative transfer necessary). With these assumptions, we can construct a system of differential equations similar to the shallow-water equations which calculate the atmospheric pressure, wind velocity, and temperature at the boundary layer. We can only reduce the vertical dependence by assuming a vertical temperature profile.

The model itself describes the conservation of mass, momentum, and energy and their interactions: the atmospheric flow is being pushed by the pressure-gradient (momentum balance) driven by the uneven evaporation and condensation (mass balance), bringing with it sensible heating and cooling (energy balance) across the planet's surface which in turns affect the evaporation/condensation. A solution is found when the pressure, temperature, and wind speed obey the conservation of mass, momentum, and energy and a steady-state flow exists.

Recent progress in these types of model have been made by including radiative transfer, in a three-band scheme (UV, optical and IR), for an SiO-dominated atmosphere \citep{Nguyen2022}. SiO absorbs strongly in the UV, which causes upper-level atmospheric heating, possibly leading to a temperature inversion \citep[][]{Ito2015}. Therefore, the updated TBL model tests different vertical temperature profiles: adiabatic, isothermal, and inverted. Finally, coupling the radiative budgets of the atmosphere and surface, we can calculate emission spectra and phase variations for K2-141\,b.

The different temperature profiles lead to significant changes to the dynamics. Making the lapse rate negative (temperature increases with height) increases the horizontal pressure gradient force which induces stronger winds. However, the energy budget is unchanged as incoming stellar flux does not depend on the temperature profile used. Therefore, the atmosphere reacts to the increased kinetic energy by lowering its thermal energy, leading to overall cooler temperatures.

\subsection{Comparison of the models}

\begin{figure}
\centering
\includegraphics[width=0.5\textwidth]{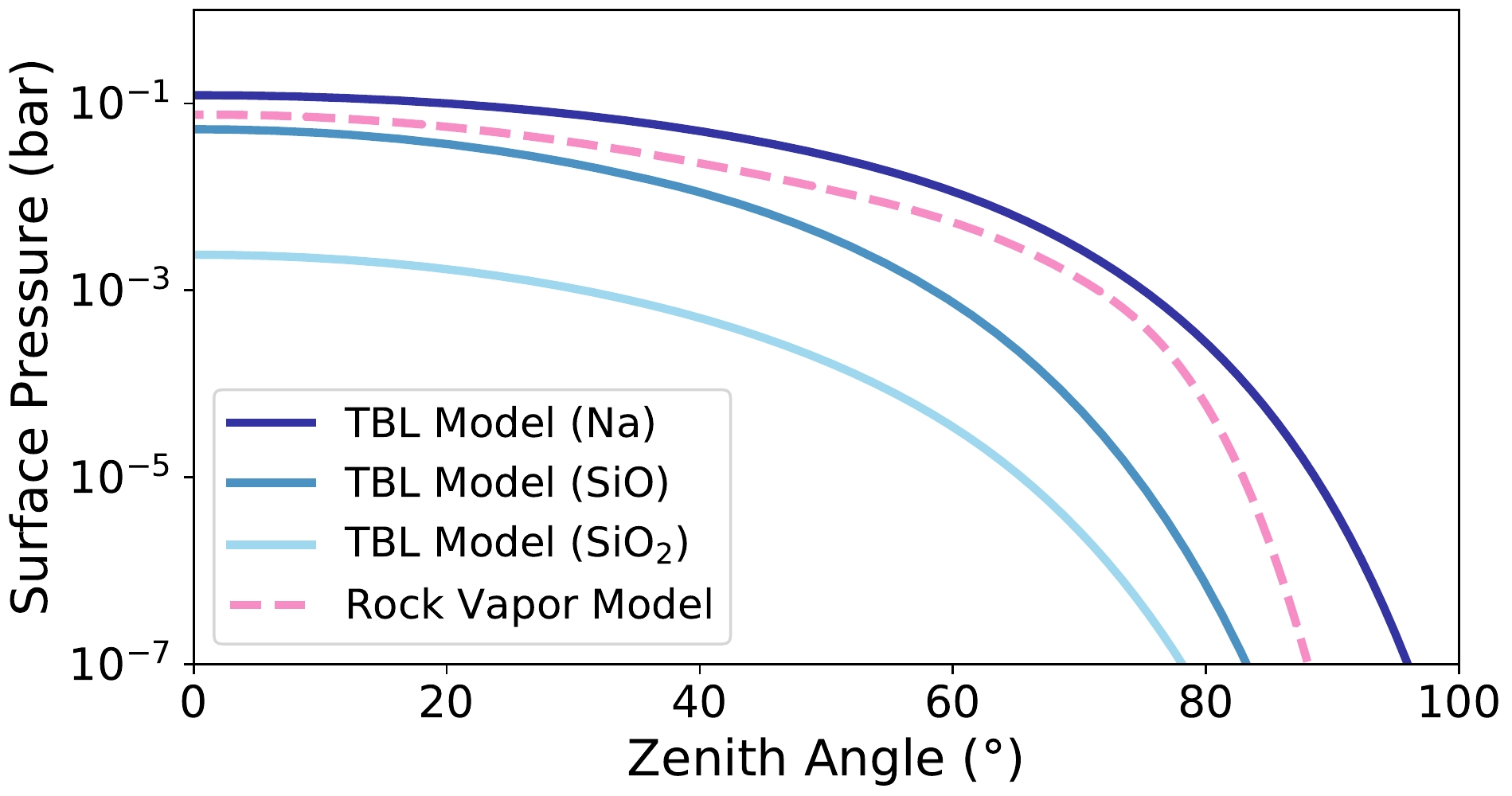}
\caption{Surface pressure as a function of the zenith angle for the two different models used in this paper. The dashed pink line is based on the pseudo-2D rock vapor model and the blue lines are based on the 1D TBL model for a Na, SiO and SiO$_2$ atmosphere. Although, the TBL models shown in this plot are based on \citet{Nguyen2020} and lack radiative transfer, this would introduce little to no changes to the surface pressure. Na is the most volatile possible component of rock vapor atmosphere, so a pure Na atmosphere has the highest surface pressure \citep{Schaefer2004}.}
\label{fig:surfacepressure}
\end{figure}

We calculated the surface pressure for both models as a function of zenith angle and show the results in Figure \ref{fig:surfacepressure}. One can see that due to the lack of atmospheric circulation in the pseudo-2D rock vapor model the surface pressure does not drop off as quickly with zenith angle as in the 1D Turbulent Boundary Layer (TBL) model. This indicates that the pseudo-2D rock vapor model a reasonably good approximation to the 1D TBL model due to the overall atmospheric circulation being low. 

\subsection{Comparison between the models and the data}

We compared both the physically motivated models and the toy models to the the measured dayside emission spectrum and the full  phase curves.

\begin{figure}
\centering
\includegraphics[width=0.5\textwidth]{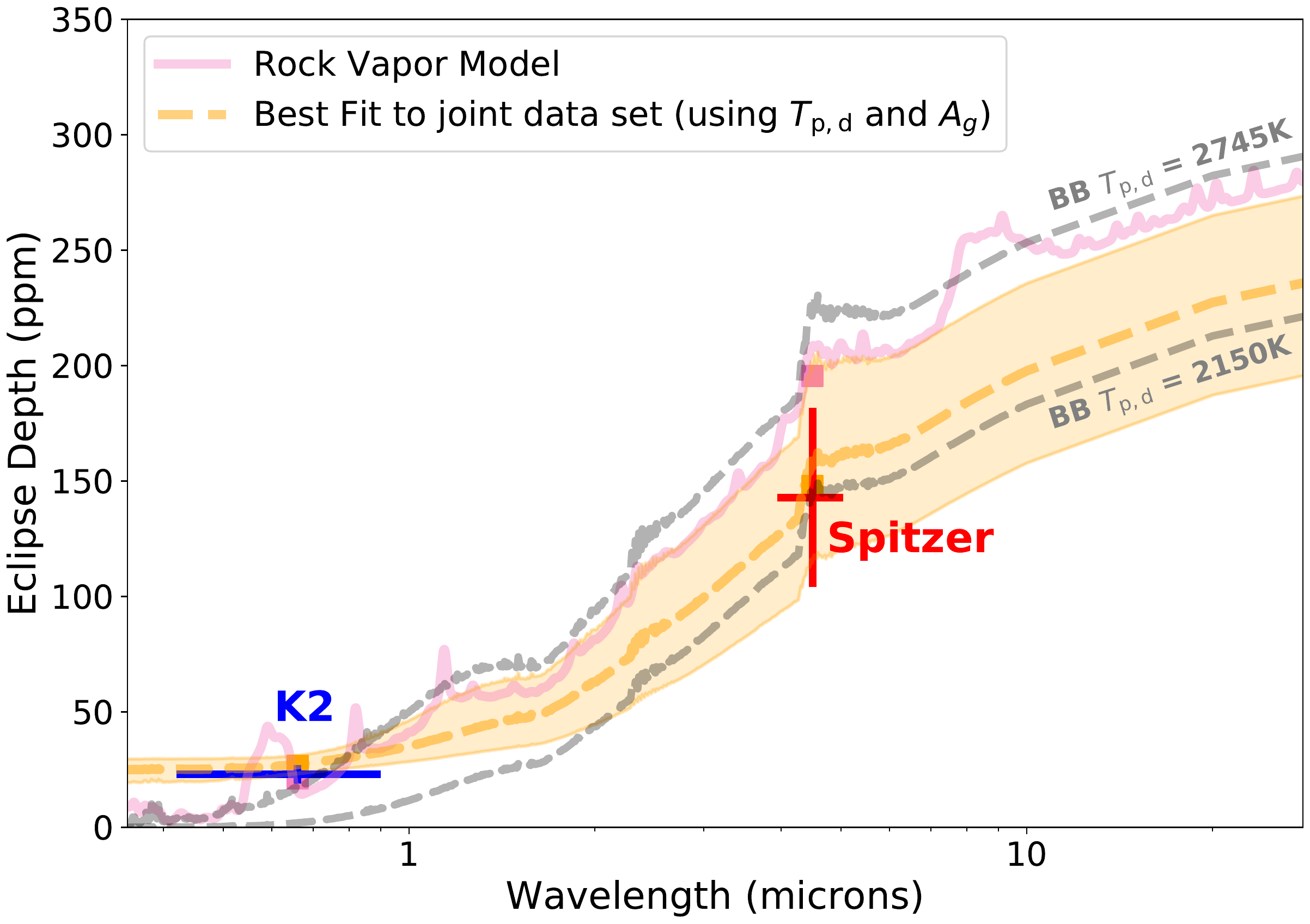}
\caption{Eclipse depths measured in the \Kepler\ and \Spitzer\ bandpasses compared to different emission spectra of the planet: the pink solid line was calculated using the pseudo-2D rock vapor model and shows thermal emissions in the optical due to Na. The dashed orange line uses the best fit dayside temperature $T_{p,d}$ and geometric albedo $A_g$ values using the joint dataset (\Spitzer\ \& K2). The orange shaded area is due to the uncertainties in $T_{p,d}$ and $A_g$. We also show black body (BB) emission spectra in gray for two exemplary dayside temperatures of K2-141\,b assuming a geometric albedo of $A_g$ = 0: $T_{p,d}$ = 2150K that corresponds to a full redistribution of heat on the planet (whole planet is isothermal) and $T_{p,d}$ = 2745K in case of instant reradiation of incoming energy (nightside temperature is zero). Any features in the black body spectra are originating from the host star because we divide the black body spectrum of the planet with a Kurucz stellar spectrum \citep{Kurucz1993}. The pink (orange) boxes show the \Spitzer\ and K2 bandpass integrated eclipse depth for the rock vapor model (sum of the thermal and reflective emission from the best fit to the joint data set).}
\label{fig:emission}
\end{figure}

The thermal emission spectrum of K2-141\,b  consists of the two broad photometric bands for K2 and \Spitzer\ IRAC Channel 2, as shown in Fig. \ref{fig:emission}. The two photometic band measurements are both consistent within two sigma with the pseudo-2D rock vapor atmosphere model and the best-fit toy model to the joint dataset, where the planet was modeled by a two temperature model and a Lambertian reflective law. As illustrated in Fig. \ref{fig:emission}, both models produce a larger eclipse depth at optical wavelengths than a single temperature blackbody. In the case of the toy model, this eclipse depth is due to reflected light from a moderately high albedo ($A_g = 0.282_{-0.078}^{0.070}$). For the rock vapor model, the eclipse depth in the K2 bandpass is dominated by thermal emission from a high-temperature inversion layer in the atmosphere. Note, that the blackbody spectra in the Figure have been divided by a Kurucz stellar spectrum \citep{Kurucz1993}; any features in the black body spectra are therefore originating from the host star.
As noted by \cite{Ito2015} for 55 Cnc\,e, the strong UV heating of the atmosphere, combined with relatively weak IR radiative cooling, leads to an inversion that extends all the way to the ground, suppressing convection.  This is a potentially important feature for interpreting infrared emission data for lava planets because the inversion makes the atmosphere considerably hotter than the planet's surface. There are strong absorption features from Na in the optical, and SiO in the infrared, so the emission in both of our photometric bands largely comes from the atmosphere rather than the surface.

\begin{figure}
\centering
\includegraphics[width=0.5\textwidth]{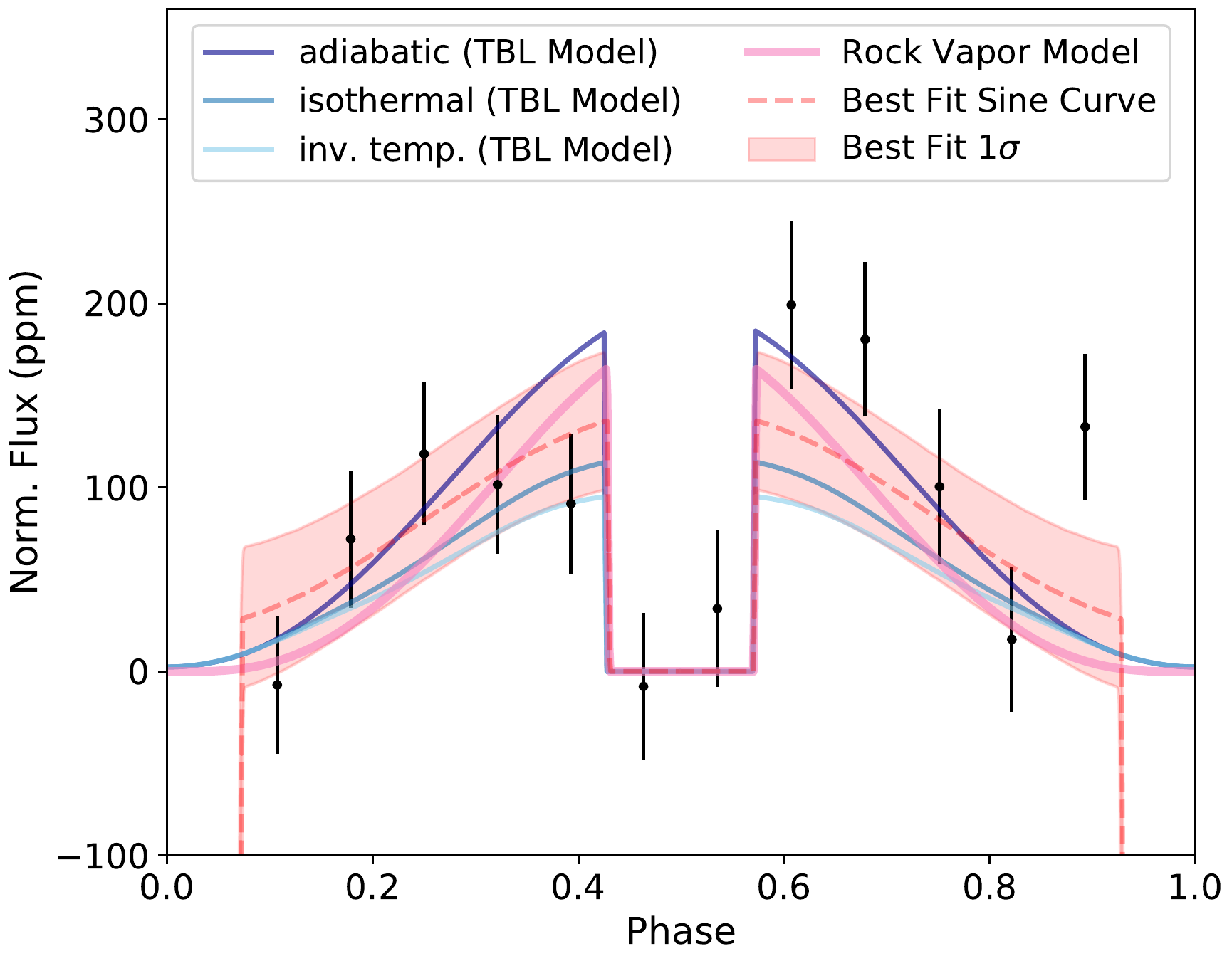}
\caption{Observed \Spitzer\ phase curve and the best fit sine curve to the \Spitzer\ data (red dashed line with the one sigma uncertainty shaded in red) compared to four theoretical phase curves: The different three blue solid lines are phase curves for the 1D TBL model assuming an adiabatic, an isothermal and an temperature inversion case. We also generated a phase curve out of the pseudo-2D rock vapor model (solid pink line). The adiabatic, isothermal and rock vapor model all fit the observations well and the temperature inversion model provides the worst fit.}
\label{fig:3dmodel_phasecurves}
\end{figure}

We also compared the measured phase curves to a range of models. For the 1D TBL model, we computed three different phase curves assuming an adiabatic, an isothermal and an temperature inversion case. We furthermore used the Open Source package \texttt{SPIDERMAN} \citep{Louden2018} to convert the emitted flux coming from each concentric ring in the pseudo-2D rock vapor model and generated a phase curve. Figure \ref{fig:3dmodel_phasecurves} shows the comparison of the \Spitzer\ data to these different phase curves. The adiabatic TBL model and rock vapor model both compare well to the data with the adiabatic model agreeing best. The temperature inversion model provides the worst fit to the data with $\Delta$BIC = 7.3 relative to the adiabatic model \citep[$\Delta$BIC > 3.2 (> 10) is a substantial (strong) evidence for the model with the lower BIC;][]{Kass1995}. The isothermal and rock vapor model have $\Delta$BIC = 3.3 and $\Delta$BIC = 2.4, respectively.

\section{Discussion} \label{sec:discussion}

What sets K2-141\,b apart from previously studied USPs is that it is the first with detected phase variation and secondary eclipse at optical and infrared wavelengths, enabling unique constraints on its atmospheric properties. By comparing the joint K2 and \Spitzer\ datasets with a range of toy and physically-motivated models, we find that a thick atmosphere is disfavored, but a rock vapor  atmosphere provides a good explanation to all available data.

\subsection{Evidence against a thick atmosphere}
One noteworthy feature of the data is that the peak brightness occurs at the substellar point. Based on  a sinusoidal model fit to the \Spitzer\ data, we found no significant offset ($\phi = -34.5_{-14.6}^{15.3}$). The observation of a thermal hotspot has been usually attributed  to a super-rotating jet that advects energy on the planet eastwards from the substellar point \citep[e.g., ][]{Showman2011}. 
Previously, an eastward offset on a small ($< 2 R_\oplus$) exoplanet was observed for 55 Cnc\,e using \Spitzer\ data. \citet{Demory2016a} analysed the shape of the thermal phase curve and measured a hotspot offset of $\phi = 41^{\circ} \pm 12^{\circ}$. This offset could be explained by a thick atmosphere and suggests a moderate mean molecular weight atmosphere with a surface pressure of a few bars \citep{Kite2016, Angelo2017, Hammond2017}. By contrast, our measured phase curve for K2-141\,b rules out a 55 Cancri e-like offset at the 3.9$\sigma$ level. The non-detection of  a significant offset in our analysis of K2-141\,b indicates that the planet does not have a thick, 55 Cnc\,e-like atmosphere.

This conclusion is further supported by the low observed nightside temperature, $T_{p,n} = 956_{-556}^{489}$K ($<1712$K at $2\sigma$) compared to the nightside temperature of $1380 \pm 400$K observed for 55 Cnc\,e \citep{Demory2016a}. Non-zero nightside temperatures are commonly also attributed by heat transport from the dayside to the nightside. To check for heat redistribution on the planet we used a toy model presented in \citet{Kreidberg2016} (see equation \ref{eq:toymodelKreidberg}) to fit the planet's thermal emission. The model uses a heat redistribution parameter, $F$, to regulate how much energy is transported from the dayside to the nightside of the planet. We fit this toy heat redistribution model to the joint (K2 \& \Spitzer) data set, and found that fixing heat redistribution parameter to $F = 0$ (i.e., no heat redistribution on the planet) is statistically preferred compared to letting $F$ vary free at $\Delta$BIC = 12.0 which is strong evidence for the model with no heat redistribution \citep{Kass1995}. Taken together, the absence of a hotspot offset and atmospheric heat redistribution support a scenario where the planet has little-to-no atmosphere.

\subsection{Evidence for a thin rock vapor atmosphere}

While a thick atmosphere is disfavored by the data, thinner atmospheres are a possibility. Thin gas-melt equilibrium atmospheres are expected for USPs \citep[e.g.][]{Miguel2011}. These atmospheres have much weaker heat circulation, but may be sufficiently optically thick that they have detectable spectral features \citep[e.g.][]{Ito2015}. To evaluate this possibility, we compared the dayside emission spectrum of K2-141  to two different models (see Figure \ref{fig:emission}). The first is a toy model based on the joint fit from Section 3.2 (a blackbody plus a reflected light component ). The second model is the physically-motivated, pseudo-2D rock vapor spectrum described in Section 4.1. We focused on the dayside spectrum alone because a full 3D model with realistic radiative transfer is beyond the scope of this paper.

Both the model spectra fit the observed eclipse depths well (within 2$\sigma$), but they have different implications for the nature of the planet's atmosphere. Both models have a larger optical eclipse depth than expected from a single-temperature blackbody. In the toy model, the  eclipse depth in the \Kepler\ bandpass  is fit by a high geometric albedo ($A_g = 0.282_{-0.078}^{0.070}$). By contrast, in the rock vapor model, the large optical eclipse depth is due to thermal emission from a hot inversion layer in the upper atmospheres, which is probed by strong absorbers at optical wavelengths.  A priori, it is challenging to say whether thermal emission or reflected light is more physically plausible. Recent lab experiments by \citet{Essack2020} have shown that lava worlds like K2-141\,b are expected to have low albedos ($A_g < 0.1$). In light of those results, a thermal inversion in a rock vapor atmosphere may be a more plausible explanation for the data. Alternatively, it is also possible that highly reflective  clouds could form in a rock vapor atmosphere; this possibility merits further theoretical investigation. Either way, whether the optical eclipse depth is due to a thermal inversion or reflective clouds, both explanations point to a thin rock vapor atmosphere rather than  a reflective surface.

\begin{figure}
\centering
\includegraphics[width=0.5\textwidth]{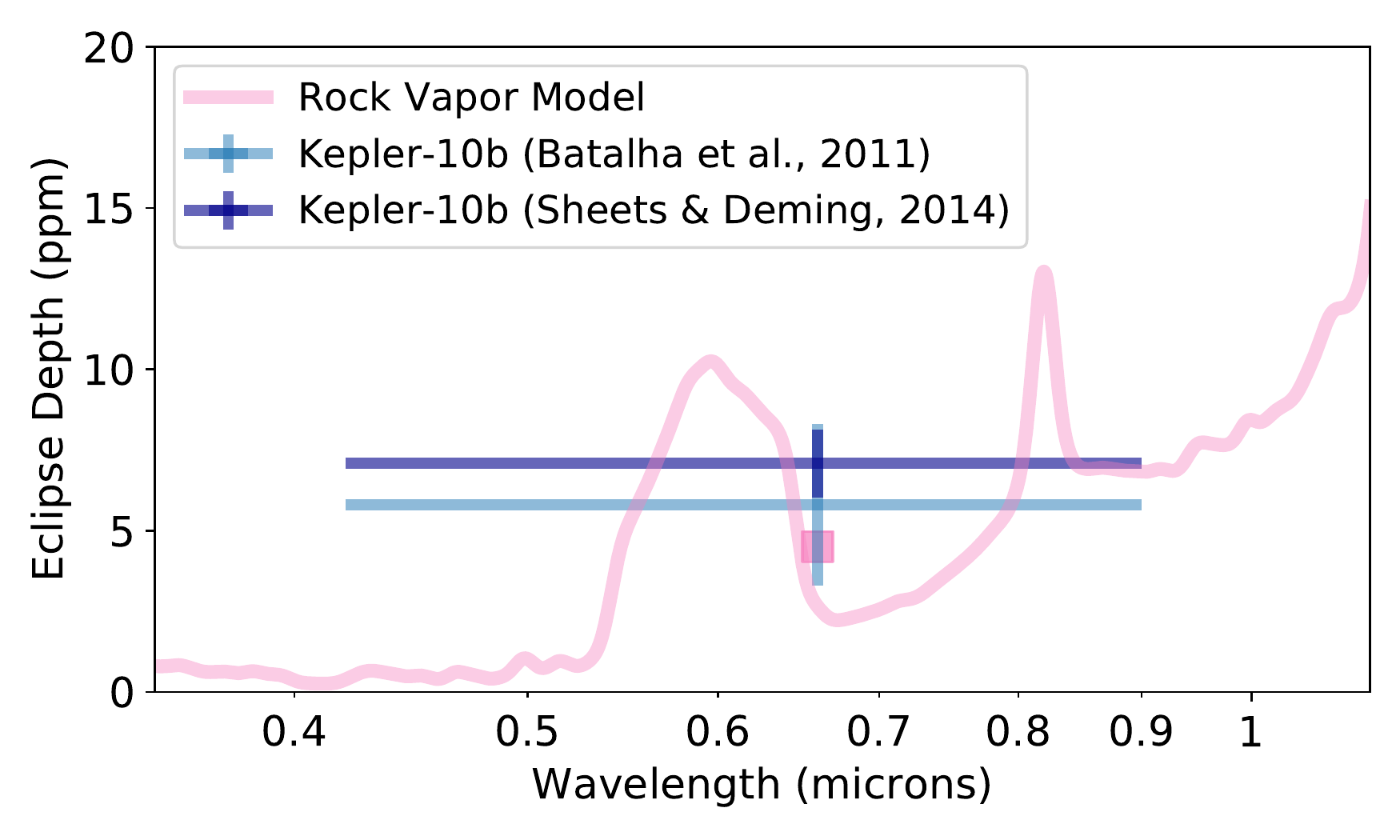}
\caption{Eclipse depth of Kepler-10\,b measured in the \Kepler\ bandpass compared to a pseudo-2D rock vapor model in pink showing emission features caused by thermal emission. The pink square is the predicted bandpass integrated eclipse depth assuming the rock vapor atmosphere model for the planet.}
\label{fig:Kepler-10b}
\end{figure}

These results shed new light on another well-known USP, Kepler-10\,b \citep{Batalha2011}, discovered by \Kepler. Kepler-10\,b also showed a surprisingly deep optical eclipse depth  attributed to a highly reflective lava \citep{Leger2011, Rouan2011}. We find that the eclipse depth may also be explained by a thermal inversion layer. Figure \ref{fig:Kepler-10b} shows our pseudo-2D rock vapor atmosphere model adjusted for the planetary and stellar parameters of Kepler-10\,b compared to the measured eclipse depth by \citet{Sheets2014}. Emission features due to the thermal inversion of Na at approximately 0.6 and 0.8 \micron\ increase the observed emission in the K2 bandpass. The spectrum agrees  well with the originally published eclipse depth. Subsequent analysis suggested that the eclipse depth may be even higher \citep{Sheets2014, Singh2021}. The thermal inversion model agrees with these values to within 2.4 and 3.6$\sigma$, respectively. Depending on the exact approach used for the data analysis, a thermal inversion can explain all or part of the observed signal. Thermal inversions are thus important to consider when interpreting the optical eclipse depths for USPs.

\section{Summary and Conclusions} \label{sec:summary}

We analyzed new \Spitzer\ observations of K2-141\,b spanning 65 hours and detected the thermal emission of the planet at 3.7$\sigma$ confidence with an eclipse depth of $f_p/f_* = 141.6_{-38.3}^{39.1}$ ppm. We fit several models to the \Spitzer\ data alone, and to the joint \Spitzer\ and \Kepler\ dataset. By fitting a sinusoid to the \Spitzer\ data we found no evidence for a hotspot offset. Our results on the hotspot offset are inconsistent with the \Spitzer\ observations of 55 Cnc\,e at a 3.9$\sigma$ level, the only other small USP planet with temperatures high enough to melt rock observed with \Spitzer. Combining the \Spitzer\ observations which are dominated by thermal emmision with the \Kepler\ observations dominated by optical emission we are able to break the degeneracy between these two emission sources. We fit a toy model described in \citet{Kreidberg2016} with the planet's heat redistribution $F$ as a free parameter and find that fixing $F=0$ is statistically preferred. We find a non-zero geometric albedo of $A_g = 0.282_{-0.078}^{0.070}$, a dayside temperature of $T_{p,d} = 2049_{-359}^{362}$K and a nightside temperature of $T_{p,n} = 956_{-556}^{489}$K (<1712K at $2\sigma$).

The planetary system containing the two confirmed planets was discovered using \Kepler\ observations collected in the K2 Campaign 12 \citep{Malavolta2018, Barragan2018}. Since then, the system has been reobserved during K2 Campaign 19 and with \Spitzer. We were able to significantly improve the ephermerides of K2-141\,b and K2-141\,c. The $3\sigma$ uncertainty on the predicted transit time in 2024 decreased from about an hour to just 2.7 minutes for planet b and from 5.2 hours to 16 minutes planet c compared to \citet{Malavolta2018}. This will help in the future to schedule observations of K2-141\,b and avoid transits or eclipses of planet c.
 
We compare the data to physically motivated models, including a pseudo-2D rock vapor atmosphere model and a 1D Turbulent Boundary Layer (TBL) model. We found that the TBL model with a adiabatic temperature pressure profile and the rock vapor model both agree well to the observed phase curve with \Spitzer. With TBL model with an isothermal T--P profile agrees worse with a $\Delta$BIC = 3.3 and the TBL model with a temperature inverted T--P profile has substential disagreement with the observations $\Delta$BIC = 7.3.

The moderately high albedo (roughly 0.3) may be due to a reflective surface, or a thermal inversion in a rock vapor atmosphere. The previous high albedo measurement for Kepler-10\,b can be also partially explained by such an inversion. A high optical emission also for other rocky planets might be therefore be explained by a thermal inversion in a rock vapor atmosphere.

The negligible hotspot offset for K2-141\,b contrasts with the large offset previously observed for 55 Cnc\,e. This suggests that the atmospheres have different properties. 55 Cnc\,e was suggested to have a moderate mean molecular weight atmosphere of a few bars \citep{Hammond2017, Angelo2017}. K2-141\,b either has a high mean molecular weight and low surface pressure or no atmosphere at all.

Future observations of ultra-short-period planets will give more insight on the nature of these extreme planets. In fact, the James Webb Space Telescope (\JWST) will observe several small (< $2\,R_\oplus$) USP planets during its Cycle 1 General Observers program: 

\begin{itemize}
    \item LHS 3844\,b with $R = 1.3\,R_\oplus$, $P$ = 11.1h, $T_{eq}$ = 805K
    \begin{itemize}
        \item three eclipses (MIRI/LRS) \citep{Kreidberg2021}
    \end{itemize}
    
    \item GJ 367 b with $R = 0.72\,R_\oplus$, $P$ = 7.7h and $T_{eq}$ = 1350K
    \begin{itemize}
        \item MIRI phase curve \citep{Zhang2021}
    \end{itemize}
    
    \item 55 Cancri e with $R = 1.9\,R_\oplus$, $P$ = 17.7h, $T_{eq}$ = 1950K
    \begin{itemize}
        \item two eclipses (NIRCam/F444W and MIRI/LRS) \citep{Hu2021}
        \item four transits (NIRCam/F444W) \citep{Brandeker2021}
    \end{itemize}

    \item  K2-141\,b with $R = 1.51\,R_\oplus$, $P$ = 6.9h and $T_{eq}$ = 2150K.
    \begin{itemize}
        \item a NIRSpec phase curve \citep{Espinoza2021}
        \item a MIRI/LRS phase curve \citep{Dang2021}
    \end{itemize}
\end{itemize}
Of these four planets observed in Cycle 1, only K2-141\,b and 55 Cnc\,e are hot enough to have a molten dayside. JWST's sensitivity and spectroscopic range is perfectly suited to study the thermal emission coming from these highly irradiated exoplanets. These planets might have detectable Na, SiO or SiO$_2$ in their atmospheres due to the evaporation of their surfaces. For example, SiO has broad features at 4 and 9 \micron\ \citep{Ito2015}. The large wavelength coverage of the MIRI/LRS instrument (\textasciitilde 5 to 12 \micron) is especially suited for probing in and out of the SiO band to determine the presence of an extended atmosphere. With \JWST\ observations already planned for K2-141\,b during Cycle 1, additionally information about the atmosphere is forthcoming. \citet{Espinoza2021} will use the NIRSpec G395H/F290LP instrument with a resolution of R = 1900 - 3700 to observe a phase curve in the near infrared (2.87 -- 5.18 \micron). The planned mid-infrared observations by \citet{Dang2021} will use MIRI's LRS mode (5 -- 12 \micron) at a resolution of R $\sim$ 100. All of these JWST observations, combined with the available data collected in the optical by K2 and in the infrared by \Spitzer\, will paint an unprecedented picture for a lava planet.

\begin{acknowledgements}

L.C. acknowledges support from the DFG Priority Programme SP1833 Grant CA 1795/3.

R.L. acknowledges support from the NASA ROSES XRP grant NNX17AC02G.

This research made use of Lightkurve, a Python package for Kepler and TESS data analysis (Lightkurve Collaboration, 2018).

This paper includes data collected by the Kepler mission and obtained from the MAST data archive at the Space Telescope Science Institute (STScI). Funding for the Kepler mission is provided by the NASA Science Mission Directorate. STScI is operated by the Association of Universities for Research in Astronomy, Inc., under NASA contract NAS 5–26555.

This work is based on observations made with the Spitzer Space Telescope, which was operated by the Jet Propulsion Laboratory, California Institute of Technology under a contract with NASA.

This research has made use of the NASA/IPAC Infrared Science Archive, which is funded by the National Aeronautics and Space Administration and operated by the California Institute of Technology.\\

We gratefully acknowledge the packages and tools that made this work possible: 
\texttt{numpy} \citep{numpy2020}, \texttt{matplotlib} \citep{matplotlib2007}, \texttt{scipy} \citep{scipy2020}, \texttt{astropy} \citep{astropy2013, astropy2018}, \texttt{ipython} \citep{ipython2007}, \texttt{batman} \citep{Kreidberg2015}, \texttt{SPIDERMAN} \citep{Louden2018}, \texttt{POET} \citep{Cubillos2013}, \texttt{corner} \citep{corner2016}, \texttt{dynesty} \citep{Speagle2020}, \texttt{MAGMA} \citep{Fegley1987,Schaefer2004}, \texttt{FastChem} \citep{Stock2018}, \texttt{HELIOS} \citep{Malik_2017,Malik_2019}, \texttt{petitRADTRANS} \citep{Molliere_2019}.\\

Finally, we thank the anonymous referee for a detailed report, which helped us to improve the quality of this paper.

\end{acknowledgements}

\newpage

\bibliographystyle{aa}
\bibliography{refs.bib}

\begin{thebibliography}{117}
\expandafter\ifx\csname natexlab\endcsname\relax\def\natexlab#1{#1}\fi

\bibitem[{{Adams} \& {Laughlin}(2006)}]{Adams2006}
{Adams}, F.~C. \& {Laughlin}, G. 2006, \apj, 649, 1004

\bibitem[{{Allard} {et~al.}(2007{\natexlab{a}}){Allard}, {Allard}, {Homeier},
  {Kielkopf}, {McCaughrean}, \& {Spiegelman}}]{Allard_2007a}
{Allard}, F., {Allard}, N.~F., {Homeier}, D., {et~al.} 2007{\natexlab{a}},
  \aap, 474, L21

\bibitem[{{Allard} {et~al.}(2007{\natexlab{b}}){Allard}, {Kielkopf}, \&
  {Allard}}]{Allard_2007b}
{Allard}, N.~F., {Kielkopf}, J.~F., \& {Allard}, F. 2007{\natexlab{b}},
  European Physical Journal D, 44, 507

\bibitem[{{Angelo} \& {Hu}(2017)}]{Angelo2017}
{Angelo}, I. \& {Hu}, R. 2017, \aj, 154, 232

\bibitem[{{Astropy Collaboration} {et~al.}(2018){Astropy Collaboration},
  {Price-Whelan}, {Sip{\H{o}}cz}, {G{\"u}nther}, {Lim}, {Crawford}, {Conseil},
  {Shupe}, {Craig}, {Dencheva}, {Ginsburg}, {VanderPlas}, {Bradley},
  {P{\'e}rez-Su{\'a}rez}, {de Val-Borro}, {Aldcroft}, {Cruz}, {Robitaille},
  {Tollerud}, {Ardelean}, {Babej}, {Bach}, {Bachetti}, {Bakanov}, {Bamford},
  {Barentsen}, {Barmby}, {Baumbach}, {Berry}, {Biscani}, {Boquien}, {Bostroem},
  {Bouma}, {Brammer}, {Bray}, {Breytenbach}, {Buddelmeijer}, {Burke},
  {Calderone}, {Cano Rodr{\'\i}guez}, {Cara}, {Cardoso}, {Cheedella}, {Copin},
  {Corrales}, {Crichton}, {D'Avella}, {Deil}, {Depagne}, {Dietrich}, {Donath},
  {Droettboom}, {Earl}, {Erben}, {Fabbro}, {Ferreira}, {Finethy}, {Fox},
  {Garrison}, {Gibbons}, {Goldstein}, {Gommers}, {Greco}, {Greenfield},
  {Groener}, {Grollier}, {Hagen}, {Hirst}, {Homeier}, {Horton}, {Hosseinzadeh},
  {Hu}, {Hunkeler}, {Ivezi{\'c}}, {Jain}, {Jenness}, {Kanarek}, {Kendrew},
  {Kern}, {Kerzendorf}, {Khvalko}, {King}, {Kirkby}, {Kulkarni}, {Kumar},
  {Lee}, {Lenz}, {Littlefair}, {Ma}, {Macleod}, {Mastropietro}, {McCully},
  {Montagnac}, {Morris}, {Mueller}, {Mumford}, {Muna}, {Murphy}, {Nelson},
  {Nguyen}, {Ninan}, {N{\"o}the}, {Ogaz}, {Oh}, {Parejko}, {Parley}, {Pascual},
  {Patil}, {Patil}, {Plunkett}, {Prochaska}, {Rastogi}, {Reddy Janga},
  {Sabater}, {Sakurikar}, {Seifert}, {Sherbert}, {Sherwood-Taylor}, {Shih},
  {Sick}, {Silbiger}, {Singanamalla}, {Singer}, {Sladen}, {Sooley},
  {Sornarajah}, {Streicher}, {Teuben}, {Thomas}, {Tremblay}, {Turner},
  {Terr{\'o}n}, {van Kerkwijk}, {de la Vega}, {Watkins}, {Weaver}, {Whitmore},
  {Woillez}, {Zabalza}, \& {Astropy Contributors}}]{astropy2018}
{Astropy Collaboration}, {Price-Whelan}, A.~M., {Sip{\H{o}}cz}, B.~M., {et~al.}
  2018, \aj, 156, 123

\bibitem[{{Astropy Collaboration} {et~al.}(2013){Astropy Collaboration},
  {Robitaille}, {Tollerud}, {Greenfield}, {Droettboom}, {Bray}, {Aldcroft},
  {Davis}, {Ginsburg}, {Price-Whelan}, {Kerzendorf}, {Conley}, {Crighton},
  {Barbary}, {Muna}, {Ferguson}, {Grollier}, {Parikh}, {Nair}, {Unther},
  {Deil}, {Woillez}, {Conseil}, {Kramer}, {Turner}, {Singer}, {Fox}, {Weaver},
  {Zabalza}, {Edwards}, {Azalee Bostroem}, {Burke}, {Casey}, {Crawford},
  {Dencheva}, {Ely}, {Jenness}, {Labrie}, {Lim}, {Pierfederici}, {Pontzen},
  {Ptak}, {Refsdal}, {Servillat}, \& {Streicher}}]{astropy2013}
{Astropy Collaboration}, {Robitaille}, T.~P., {Tollerud}, E.~J., {et~al.} 2013,
  \aap, 558, A33

\bibitem[{{Auvergne} {et~al.}(2009){Auvergne}, {Bodin}, {Boisnard}, {Buey},
  {Chaintreuil}, {Epstein}, {Jouret}, {Lam-Trong}, {Levacher}, {Magnan},
  {Perez}, {Plasson}, {Plesseria}, {Peter}, {Steller}, {Tiph{\`e}ne}, {Baglin},
  {Agogu{\'e}}, {Appourchaux}, {Barbet}, {Beaufort}, {Bellenger}, {Berlin},
  {Bernardi}, {Blouin}, {Boumier}, {Bonneau}, {Briet}, {Butler}, {Cautain},
  {Chiavassa}, {Costes}, {Cuvilho}, {Cunha-Parro}, {de Oliveira Fialho},
  {Decaudin}, {Defise}, {Djalal}, {Docclo}, {Drummond}, {Dupuis}, {Exil},
  {Faur{\'e}}, {Gaboriaud}, {Gamet}, {Gavalda}, {Grolleau}, {Gueguen},
  {Guivarc'h}, {Guterman}, {Hasiba}, {Huntzinger}, {Hustaix}, {Imbert},
  {Jeanville}, {Johlander}, {Jorda}, {Journoud}, {Karioty}, {Kerjean},
  {Lafond}, {Lapeyrere}, {Landiech}, {Larqu{\'e}}, {Laudet}, {Le Merrer},
  {Leporati}, {Leruyet}, {Levieuge}, {Llebaria}, {Martin}, {Mazy}, {Mesnager},
  {Michel}, {Moalic}, {Monjoin}, {Naudet}, {Neukirchner}, {Nguyen-Kim},
  {Ollivier}, {Orcesi}, {Ottacher}, {Oulali}, {Parisot}, {Perruchot},
  {Piacentino}, {Pinheiro da Silva}, {Platzer}, {Pontet}, {Pradines},
  {Quentin}, {Rohbeck}, {Rolland}, {Rollenhagen}, {Romagnan}, {Russ}, {Samadi},
  {Schmidt}, {Schwartz}, {Sebbag}, {Smit}, {Sunter}, {Tello}, {Toulouse},
  {Ulmer}, {Vandermarcq}, {Vergnault}, {Wallner}, {Waultier}, \&
  {Zanatta}}]{Auvergne2009}
{Auvergne}, M., {Bodin}, P., {Boisnard}, L., {et~al.} 2009, \aap, 506, 411

\bibitem[{{Barrag{\'a}n} {et~al.}(2018){Barrag{\'a}n}, {Gandolfi}, {Dai},
  {Livingston}, {Persson}, {Hirano}, {Narita}, {Csizmadia}, {Winn}, {Nespral},
  {Prieto-Arranz}, {Smith}, {Nowak}, {Albrecht}, {Antoniciello}, {Bo Justesen},
  {Cabrera}, {Cochran}, {Deeg}, {Eigmuller}, {Endl}, {Erikson}, {Fridlund},
  {Fukui}, {Grziwa}, {Guenther}, {Hatzes}, {Hidalgo}, {Johnson}, {Korth},
  {Palle}, {Patzold}, {Rauer}, {Tanaka}, \& {Van Eylen}}]{Barragan2018}
{Barrag{\'a}n}, O., {Gandolfi}, D., {Dai}, F., {et~al.} 2018, \aap, 612, A95

\bibitem[{{Barton} {et~al.}(2013){Barton}, {Yurchenko}, \&
  {Tennyson}}]{Barton_2013}
{Barton}, E.~J., {Yurchenko}, S.~N., \& {Tennyson}, J. 2013, \mnras, 434, 1469

\bibitem[{{Batalha} {et~al.}(2017){Batalha}, {Mandell}, {Pontoppidan},
  {Stevenson}, {Lewis}, {Kalirai}, {Earl}, {Greene}, {Albert}, \&
  {Nielsen}}]{Batalha2017}
{Batalha}, N.~E., {Mandell}, A., {Pontoppidan}, K., {et~al.} 2017, \pasp, 129,
  064501

\bibitem[{{Batalha} {et~al.}(2011){Batalha}, {Borucki}, {Bryson}, {Buchhave},
  {Caldwell}, {Christensen-Dalsgaard}, {Ciardi}, {Dunham}, {Fressin},
  {Gautier}, {Gilliland}, {Haas}, {Howell}, {Jenkins}, {Kjeldsen}, {Koch},
  {Latham}, {Lissauer}, {Marcy}, {Rowe}, {Sasselov}, {Seager}, {Steffen},
  {Torres}, {Basri}, {Brown}, {Charbonneau}, {Christiansen}, {Clarke},
  {Cochran}, {Dupree}, {Fabrycky}, {Fischer}, {Ford}, {Fortney}, {Girouard},
  {Holman}, {Johnson}, {Isaacson}, {Klaus}, {Machalek}, {Moorehead},
  {Morehead}, {Ragozzine}, {Tenenbaum}, {Twicken}, {Quinn}, {VanCleve},
  {Walkowicz}, {Welsh}, {Devore}, \& {Gould}}]{Batalha2011}
{Batalha}, N.~M., {Borucki}, W.~J., {Bryson}, S.~T., {et~al.} 2011, \apj, 729,
  27

\bibitem[{{Borucki} {et~al.}(2010){Borucki}, {Koch}, {Basri}, {Batalha},
  {Brown}, {Caldwell}, {Caldwell}, {Christensen-Dalsgaard}, {Cochran},
  {DeVore}, {Dunham}, {Dupree}, {Gautier}, {Geary}, {Gilliland}, {Gould},
  {Howell}, {Jenkins}, {Kondo}, {Latham}, {Marcy}, {Meibom}, {Kjeldsen},
  {Lissauer}, {Monet}, {Morrison}, {Sasselov}, {Tarter}, {Boss}, {Brownlee},
  {Owen}, {Buzasi}, {Charbonneau}, {Doyle}, {Fortney}, {Ford}, {Holman},
  {Seager}, {Steffen}, {Welsh}, {Rowe}, {Anderson}, {Buchhave}, {Ciardi},
  {Walkowicz}, {Sherry}, {Horch}, {Isaacson}, {Everett}, {Fischer}, {Torres},
  {Johnson}, {Endl}, {MacQueen}, {Bryson}, {Dotson}, {Haas}, {Kolodziejczak},
  {Van Cleve}, {Chandrasekaran}, {Twicken}, {Quintana}, {Clarke}, {Allen},
  {Li}, {Wu}, {Tenenbaum}, {Verner}, {Bruhweiler}, {Barnes}, \&
  {Prsa}}]{Borucki2010}
{Borucki}, W.~J., {Koch}, D., {Basri}, G., {et~al.} 2010, Science, 327, 977

\bibitem[{{Bourrier} {et~al.}(2018){Bourrier}, {Dumusque}, {Dorn}, {Henry},
  {Astudillo-Defru}, {Rey}, {Benneke}, {H{\'e}brard}, {Lovis}, {Demory},
  {Moutou}, \& {Ehrenreich}}]{Bourrier2018}
{Bourrier}, V., {Dumusque}, X., {Dorn}, C., {et~al.} 2018, \aap, 619, A1

\bibitem[{{Brandeker} {et~al.}(2021){Brandeker}, {Alibert}, {Bourrier},
  {Delrez}, {Demory}, {Ehrenreich}, {Fridlund}, {Heng}, {Hooton}, {Hoyer},
  {Hu}, {Kitzmann}, {Lendl}, \& {Persson}}]{Brandeker2021}
{Brandeker}, A., {Alibert}, Y., {Bourrier}, V., {et~al.} 2021, {Is it raining
  lava in the evening on 55 Cancri e?}, JWST Proposal. Cycle 1

\bibitem[{{Castan} \& {Menou}(2011)}]{Castan2011}
{Castan}, T. \& {Menou}, K. 2011, \apjl, 743, L36

\bibitem[{{Challener} {et~al.}(2020){Challener}, {Harrington}, {Jenkins},
  {Kurtovic}, {Ramirez}, {McIntyre}, {Himes}, {Rodr{\'\i}guez},
  {Anglada-Escud{\'e}}, {Dreizler}, {Ofir}, {Pe{\~n}a Rojas}, {Ribas}, {Rojo},
  {Kipping}, {Butler}, {Amado}, {Rodr{\'\i}guez-L{\'o}pez}, {Palle}, \&
  {Murgas}}]{Challener2020}
{Challener}, R.~C., {Harrington}, J., {Jenkins}, J., {et~al.} 2020, arXiv
  e-prints, arXiv:2011.05270

\bibitem[{{Charbonneau} {et~al.}(2005){Charbonneau}, {Allen}, {Megeath},
  {Torres}, {Alonso}, {Brown}, {Gilliland}, {Latham}, {Mandushev}, {O'Donovan},
  \& {Sozzetti}}]{Charbonneau2005}
{Charbonneau}, D., {Allen}, L.~E., {Megeath}, S.~T., {et~al.} 2005, \apj, 626,
  523

\bibitem[{{Crida} {et~al.}(2018{\natexlab{a}}){Crida}, {Ligi}, {Dorn}, {Borsa},
  \& {Lebreton}}]{Crida2018b}
{Crida}, A., {Ligi}, R., {Dorn}, C., {Borsa}, F., \& {Lebreton}, Y.
  2018{\natexlab{a}}, Research Notes of the American Astronomical Society, 2,
  172

\bibitem[{{Crida} {et~al.}(2018{\natexlab{b}}){Crida}, {Ligi}, {Dorn}, \&
  {Lebreton}}]{Crida2018a}
{Crida}, A., {Ligi}, R., {Dorn}, C., \& {Lebreton}, Y. 2018{\natexlab{b}},
  \apj, 860, 122

\bibitem[{{Cubillos} {et~al.}(2013){Cubillos}, {Harrington}, {Madhusudhan},
  {Stevenson}, {Hardy}, {Blecic}, {Anderson}, {Hardin}, \&
  {Campo}}]{Cubillos2013}
{Cubillos}, P., {Harrington}, J., {Madhusudhan}, N., {et~al.} 2013, \apj, 768,
  42

\bibitem[{{Dai} {et~al.}(2019){Dai}, {Masuda}, {Winn}, \& {Zeng}}]{Dai2019}
{Dai}, F., {Masuda}, K., {Winn}, J.~N., \& {Zeng}, L. 2019, \apj, 883, 79

\bibitem[{{Dang} {et~al.}(2021){Dang}, {Cowan}, {Hammond}, {Kreidberg}, {Lupu},
  {Miguel}, {Nguyen}, {Pierrehumbert}, {Zieba}, \& {Zilinskas}}]{Dang2021}
{Dang}, L., {Cowan}, N.~B., {Hammond}, M., {et~al.} 2021, {A Hell of a Phase
  Curve: Mapping the Surface and Atmosphere of a Lava Planet K2-141b}, JWST
  Proposal. Cycle 1

\bibitem[{{Dawson} \& {Fabrycky}(2010)}]{Dawson2010}
{Dawson}, R.~I. \& {Fabrycky}, D.~C. 2010, \apj, 722, 937

\bibitem[{{Dawson} \& {Johnson}(2018)}]{Dawson2018}
{Dawson}, R.~I. \& {Johnson}, J.~A. 2018, \araa, 56, 175

\bibitem[{{Demory}(2014)}]{Demory2014}
{Demory}, B.-O. 2014, \apjl, 789, L20

\bibitem[{{Demory} {et~al.}(2016{\natexlab{a}}){Demory}, {Gillon}, {de Wit},
  {Madhusudhan}, {Bolmont}, {Heng}, {Kataria}, {Lewis}, {Hu}, {Krick},
  {Stamenkovi{\'c}}, {Benneke}, {Kane}, \& {Queloz}}]{Demory2016a}
{Demory}, B.-O., {Gillon}, M., {de Wit}, J., {et~al.} 2016{\natexlab{a}}, \nat,
  532, 207

\bibitem[{{Demory} {et~al.}(2011){Demory}, {Gillon}, {Deming}, {Valencia},
  {Seager}, {Benneke}, {Lovis}, {Cubillos}, {Harrington}, {Stevenson}, {Mayor},
  {Pepe}, {Queloz}, {S{\'e}gransan}, \& {Udry}}]{Demory2011}
{Demory}, B.~O., {Gillon}, M., {Deming}, D., {et~al.} 2011, \aap, 533, A114

\bibitem[{{Demory} {et~al.}(2016{\natexlab{b}}){Demory}, {Gillon},
  {Madhusudhan}, \& {Queloz}}]{Demory2016b}
{Demory}, B.-O., {Gillon}, M., {Madhusudhan}, N., \& {Queloz}, D.
  2016{\natexlab{b}}, \mnras, 455, 2018

\bibitem[{{Dorn} {et~al.}(2019){Dorn}, {Harrison}, {Bonsor}, \&
  {Hands}}]{Dorn2019}
{Dorn}, C., {Harrison}, J.~H.~D., {Bonsor}, A., \& {Hands}, T.~O. 2019, \mnras,
  484, 712

\bibitem[{{Dumusque} {et~al.}(2014){Dumusque}, {Bonomo}, {Haywood},
  {Malavolta}, {S{\'e}gransan}, {Buchhave}, {Collier Cameron}, {Latham},
  {Molinari}, {Pepe}, {Udry}, {Charbonneau}, {Cosentino}, {Dressing},
  {Figueira}, {Fiorenzano}, {Gettel}, {Harutyunyan}, {Horne}, {Lopez-Morales},
  {Lovis}, {Mayor}, {Micela}, {Motalebi}, {Nascimbeni}, {Phillips}, {Piotto},
  {Pollacco}, {Queloz}, {Rice}, {Sasselov}, {Sozzetti}, {Szentgyorgyi}, \&
  {Watson}}]{Dumusque2014}
{Dumusque}, X., {Bonomo}, A.~S., {Haywood}, R.~D., {et~al.} 2014, \apj, 789,
  154

\bibitem[{{Espinoza} {et~al.}(2021){Espinoza}, {Bello-Arufe}, {Buchhave},
  {Burgasser}, {Demory}, {Diamond-Lowe}, {Fisher}, {Gibson}, {Guzman Mesa},
  {Heng}, {Hoeijmakers}, {Hooton}, {Hu}, {Kitzmann}, {Kozakis},
  {Lopez-Morales}, {Malik}, {Mendonca}, {Miguel}, {Morris}, \&
  {Rathcke}}]{Espinoza2021}
{Espinoza}, N., {Bello-Arufe}, A., {Buchhave}, L.~A., {et~al.} 2021, {The first
  near-infrared spectroscopic phase-curve of a super-Earth}, JWST Proposal.
  Cycle 1

\bibitem[{{Essack} {et~al.}(2020){Essack}, {Seager}, \&
  {Pajusalu}}]{Essack2020}
{Essack}, Z., {Seager}, S., \& {Pajusalu}, M. 2020, \apj, 898, 160

\bibitem[{{Fazio} {et~al.}(2004){Fazio}, {Hora}, {Allen}, {Ashby}, {Barmby},
  {Deutsch}, {Huang}, {Kleiner}, {Marengo}, {Megeath}, {Melnick}, {Pahre},
  {Patten}, {Polizotti}, {Smith}, {Taylor}, {Wang}, {Willner}, {Hoffmann},
  {Pipher}, {Forrest}, {McMurty}, {McCreight}, {McKelvey}, {McMurray}, {Koch},
  {Moseley}, {Arendt}, {Mentzell}, {Marx}, {Losch}, {Mayman}, {Eichhorn},
  {Krebs}, {Jhabvala}, {Gezari}, {Fixsen}, {Flores}, {Shakoorzadeh}, {Jungo},
  {Hakun}, {Workman}, {Karpati}, {Kichak}, {Whitley}, {Mann}, {Tollestrup},
  {Eisenhardt}, {Stern}, {Gorjian}, {Bhattacharya}, {Carey}, {Nelson},
  {Glaccum}, {Lacy}, {Lowrance}, {Laine}, {Reach}, {Stauffer}, {Surace},
  {Wilson}, {Wright}, {Hoffman}, {Domingo}, \& {Cohen}}]{Fazio2004}
{Fazio}, G.~G., {Hora}, J.~L., {Allen}, L.~E., {et~al.} 2004, \apjs, 154, 10

\bibitem[{{Fegley} \& {Cameron}(1987)}]{Fegley1987}
{Fegley}, B. \& {Cameron}, A.~G.~W. 1987, Earth and Planetary Science Letters,
  82, 207

\bibitem[{{Fischer} {et~al.}(2008){Fischer}, {Marcy}, {Butler}, {Vogt},
  {Laughlin}, {Henry}, {Abouav}, {Peek}, {Wright}, {Johnson}, {McCarthy}, \&
  {Isaacson}}]{Fischer2008}
{Fischer}, D.~A., {Marcy}, G.~W., {Butler}, R.~P., {et~al.} 2008, \apj, 675,
  790

\bibitem[{{Fischer} \& {Valenti}(2005)}]{Fischer2005}
{Fischer}, D.~A. \& {Valenti}, J. 2005, \apj, 622, 1102

\bibitem[{Foreman-Mackey(2016)}]{corner2016}
Foreman-Mackey, D. 2016, The Journal of Open Source Software, 1, 24

\bibitem[{Gelman \& Rubin(1992)}]{Gelman1992}
Gelman, A. \& Rubin, D.~B. 1992, Statist. Sci., 7, 457

\bibitem[{{Hammond} \& {Pierrehumbert}(2017)}]{Hammond2017}
{Hammond}, M. \& {Pierrehumbert}, R.~T. 2017, \apj, 849, 152

\bibitem[{{Harrington} {et~al.}(2007){Harrington}, {Luszcz}, {Seager},
  {Deming}, \& {Richardson}}]{Harrington2007}
{Harrington}, J., {Luszcz}, S., {Seager}, S., {Deming}, D., \& {Richardson},
  L.~J. 2007, \nat, 447, 691

\bibitem[{{Harris} {et~al.}(2020){Harris}, {Millman}, {van der Walt},
  {Gommers}, {Virtanen}, {Cournapeau}, {Wieser}, {Taylor}, {Berg}, {Smith},
  {Kern}, {Picus}, {Hoyer}, {van Kerkwijk}, {Brett}, {Haldane}, {del R{\'\i}o},
  {Wiebe}, {Peterson}, {G{\'e}rard-Marchant}, {Sheppard}, {Reddy}, {Weckesser},
  {Abbasi}, {Gohlke}, \& {Oliphant}}]{numpy2020}
{Harris}, C.~R., {Millman}, K.~J., {van der Walt}, S.~J., {et~al.} 2020, \nat,
  585, 357

\bibitem[{{Howell} {et~al.}(2014){Howell}, {Sobeck}, {Haas}, {Still},
  {Barclay}, {Mullally}, {Troeltzsch}, {Aigrain}, {Bryson}, {Caldwell},
  {Chaplin}, {Cochran}, {Huber}, {Marcy}, {Miglio}, {Najita}, {Smith},
  {Twicken}, \& {Fortney}}]{Howell2014}
{Howell}, S.~B., {Sobeck}, C., {Haas}, M., {et~al.} 2014, \pasp, 126, 398

\bibitem[{{Hu} {et~al.}(2021){Hu}, {Brandeker}, {Damiano}, {Demory},
  {Dragomir}, {Ito}, {Knutson}, {Miguel}, \& {Zhang}}]{Hu2021}
{Hu}, R., {Brandeker}, A., {Damiano}, M., {et~al.} 2021, {Determining the
  Atmospheric Composition of the Super-Earth 55 Cancri e}, JWST Proposal. Cycle
  1

\bibitem[{{Hu} {et~al.}(2015){Hu}, {Demory}, {Seager}, {Lewis}, \&
  {Showman}}]{Hu2015}
{Hu}, R., {Demory}, B.-O., {Seager}, S., {Lewis}, N., \& {Showman}, A.~P. 2015,
  \apj, 802, 51

\bibitem[{{Hunter}(2007)}]{matplotlib2007}
{Hunter}, J.~D. 2007, Computing in Science and Engineering, 9, 90

\bibitem[{{Husser} {et~al.}(2013){Husser}, {Wende-von Berg}, {Dreizler},
  {Homeier}, {Reiners}, {Barman}, \& {Hauschildt}}]{Husser_2013}
{Husser}, T.~O., {Wende-von Berg}, S., {Dreizler}, S., {et~al.} 2013, Astronomy
  and Astrophysics, 553, A6

\bibitem[{{Ito} {et~al.}(2015){Ito}, {Ikoma}, {Kawahara}, {Nagahara},
  {Kawashima}, \& {Nakamoto}}]{Ito2015}
{Ito}, Y., {Ikoma}, M., {Kawahara}, H., {et~al.} 2015, \apj, 801, 144

\bibitem[{{Jackson} {et~al.}(2013){Jackson}, {Stark}, {Adams}, {Chambers}, \&
  {Deming}}]{Jackson2013}
{Jackson}, B., {Stark}, C.~C., {Adams}, E.~R., {Chambers}, J., \& {Deming}, D.
  2013, \apj, 779, 165

\bibitem[{Kass \& Raftery(1995)}]{Kass1995}
Kass, R.~E. \& Raftery, A.~E. 1995, Journal of the American Statistical
  Association, 90, 773

\bibitem[{{Kipping} \& {Jansen}(2020)}]{Kipping2020}
{Kipping}, D. \& {Jansen}, T. 2020, Research Notes of the American Astronomical
  Society, 4, 170

\bibitem[{{Kite} {et~al.}(2016){Kite}, {Fegley}, {Schaefer}, \&
  {Gaidos}}]{Kite2016}
{Kite}, E.~S., {Fegley}, Bruce, J., {Schaefer}, L., \& {Gaidos}, E. 2016, \apj,
  828, 80

\bibitem[{{Koll} {et~al.}(2019){Koll}, {Malik}, {Mansfield}, {Kempton}, {Kite},
  {Abbot}, \& {Bean}}]{Koll2019}
{Koll}, D. D.~B., {Malik}, M., {Mansfield}, M., {et~al.} 2019, \apj, 886, 140

\bibitem[{{Kopal}(1954)}]{Kopal1954}
{Kopal}, Z. 1954, \mnras, 114, 101

\bibitem[{{Kreidberg}(2015)}]{Kreidberg2015}
{Kreidberg}, L. 2015, \pasp, 127, 1161

\bibitem[{{Kreidberg} {et~al.}(2021){Kreidberg}, {Hu}, {Kite}, {Koll}, {Malik},
  {Morley}, {Schaefer}, \& {Whittaker}}]{Kreidberg2021}
{Kreidberg}, L., {Hu}, R., {Kite}, E.~S., {et~al.} 2021, {A Search for
  Signatures of Volcanism and Geodynamics on the Hot Rocky Exoplanet LHS
  3844b}, JWST Proposal. Cycle 1

\bibitem[{{Kreidberg} {et~al.}(2019){Kreidberg}, {Koll}, {Morley}, {Hu},
  {Schaefer}, {Deming}, {Stevenson}, {Dittmann}, {Vanderburg}, {Berardo},
  {Guo}, {Stassun}, {Crossfield}, {Charbonneau}, {Latham}, {Loeb}, {Ricker},
  {Seager}, \& {Vanderspek}}]{Kreidberg2019}
{Kreidberg}, L., {Koll}, D. D.~B., {Morley}, C., {et~al.} 2019, \nat, 573, 87

\bibitem[{{Kreidberg} \& {Loeb}(2016)}]{Kreidberg2016}
{Kreidberg}, L. \& {Loeb}, A. 2016, \apjl, 832, L12

\bibitem[{{Kreidberg} {et~al.}(2018){Kreidberg}, {Lopez}, {Cowan}, {Lupu},
  {Stevenson}, {Louden}, \& {Malavolta}}]{Kreidberg2018}
{Kreidberg}, L., {Lopez}, E., {Cowan}, N., {et~al.} 2018, {Taking the
  Temperature of a Lava Planet}, Spitzer Proposal

\bibitem[{{Kurucz}(1992)}]{Kurucz_1992}
{Kurucz}, R.~L. 1992, \rmxaa, 23, 45

\bibitem[{{Kurucz}(1993)}]{Kurucz1993}
{Kurucz}, R.~L. 1993, {SYNTHE spectrum synthesis programs and line data}

\bibitem[{{Lanotte} {et~al.}(2014){Lanotte}, {Gillon}, {Demory}, {Fortney},
  {Astudillo}, {Bonfils}, {Magain}, {Delfosse}, {Forveille}, {Lovis}, {Mayor},
  {Neves}, {Pepe}, {Queloz}, {Santos}, \& {Udry}}]{Lanotte2014}
{Lanotte}, A.~A., {Gillon}, M., {Demory}, B.~O., {et~al.} 2014, \aap, 572, A73

\bibitem[{{L{\'e}ger} {et~al.}(2011){L{\'e}ger}, {Grasset}, {Fegley}, {Codron},
  {Albarede}, {Barge}, {Barnes}, {Cance}, {Carpy}, {Catalano}, {Cavarroc},
  {Demangeon}, {Ferraz-Mello}, {Gabor}, {Grie{\ss}meier}, {Leibacher},
  {Libourel}, {Maurin}, {Raymond}, {Rouan}, {Samuel}, {Schaefer}, {Schneider},
  {Schuller}, {Selsis}, \& {Sotin}}]{Leger2011}
{L{\'e}ger}, A., {Grasset}, O., {Fegley}, B., {et~al.} 2011, \icarus, 213, 1

\bibitem[{{Liddle}(2007)}]{Liddle2007}
{Liddle}, A.~R. 2007, \mnras, 377, L74

\bibitem[{{Lightkurve Collaboration} {et~al.}(2018){Lightkurve Collaboration},
  {Cardoso}, {Hedges}, {Gully-Santiago}, {Saunders}, {Cody}, {Barclay}, {Hall},
  {Sagear}, {Turtelboom}, {Zhang}, {Tzanidakis}, {Mighell}, {Coughlin}, {Bell},
  {Berta-Thompson}, {Williams}, {Dotson}, \& {Barentsen}}]{Lightkurve2018}
{Lightkurve Collaboration}, {Cardoso}, J.~V.~d.~M., {Hedges}, C., {et~al.}
  2018, {Lightkurve: Kepler and TESS time series analysis in Python},
  Astrophysics Source Code Library

\bibitem[{{Lopez}(2017)}]{Lopez2017}
{Lopez}, E.~D. 2017, \mnras, 472, 245

\bibitem[{{Louden} \& {Kreidberg}(2018)}]{Louden2018}
{Louden}, T. \& {Kreidberg}, L. 2018, \mnras, 477, 2613

\bibitem[{{Lundkvist} {et~al.}(2016){Lundkvist}, {Kjeldsen}, {Albrecht},
  {Davies}, {Basu}, {Huber}, {Justesen}, {Karoff}, {Silva Aguirre}, {van
  Eylen}, {Vang}, {Arentoft}, {Barclay}, {Bedding}, {Campante}, {Chaplin},
  {Christensen-Dalsgaard}, {Elsworth}, {Gilliland}, {Handberg}, {Hekker},
  {Kawaler}, {Lund}, {Metcalfe}, {Miglio}, {Rowe}, {Stello}, {Tingley}, \&
  {White}}]{Lundkvist2016}
{Lundkvist}, M.~S., {Kjeldsen}, H., {Albrecht}, S., {et~al.} 2016, Nature
  Communications, 7, 11201

\bibitem[{{Lust} {et~al.}(2014){Lust}, {Britt}, {Harrington}, {Nymeyer},
  {Stevenson}, {Ross}, {Bowman}, \& {Fraine}}]{Lust2014}
{Lust}, N.~B., {Britt}, D., {Harrington}, J., {et~al.} 2014, \pasp, 126, 1092

\bibitem[{{Malavolta} {et~al.}(2018){Malavolta}, {Mayo}, {Louden}, {Rajpaul},
  {Bonomo}, {Buchhave}, {Kreidberg}, {Kristiansen}, {Lopez-Morales}, {Mortier},
  {Vand erburg}, {Coffinet}, {Ehrenreich}, {Lovis}, {Bouchy}, {Charbonneau},
  {Ciardi}, {Collier Cameron}, {Cosentino}, {Crossfield}, {Damasso},
  {Dressing}, {Dumusque}, {Everett}, {Figueira}, {Fiorenzano}, {Gonzales},
  {Haywood}, {Harutyunyan}, {Hirsch}, {Howell}, {Johnson}, {Latham}, {Lopez},
  {Mayor}, {Micela}, {Molinari}, {Nascimbeni}, {Pepe}, {Phillips}, {Piotto},
  {Rice}, {Sasselov}, {S{\'e}gransan}, {Sozzetti}, {Udry}, \&
  {Watson}}]{Malavolta2018}
{Malavolta}, L., {Mayo}, A.~W., {Louden}, T., {et~al.} 2018, \aj, 155, 107

\bibitem[{Malik {et~al.}(2017)Malik, Grosheintz, Mendon{\c{c}}a, Grimm, Lavie,
  Kitzmann, Tsai, Burrows, Kreidberg, Bedell, Bean, Stevenson, \&
  Heng}]{Malik_2017}
Malik, M., Grosheintz, L., Mendon{\c{c}}a, J.~M., {et~al.} 2017, The
  Astronomical Journal, 153, 56

\bibitem[{{Malik} {et~al.}(2019){Malik}, {Kitzmann}, {Mendon{\c{c}}a}, {Grimm},
  {Marleau}, {Linder}, {Tsai}, \& {Heng}}]{Malik_2019}
{Malik}, M., {Kitzmann}, D., {Mendon{\c{c}}a}, J.~M., {et~al.} 2019, The
  Astronomical Journal, 157, 170

\bibitem[{{Mandel} \& {Agol}(2002)}]{Mandel2002}
{Mandel}, K. \& {Agol}, E. 2002, \apjl, 580, L171

\bibitem[{{May} \& {Stevenson}(2020)}]{May2020}
{May}, E.~M. \& {Stevenson}, K.~B. 2020, \aj, 160, 140

\bibitem[{{Mayorga} {et~al.}(2016){Mayorga}, {Jackiewicz}, {Rages}, {West},
  {Knowles}, {Lewis}, \& {Marley}}]{Mayorga2016}
{Mayorga}, L.~C., {Jackiewicz}, J., {Rages}, K., {et~al.} 2016, \aj, 152, 209

\bibitem[{{McEwen} {et~al.}(1998){McEwen}, {Keszthelyi}, {Spencer}, {Schubert},
  {Matson}, {Lopes-Gautier}, {Klaasen}, {Johnson}, {Head}, {Geissler},
  {Fagents}, {Davies}, {Carr}, {Breneman}, \& {Belton}}]{McEwen1998}
{McEwen}, A.~S., {Keszthelyi}, L., {Spencer}, J.~R., {et~al.} 1998, Science,
  281, 87

\bibitem[{{Mendon{\c{c}}a} {et~al.}(2018){Mendon{\c{c}}a}, {Malik}, {Demory},
  \& {Heng}}]{Mendonca2018}
{Mendon{\c{c}}a}, J.~M., {Malik}, M., {Demory}, B.-O., \& {Heng}, K. 2018, \aj,
  155, 150

\bibitem[{{Miguel} {et~al.}(2011){Miguel}, {Kaltenegger}, {Fegley}, \&
  {Schaefer}}]{Miguel2011}
{Miguel}, Y., {Kaltenegger}, L., {Fegley}, B., \& {Schaefer}, L. 2011, \apjl,
  742, L19

\bibitem[{{Modirrousta-Galian} {et~al.}(2021){Modirrousta-Galian}, {Ito}, \&
  {Micela}}]{ModirroustaGalian2021}
{Modirrousta-Galian}, D., {Ito}, Y., \& {Micela}, G. 2021, \icarus, 358, 114175

\bibitem[{{Molli{\`e}re} {et~al.}(2019){Molli{\`e}re}, {Wardenier}, {van
  Boekel}, {Henning}, {Molaverdikhani}, \& {Snellen}}]{Molliere_2019}
{Molli{\`e}re}, P., {Wardenier}, J.~P., {van Boekel}, R., {et~al.} 2019, \aap,
  627, A67

\bibitem[{{Morris} {et~al.}(2021){Morris}, {Delrez}, {Brandeker}, {Cameron},
  {Simon}, {Futyan}, {Olofsson}, {Hoyer}, {Fortier}, {Demory}, {Lendl},
  {Wilson}, {Oshagh}, {Heng}, {Ehrenreich}, {Sulis}, {Alibert}, {Alonso},
  {Anglada Escud{\'e}}, {Barrado}, {Barros}, {Baumjohann}, {Beck}, {Beck},
  {Bekkelien}, {Benz}, {Bergomi}, {Billot}, {Bonfils}, {Bourrier}, {Broeg},
  {B{\'a}rczy}, {Cabrera}, {Charnoz}, {Davies}, {De Miguel Ferreras},
  {Deleuil}, {Deline}, {Demangeon}, {Erikson}, {Floren}, {Fossati}, {Fridlund},
  {Gandolfi}, {Garc{\'\i}a Mu{\~n}oz}, {Gillon}, {Guedel}, {Guterman}, {Isaak},
  {Kiss}, {Laskar}, {Lecavelier des Etangs}, {Lieder}, {Lovis}, {Magrin},
  {Maxted}, {Nascimbeni}, {Ottensamer}, {Pagano}, {Pall{\'e}}, {Peter},
  {Piotto}, {Pizarro Rubio}, {Pollacco}, {Pozuelos}, {Queloz}, {Ragazzoni},
  {Rando}, {Rauer}, {Ribas}, {Santos}, {Scandariato}, {Smith}, {Sousa},
  {Steller}, {Szab{\'o}}, {S{\'e}gransan}, {Thomas}, {Udry}, {Ulmer}, {Van
  Grootel}, \& {Walton}}]{Morris2021}
{Morris}, B.~M., {Delrez}, L., {Brandeker}, A., {et~al.} 2021, \aap, 653, A173

\bibitem[{{Nguyen} {et~al.}(in press 2022){Nguyen}, {Cowan}, {Pierrehumbert},
  {Lupu}, \& {Moores}}]{Nguyen2022}
{Nguyen}, T.~G., {Cowan}, N., {Pierrehumbert}, R., {Lupu}, R., \& {Moores}, J.
  in press 2022, \mnras

\bibitem[{{Nguyen} {et~al.}(2020){Nguyen}, {Cowan}, {Banerjee}, \&
  {Moores}}]{Nguyen2020}
{Nguyen}, T.~G., {Cowan}, N.~B., {Banerjee}, A., \& {Moores}, J.~E. 2020,
  \mnras, 499, 4605

\bibitem[{{Perez} \& {Granger}(2007)}]{ipython2007}
{Perez}, F. \& {Granger}, B.~E. 2007, Computing in Science and Engineering, 9,
  21

\bibitem[{{Pont} {et~al.}(2006){Pont}, {Zucker}, \& {Queloz}}]{Pont2006}
{Pont}, F., {Zucker}, S., \& {Queloz}, D. 2006, \mnras, 373, 231

\bibitem[{{Ricker} {et~al.}(2014){Ricker}, {Winn}, {Vanderspek}, {Latham},
  {Bakos}, {Bean}, {Berta-Thompson}, {Brown}, {Buchhave}, {Butler}, {Butler},
  {Chaplin}, {Charbonneau}, {Christensen-Dalsgaard}, {Clampin}, {Deming},
  {Doty}, {De Lee}, {Dressing}, {Dunham}, {Endl}, {Fressin}, {Ge}, {Henning},
  {Holman}, {Howard}, {Ida}, {Jenkins}, {Jernigan}, {Johnson}, {Kaltenegger},
  {Kawai}, {Kjeldsen}, {Laughlin}, {Levine}, {Lin}, {Lissauer}, {MacQueen},
  {Marcy}, {McCullough}, {Morton}, {Narita}, {Paegert}, {Palle}, {Pepe},
  {Pepper}, {Quirrenbach}, {Rinehart}, {Sasselov}, {Sato}, {Seager},
  {Sozzetti}, {Stassun}, {Sullivan}, {Szentgyorgyi}, {Torres}, {Udry}, \&
  {Villasenor}}]{Ricker2014}
{Ricker}, G.~R., {Winn}, J.~N., {Vanderspek}, R., {et~al.} 2014, in Society of
  Photo-Optical Instrumentation Engineers (SPIE) Conference Series, Vol. 9143,
  Space Telescopes and Instrumentation 2014: Optical, Infrared, and Millimeter
  Wave, ed. J.~{Oschmann}, Jacobus~M., M.~{Clampin}, G.~G. {Fazio}, \& H.~A.
  {MacEwen}, 914320

\bibitem[{{Rossi} \& {Pascale}(1985)}]{Rossi_1985}
{Rossi}, F. \& {Pascale}, J. 1985, \pra, 32, 2657

\bibitem[{{Rouan} {et~al.}(2011){Rouan}, {Deeg}, {Demangeon}, {Samuel},
  {Cavarroc}, {Fegley}, \& {L{\'e}ger}}]{Rouan2011}
{Rouan}, D., {Deeg}, H.~J., {Demangeon}, O., {et~al.} 2011, \apjl, 741, L30

\bibitem[{{Ryabchikova} {et~al.}(2015){Ryabchikova}, {Piskunov}, {Kurucz},
  {Stempels}, {Heiter}, {Pakhomov}, \& {Barklem}}]{Ryab_2015}
{Ryabchikova}, T., {Piskunov}, N., {Kurucz}, R.~L., {et~al.} 2015, \physscr,
  90, 054005

\bibitem[{{Sanchis-Ojeda} {et~al.}(2014){Sanchis-Ojeda}, {Rappaport}, {Winn},
  {Kotson}, {Levine}, \& {El Mellah}}]{Sanchis-Ojeda2014}
{Sanchis-Ojeda}, R., {Rappaport}, S., {Winn}, J.~N., {et~al.} 2014, \apj, 787,
  47

\bibitem[{{Sanchis-Ojeda} {et~al.}(2013){Sanchis-Ojeda}, {Rappaport}, {Winn},
  {Levine}, {Kotson}, {Latham}, \& {Buchhave}}]{Sanchis-Ojeda2013}
{Sanchis-Ojeda}, R., {Rappaport}, S., {Winn}, J.~N., {et~al.} 2013, \apj, 774,
  54

\bibitem[{{Schaefer} \& {Fegley}(2004)}]{Schaefer2004}
{Schaefer}, L. \& {Fegley}, B. 2004, \icarus, 169, 216

\bibitem[{{Schaefer} \& {Fegley}(2010)}]{Schaefer2010}
{Schaefer}, L. \& {Fegley}, B. 2010, \icarus, 208, 438

\bibitem[{{Schwarz}(1978)}]{Schwarz1978}
{Schwarz}, G. 1978, Annals of Statistics, 6, 461

\bibitem[{{Seager}(2010)}]{Seager2010}
{Seager}, S. 2010, {Exoplanet Atmospheres: Physical Processes}

\bibitem[{{Sheets} \& {Deming}(2014)}]{Sheets2014}
{Sheets}, H.~A. \& {Deming}, D. 2014, \apj, 794, 133

\bibitem[{{Showman} \& {Polvani}(2011)}]{Showman2011}
{Showman}, A.~P. \& {Polvani}, L.~M. 2011, \apj, 738, 71

\bibitem[{{Singh} {et~al.}(2021){Singh}, {Bonomo}, {Scandariato}, {Cibrario},
  {Barbato}, {Fossati}, {Pagano}, \& {Sozzetti}}]{Singh2021}
{Singh}, V., {Bonomo}, A.~S., {Scandariato}, G., {et~al.} 2021, arXiv e-prints,
  arXiv:2111.05716

\bibitem[{{Speagle}(2020)}]{Speagle2020}
{Speagle}, J.~S. 2020, \mnras, 493, 3132

\bibitem[{{Stevenson} {et~al.}(2012){Stevenson}, {Harrington}, {Fortney},
  {Loredo}, {Hardy}, {Nymeyer}, {Bowman}, {Cubillos}, {Bowman}, \&
  {Hardin}}]{Stevenson2012}
{Stevenson}, K.~B., {Harrington}, J., {Fortney}, J.~J., {et~al.} 2012, \apj,
  754, 136

\bibitem[{Stock {et~al.}(2018)Stock, Kitzmann, Patzer, \& Sedlmayr}]{Stock2018}
Stock, J.~W., Kitzmann, D., Patzer, A. B.~C., \& Sedlmayr, E. 2018, Monthly
  Notices of the Royal Astronomical Society, 479, 865

\bibitem[{{Sulis} {et~al.}(2019){Sulis}, {Dragomir}, {Lendl}, {Bourrier},
  {Demory}, {Fossati}, {Cubillos}, {Guenther}, {Kane}, {Kuschnig}, {Matthews},
  {Moffat}, {Rowe}, {Sasselov}, {Weiss}, \& {Winn}}]{Sulis2019}
{Sulis}, S., {Dragomir}, D., {Lendl}, M., {et~al.} 2019, \aap, 631, A129

\bibitem[{{Tamburo} {et~al.}(2018){Tamburo}, {Mandell}, {Deming}, \&
  {Garhart}}]{Tamburo2018}
{Tamburo}, P., {Mandell}, A., {Deming}, D., \& {Garhart}, E. 2018, \aj, 155,
  221

\bibitem[{{Ter Braak}(2006)}]{Braak2006}
{Ter Braak}, C. J.~F. 2006, Statistics and Computing, 16, 239

\bibitem[{{Van Eylen} {et~al.}(2018){Van Eylen}, {Agentoft}, {Lundkvist},
  {Kjeldsen}, {Owen}, {Fulton}, {Petigura}, \& {Snellen}}]{VanEylen2018}
{Van Eylen}, V., {Agentoft}, C., {Lundkvist}, M.~S., {et~al.} 2018, \mnras,
  479, 4786

\bibitem[{{Vanderburg} {et~al.}(2017){Vanderburg}, {Becker}, {Buchhave},
  {Mortier}, {Lopez}, {Malavolta}, {Haywood}, {Latham}, {Charbonneau},
  {L{\'o}pez-Morales}, {Adams}, {Bonomo}, {Bouchy}, {Collier Cameron},
  {Cosentino}, {Di Fabrizio}, {Dumusque}, {Fiorenzano}, {Harutyunyan},
  {Johnson}, {Lorenzi}, {Lovis}, {Mayor}, {Micela}, {Molinari}, {Pedani},
  {Pepe}, {Piotto}, {Phillips}, {Rice}, {Sasselov}, {S{\'e}gransan},
  {Sozzetti}, {Udry}, \& {Watson}}]{Vanderburg2017}
{Vanderburg}, A., {Becker}, J.~C., {Buchhave}, L.~A., {et~al.} 2017, \aj, 154,
  237

\bibitem[{{Vanderburg} \& {Johnson}(2014)}]{Vanderburg2014}
{Vanderburg}, A. \& {Johnson}, J.~A. 2014, \pasp, 126, 948

\bibitem[{{Vanderburg} {et~al.}(2016){Vanderburg}, {Latham}, {Buchhave},
  {Bieryla}, {Berlind}, {Calkins}, {Esquerdo}, {Welsh}, \&
  {Johnson}}]{Vanderburg2016}
{Vanderburg}, A., {Latham}, D.~W., {Buchhave}, L.~A., {et~al.} 2016, \apjs,
  222, 14

\bibitem[{{Vanderspek} {et~al.}(2019){Vanderspek}, {Huang}, {Vanderburg},
  {Ricker}, {Latham}, {Seager}, {Winn}, {Jenkins}, {Burt}, {Dittmann},
  {Newton}, {Quinn}, {Shporer}, {Charbonneau}, {Irwin}, {Ment}, {Winters},
  {Collins}, {Evans}, {Gan}, {Hart}, {Jensen}, {Kielkopf}, {Mao}, {Waalkes},
  {Bouchy}, {Marmier}, {Nielsen}, {Ottoni}, {Pepe}, {S{\'e}gransan}, {Udry},
  {Henry}, {Paredes}, {James}, {Hinojosa}, {Silverstein}, {Palle},
  {Berta-Thompson}, {Crossfield}, {Davies}, {Dragomir}, {Fausnaugh}, {Glidden},
  {Pepper}, {Morgan}, {Rose}, {Twicken}, {Villase{\~n}or}, {Yu}, {Bakos},
  {Bean}, {Buchhave}, {Christensen-Dalsgaard}, {Christiansen}, {Ciardi},
  {Clampin}, {De Lee}, {Deming}, {Doty}, {Jernigan}, {Kaltenegger}, {Lissauer},
  {McCullough}, {Narita}, {Paegert}, {Pal}, {Rinehart}, {Sasselov}, {Sato},
  {Sozzetti}, {Stassun}, \& {Torres}}]{Vanderspek2019}
{Vanderspek}, R., {Huang}, C.~X., {Vanderburg}, A., {et~al.} 2019, \apjl, 871,
  L24

\bibitem[{{Virtanen} {et~al.}(2020){Virtanen}, {Gommers}, {Oliphant},
  {Haberland}, {Reddy}, {Cournapeau}, {Burovski}, {Peterson}, {Weckesser},
  {Bright}, {van der Walt}, {Brett}, {Wilson}, {Millman}, {Mayorov}, {Nelson},
  {Jones}, {Kern}, {Larson}, {Carey}, {Polat}, {Feng}, {Moore}, {VanderPlas},
  {Laxalde}, {Perktold}, {Cimrman}, {Henriksen}, {Quintero}, {Harris},
  {Archibald}, {Ribeiro}, {Pedregosa}, {van Mulbregt}, \& {SciPy 1. 0
  Contributors}}]{scipy2020}
{Virtanen}, P., {Gommers}, R., {Oliphant}, T.~E., {et~al.} 2020, Nature
  Methods, 17, 261

\bibitem[{{Winn} {et~al.}(2007){Winn}, {Holman}, {Bakos}, {P{\'a}l}, {Johnson},
  {Williams}, {Shporer}, {Mazeh}, {Fernandez}, {Latham}, \&
  {Gillon}}]{Winn2007}
{Winn}, J.~N., {Holman}, M.~J., {Bakos}, G.~{\'A}., {et~al.} 2007, \aj, 134,
  1707

\bibitem[{{Winn} {et~al.}(2008){Winn}, {Holman}, {Torres}, {McCullough},
  {Johns-Krull}, {Latham}, {Shporer}, {Mazeh}, {Garcia-Melendo}, {Foote},
  {Esquerdo}, \& {Everett}}]{Winn2008}
{Winn}, J.~N., {Holman}, M.~J., {Torres}, G., {et~al.} 2008, \apj, 683, 1076

\bibitem[{{Winn} {et~al.}(2011){Winn}, {Matthews}, {Dawson}, {Fabrycky},
  {Holman}, {Kallinger}, {Kuschnig}, {Sasselov}, {Dragomir}, {Guenther},
  {Moffat}, {Rowe}, {Rucinski}, \& {Weiss}}]{Winn2011}
{Winn}, J.~N., {Matthews}, J.~M., {Dawson}, R.~I., {et~al.} 2011, \apjl, 737,
  L18

\bibitem[{{Winn} {et~al.}(2018){Winn}, {Sanchis-Ojeda}, \&
  {Rappaport}}]{Winn2018}
{Winn}, J.~N., {Sanchis-Ojeda}, R., \& {Rappaport}, S. 2018, \nar, 83, 37

\bibitem[{{Winn} {et~al.}(2017){Winn}, {Sanchis-Ojeda}, {Rogers}, {Petigura},
  {Howard}, {Isaacson}, {Marcy}, {Schlaufman}, {Cargile}, \& {Hebb}}]{Winn2017}
{Winn}, J.~N., {Sanchis-Ojeda}, R., {Rogers}, L., {et~al.} 2017, \aj, 154, 60

\bibitem[{{Zeng} {et~al.}(2019){Zeng}, {Jacobsen}, {Sasselov}, {Petaev},
  {Vanderburg}, {Lopez-Morales}, {Perez-Mercader}, {Mattsson}, {Li}, {Heising},
  {Bonomo}, {Damasso}, {Berger}, {Cao}, {Levi}, \& {Wordsworth}}]{Zeng2019}
{Zeng}, L., {Jacobsen}, S.~B., {Sasselov}, D.~D., {et~al.} 2019, Proceedings of
  the National Academy of Science, 116, 9723

\bibitem[{{Zhang} {et~al.}(2021){Zhang}, {Dai}, {Hu}, {Knutson}, \&
  {Lam}}]{Zhang2021}
{Zhang}, M., {Dai}, F., {Hu}, R., {Knutson}, H.~A., \& {Lam}, K. 2021, {The
  First and Only Multi-wavelength Map of an Ultra-short-period sub-Earth}, JWST
  Proposal. Cycle 1

\bibitem[{{Zilinskas} {et~al.}(2021){Zilinskas}, {Miguel}, {Lyu}, \&
  {Bax}}]{Zilinskas2021}
{Zilinskas}, M., {Miguel}, Y., {Lyu}, Y., \& {Bax}, M. 2021, \mnras, 500, 2197

\end{thebibliography}

\appendix

\section{Updated Ephemeris for K2-141\,c}
\label{sec:K2-141c_ephemeris}

\begin{figure}[h]
\centering
\includegraphics[width=0.49\textwidth]{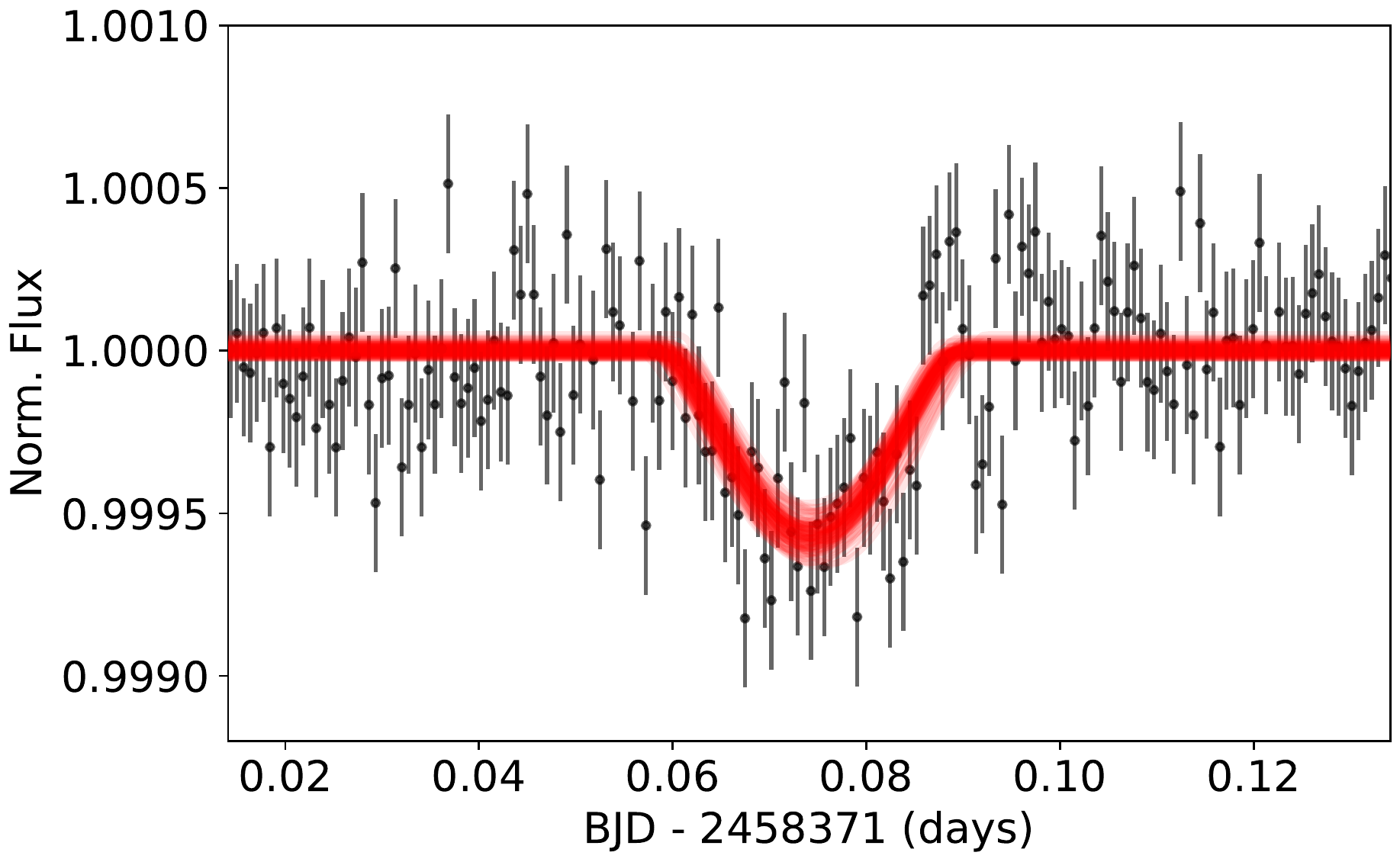}
\caption{The transit of K2-141c in C19 with 100 random draws from the posterior in red.}
\label{fig:K2-141c_fit}
\end{figure}

\begin{figure}[h]
\centering
\includegraphics[width=0.49\textwidth]{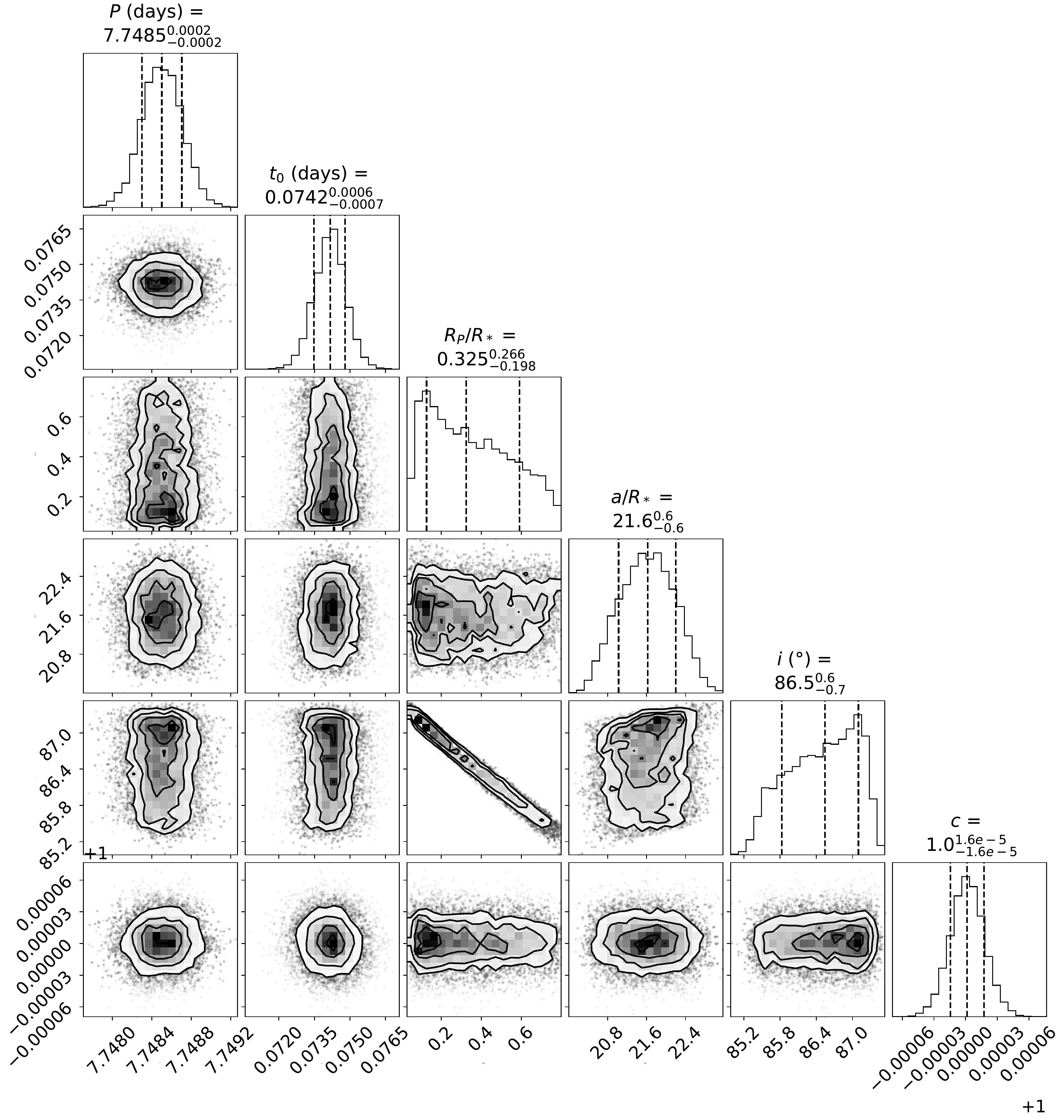}
\caption{Corner plot of the fitted transit model based on the single transit of K2-141\,c in K2 C19 to update the ephemeris of the planet. Gaussian Priors were used on $P$, $a/R_*$ and $\cos i$ based on the values reported in \citet{Malavolta2018} who only used the observations in C12. We used the resulting transit time $t_0$ to recalculate the orbital period $P$. The transit time in this plot is expressed as BJD$_{\rm{TDB}}$ - 2458371.0 d.}
\label{fig:K2-141c_corner}
\end{figure}

We used the one transit of K2-141\,c which occurred in the continuous 8 days of K2 C19 (see Figure \ref{fig:C19}) to improve the ephemeris of the planet. Our total model uses the transit model implemented in \texttt{BATMAN} \citep{Kreidberg2015} multiplied by a constant. The fit model has 5 free parameters: the orbital period $P$, the time of central transit $t_0$, the radius of planet in units of stellar radii $R_p/R_*$, the semi-major axis in units of stellar radii $a/R_*$, the cosine of the inclination $\cos i$ and a constant $c$. We fixed the eccentricity $ecc$ and the argument of periastron $\omega$ to zero. We used the same values for the limb-darkening coefficients $u_1$ and $u_2$ as in our analysis of K2-141\,b (see Section \ref{sec:analysis}). The Parameters and their uncertainties were estimated using the Nested Sampling package \texttt{dynesty} \citep{Speagle2020}. We used Gaussian priors based on the values reported in \citet{Malavolta2018} for $P$, $a/R_*$ and $\cos i$. The final fit and the pairs plot of the posteriors can be found in Figure \ref{fig:K2-141c_fit} and Figure \ref{fig:K2-141c_corner}, respectively. The resulting $t_0$ was used to recalculate the orbital period $P$ following the following approach:

\begin{equation}
\label{eq:ephemeris1}
    \frac{t_{0, new}- t_{0, Lit.}}{P_{Lit.}} = n_{old}
\end{equation}\\
[-4.0ex]
\begin{equation}
\label{eq:ephemeris2}
    n_{old} \approx n_{new}, \textrm{with}~ n_{new} \in \mathbb{Z}
\end{equation}\\
[-4.0ex]
\begin{equation}
\label{eq:ephemeris3}
    (t_{0, Lit.} - t_{0, new}) * n_{new} = P_{new}
\end{equation}

Firstly, we take the difference between our newly determined transit time $t_{0, new}$ and the value $t_{0, Lit.}$ reported in \citet{Malavolta2018} and divide this value by the orbital period (see Equation \ref{eq:ephemeris1}). This will equal to the number of elapsed orbits between the two transit times and be a number $n_{old}$ really close to a full integer $n_{new}$ (see Equation \ref{eq:ephemeris2}). Finally, we can use the newly determined $n_{new}$ to update the orbital period (see Equation \ref{eq:ephemeris3}). Our updated ephemeris for K2-141\,c is listed in Table \ref{tab:K2-141c_ephemeris}. We could improve the uncertainties on the orbital period on the transit time for the planet, so that the $3\sigma$ uncertainty on the predicted transit time in 2024 was reduced from 5.2 hours to 16 minutes compared to \citet{Malavolta2018}. This will make it especially easier in the future to schedule observations of K2-141\,b and avoid transits or eclipses of planet c.

\begin{table}[h]
\centering
\caption{Updated ephemeris for K2-141\,c and the $3\sigma$ uncertainty on the predicted transit time in 2022 and 2024.}
\label{tab:K2-141c_ephemeris}

\renewcommand{\arraystretch}{1.6}

\begin{tabular}{l|l|l}\hline\hline
K2-141\,c              & \tablefootmark{(1)}Discovery       & Updated                              \\\hline
$P$ (d)               & $7.74850 \pm 0.00022$  & $7.7489943_{-1.49e-05}^{+1.48e-05}$  \\
\tablefootmark{(2)}$t_0$ (d)     & $7751.1546 \pm 0.0010$ & $8371.07415_{-0.000652}^{+0.000632}$ \\
3$\sigma_{\rm{2022}}$ & 3.7 hours              & 10 minutes                           \\
3$\sigma_{\rm{2024}}$ & 5.2 hours              & 16 minutes                          
\end{tabular}
\renewcommand{\arraystretch}{1}
\tablefoot{\\
\tablefoottext{1}{Based on \citet{Malavolta2018}}\\
\tablefoottext{2}{Expressed as BJD$_{\rm{TDB}}$ - 2450000.0 d}
}

\end{table}

We only fitted for the single transit which occured in K2 C19. This lead to a better ephemeris, but we were not able to improve other orbital parameters like $a/R_*$, $\cos i$ or the planet's size $R_p/R_*$. Our resulting radius of K2-141\,c in units of stellar radii $R_p/R_*$ is consistent with the value reported in \citet{Malavolta2018}. It is, however, strongly correlated with the cosine of the inclination $\cos i$ due to the grazing transit geometry of the planet. This can be also seen in Figure \ref{fig:K2-141c_fit} as the duration of the transit is short and V-shaped.

\clearpage
\onecolumn

\section{Additional Tables}
\subsection{Rescaling of uncertainties}

\begin{table}[!h]
\centering
\caption{$\chi^2$ values for each \Spitzer\ AOR before rescaling them to unity.}
\label{tab:rms_all_AORS}

\renewcommand{\arraystretch}{1.2}

\begin{tabular}{c|c|c|c|c}\hline\hline
AOR & $\chi^2_{\text{old}}$ & rms$_{\text{obs}}$ (ppm)& rms$_{\text{phot}}$ (ppm)& $\beta$\\\hline
1   & 1.155   & 3477    & 3236         & 1.129     \\
2   & 1.082   & 3463    & 3330         & 0.996      \\
3   & 1.103   & 3471    & 3306         & 1.369     \\
4   & 1.106   & 3466    & 3296         & 1.184     \\
5   & 1.098   & 3485    & 3327         & 0.976     \\
6   & 1.102   & 3452    & 3288         & 0.927     \\\hline
All & ---     & 3471    & ---         & 1.132     
\end{tabular}
\renewcommand{\arraystretch}{1}
\tablefoot{
 The values in this table are based on the residuals of the full dataset (\Spitzer\ and \Kepler) and the Toy model without redistribution fit. The photon noise-limited root-mean-square (rms) was calculated like the following: \(\text{rms}_{\text{phot}} = \text{rms}_{\text{obs}}/\sqrt{\chi^2_{\text{old}}}\), where rms$_{\text{obs}}$ is the rms of the residuals. $\beta$ describes the ratio between the achieved standard deviation ($\text{rms}_{\text{obs}}$) of the binned residuals and the standard deviation in absence of red noise. It was calculated using the ``time-averaging'' method \citep{Pont2006, Winn2007, Winn2008} by calculating median values of this ratio for binnings around the transit duration. The Allan deviation plots for each AOR can be found in Section \ref{sec:all_allan_spitzer}.
}

\end{table}

\clearpage

\subsection{Parameters of the fitted models}

\begin{table*}[!h]
\centering
\caption{All free parameters used in the models which were fitted to the \Spitzer\ data alone.}
\renewcommand{\arraystretch}{1.6}
\begin{tabular}{c|c|c}\hline\hline
Model & Free Parameters & $\Delta$BIC \\\hline\hline
Sinusoidal Model ($\phi$ = 0) & \begin{tabular}[c]{@{}c@{}}$t_0$, $R_p/R_*$, $a/R_*$, $\cos i$ \\ $A$, $f_p/f_*$,           \\ \cspitzer,\\ \cubicgwspitzer \end{tabular} & 0 \\\hline
Sinusoidal Model ($\phi$ free)& \begin{tabular}[c]{@{}c@{}}$t_0$, $R_p/R_*$, $a/R_*$, $\cos i$ \\ $A$, $f_p/f_*$, $\phi$,   \\ \cspitzer,\\ \cubicgwspitzer \end{tabular} & 8.8 \\\hline
Two Temp. Model         & \begin{tabular}[c]{@{}c@{}}$t_0$, $R_p/R_*$, $a/R_*$, $\cos i$,\\ $T_*$, $T_{p,n}$, $T_{p,d}$, \\ \cspitzer,\\ \cubicgwspitzer \end{tabular} & 9.6
\end{tabular}
\renewcommand{\arraystretch}{1}
\label{tab:Spitzer_models}
\end{table*}

\begin{table*}[!h]
\centering
\caption{All free parameters used in the models which were fitted to the joint \Kepler\ and \Spitzer\ dataset.}
\renewcommand{\arraystretch}{1.6}
\begin{tabular}{c|c|c}\hline\hline
Model    & Free Parameters & $\Delta$BIC \\\hline\hline
Toy Model ($F$ = 0)   & \begin{tabular}[c]{@{}c@{}}$t_0$, $R_p/R_*$, $a/R_*$, $\cos i$, $P$,\\ $T_*$, $A_g$,      \\ \cjoint,\\ \cubicgwspitzer\end{tabular}                 & 0 \\\hline
Toy Model ($F$ free)& \begin{tabular}[c]{@{}c@{}}$t_0$, $R_p/R_*$, $a/R_*$, $\cos i$, $P$,\\ $T_*$, $F$, $A_g$, \\ \cjoint,\\ \cubicgwspitzer\end{tabular}           & 12.0 \\\hline
Two Temp. Model           & \begin{tabular}[c]{@{}c@{}}$t_0$, $R_p/R_*$, $a/R_*$, $\cos i$, $P$,\\ $T_*$, $T_{p,d}$, $T_{p,n}$, $A_g$, \\ \cjoint,\\ \cubicgwspitzer\end{tabular} & 22.2
\end{tabular}
\renewcommand{\arraystretch}{1}
\label{tab:joint_models}
\end{table*}

\begin{landscape}

\begin{table*}[!h]
\centering
\caption{Best fit systematic parameters for all 6 models.}
\renewcommand{\arraystretch}{1.6}
\begin{tabular}{c|cccccc}\hline\hline
 & \multicolumn{6}{c}{Model Name} \\
Parameter             & Sin. M. ($\phi$ = 0) & Sin. M. ($\phi$ free)& Two Temp. & Toy Model ($F$ = 0)         & Toy Model ($F$ free)   & Two Temp. Model\\ \hline
$c_{\textrm{AOR1}}$   & $75412_{-391}^{+386}$             & $75414_{-396}^{+400}$            & $75430_{-415}^{+392}$            & $75427_{-385}^{+389}$             & $75420_{-420}^{+421}$               & $75414_{-398}^{+396}$\\
$v_{\textrm{AOR2}}$   & $-130.8_{-30.1}^{+29.3}$          & $-120.7_{-29.5}^{+29.9}$         & $-131.8_{-29.2}^{+30.0}$         & $-128.5_{-29.3}^{+29.2}$          & $-133.8_{-29.7}^{+29.1}$            & $-128.8_{-29.7}^{+29.8}$\\
$c_{\textrm{AOR2}}$   & $1.601e+05_{-19270}^{+19730}$     & $1.5344e+05_{-19630}^{+19420}$   & $1.6074e+05_{-19600}^{+19190}$   & $1.586e+05_{-19180}^{+19180}$     & $1.6209e+05_{-19090}^{+19610}$      & $1.5883e+05_{-19590}^{+19470}$\\
$c_{\textrm{AOR3}}$   & $77386_{-445}^{+435}$             & $77296_{-448}^{+465}$            & $77397_{-470}^{+483}$            & $77407_{-445}^{+467}$             & $77393_{-422}^{+448}$               & $77381_{-472}^{+461}$\\
$c_{\textrm{AOR4}}$   & $74910.5_{-120.8}^{+123.5}$       & $74905.4_{-121.3}^{+122.5}$      & $74921.8_{-125.4}^{+123.5}$      & $74917.1_{-123.2}^{+124.9}$       & $74932.3_{-129.8}^{+122.2}$         & $74912.4_{-117.8}^{+123.1}$\\
$c_{\textrm{AOR5}}$   & $80275_{-270}^{+268}$             & $80249_{-264}^{+275}$            & $80278_{-276}^{+276}$            & $80248_{-275}^{+277}$             & $80260_{-263}^{+263}$               & $80265_{-271}^{+290}$\\
$c_{\textrm{AOR6}}$   & $75321_{-243}^{+243}$             & $75299_{-243}^{+253}$            & $75312_{-242}^{+243}$            & $75330_{-240}^{+244}$             & $75348_{-235}^{+230}$               & $75324_{-249}^{+240}$\\
$c_{\textrm{K2C12}}$  & ---                               & ---                              & ---                              & $1_{-1.485e-06}^{+1.481e-06}$     & $0.99999_{-1.459e-06}^{+1.465e-06}$ & $0.99999_{-1.498e-06}^{+1.513e-06}$\\
$c_{\textrm{K2C19}}$  & ---                               & ---                              & ---                              & $0.99999_{-2.44e-06}^{+2.54e-06}$ & $0.99999_{-2.49e-06}^{+2.5e-06}$    & $0.99999_{-2.59e-06}^{+2.49e-06}$\\
$Y_{\textrm{1, AOR1}}$& $-0.06349_{-0.00852}^{+0.00869}$  & $-0.06353_{-0.00880}^{+0.00881}$ & $-0.06388_{-0.00864}^{+0.00924}$ & $-0.06384_{-0.00857}^{+0.00855}$  & $-0.06366_{-0.00927}^{+0.00936}$    & $-0.06353_{-0.00871}^{+0.00884}$\\
$X_{\textrm{3, AOR2}}$& $-0.094898_{-0.01582}^{+0.01552}$ & $-0.09531_{-0.01518}^{+0.01565}$ & $-0.09523_{-0.01578}^{+0.01631}$ & $-0.09460_{-0.01562}^{+0.01606}$  & $-0.09470_{-0.01550}^{+0.01595}$    & $-0.09462_{-0.01607}^{+0.01542}$\\
$X_{\textrm{1, AOR3}}$& $-0.09822_{-0.01005}^{+0.01039}$  & $-0.09612_{-0.01077}^{+0.01049}$ & $-0.09847_{-0.01114}^{+0.01098}$ & $-0.09872_{-0.01076}^{+0.01038}$  & $-0.09836_{-0.01034}^{+0.00984}$    & $-0.09809_{-0.01063}^{+0.01100}$\\
$X_{\textrm{3, AOR4}}$& $-0.14612_{-0.01132}^{+0.01114}$  & $-0.14570_{-0.01124}^{+0.01115}$ & $-0.14717_{-0.01133}^{+0.01156}$ & $-0.14681_{-0.01145}^{+0.01134}$  & $-0.14809_{-0.01120}^{+0.01193}$    & $-0.14635_{-0.01128}^{+0.01083}$\\
$X_{\textrm{1, AOR5}}$& $-0.14020_{-0.00769}^{+0.00776}$  & $-0.13899_{-0.00756}^{+0.0078}$  & $-0.14013_{-0.00787}^{+0.00762}$ & $-0.13967_{-0.00796}^{+0.00791}$  & $-0.14041_{-0.00774}^{+0.00757}$    & $-0.14005_{-0.0078}^{+0.00774}$\\
$Y_{\textrm{3, AOR5}}$& $-0.06069_{-0.01643}^{+0.01649}$  & $-0.06409_{-0.01625}^{+0.01678}$ & $-0.06114_{-0.01607}^{+0.01659}$ & $-0.06047_{-0.01724}^{+0.01671}$  & $-0.05830_{-0.01686}^{+0.01666}$    & $-0.060765_{-0.01633}^{+0.01637}$\\
$X_{\textrm{3, AOR6}}$& $-0.0893_{-0.0207}^{+0.0208}$     & $-0.08745_{-0.0211}^{+0.0208}$   & $-0.0884_{-0.0214}^{+0.0210}$    & $-0.0899_{-0.0213}^{+0.0220}$     & $-0.0894_{-0.0205}^{+0.0207}$       & $-0.0888_{-0.0221}^{+0.0213}$\\
$Y_{\textrm{3, AOR6}}$& $-0.0836_{-0.0201}^{+0.0201}$     & $-0.0839_{-0.0200}^{+0.0205}$    & $-0.08393_{-0.0204}^{+0.0209}$   & $-0.08447_{-0.01994}^{+0.02029}$  & $-0.08561_{-0.02042}^{+0.01998}$    & $-0.08352_{-0.02018}^{+0.01934}$\\
\end{tabular}
\renewcommand{\arraystretch}{1}

\label{tab:systematic_values}
\end{table*}

\end{landscape}

\clearpage
\section{Additional Plots}

\subsection{Systematics}

\begin{figure*}[!h]
\centering
\includegraphics[width=1.0\textwidth]{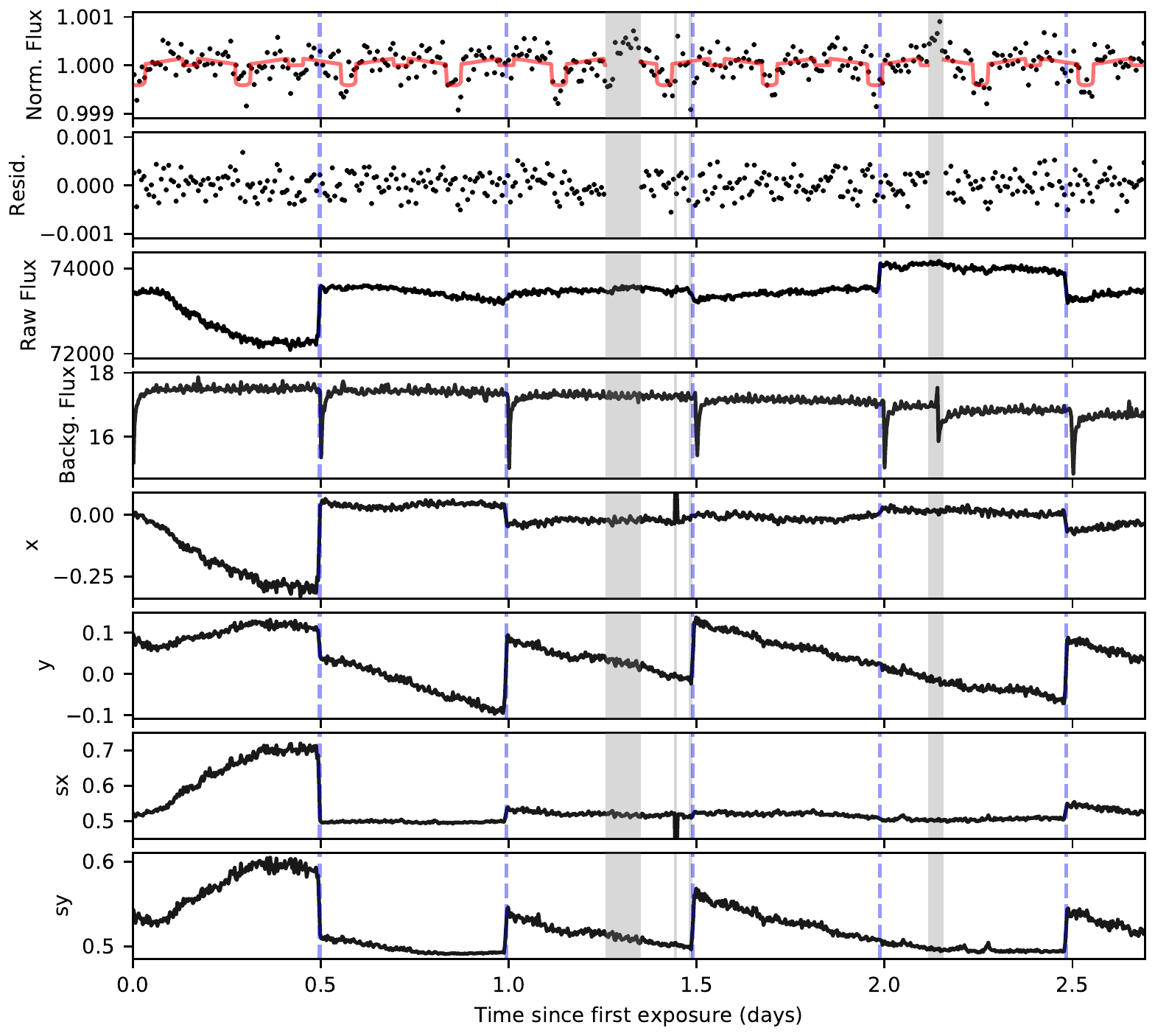}
\caption{Diagnostic plot of the full \Spitzer\ observations: The vertical, dashed blue lines indicate the start of a new Astronomical Observation Request (AOR). The data in the vertical, grey shaded regions were removed for the fit due to systematic effects. The red line in the top panel shows the best fit model of the \Spitzer\ data. The difference between the model and the normalized flux can be seen the the panel below. The data has been binned downed to 10 minutes in the top two panels and to 4 minutes in the lower panels. The observed raw flux in $\mu$Jy/pixel is shown in the third panel. The background flux in the fourth panel is showing changes at the beginning of every AOR as expected due to changes in pointing. An outlier segment in AOR5 which was manually removed from the dataset has been able to be attributed to a strong cosmic ray hit on the detector. The 2D images showing this event can be found in Fig. \ref{fig:frames}. x and y are the pixel position of the target relative to the ``sweet spot''. sx and sy describe the Gaussian widths of the star's point spread function.}
\label{fig:systematics}
\end{figure*}

\begin{figure*}[!h]
\centering
\includegraphics[width=1.0\textwidth]{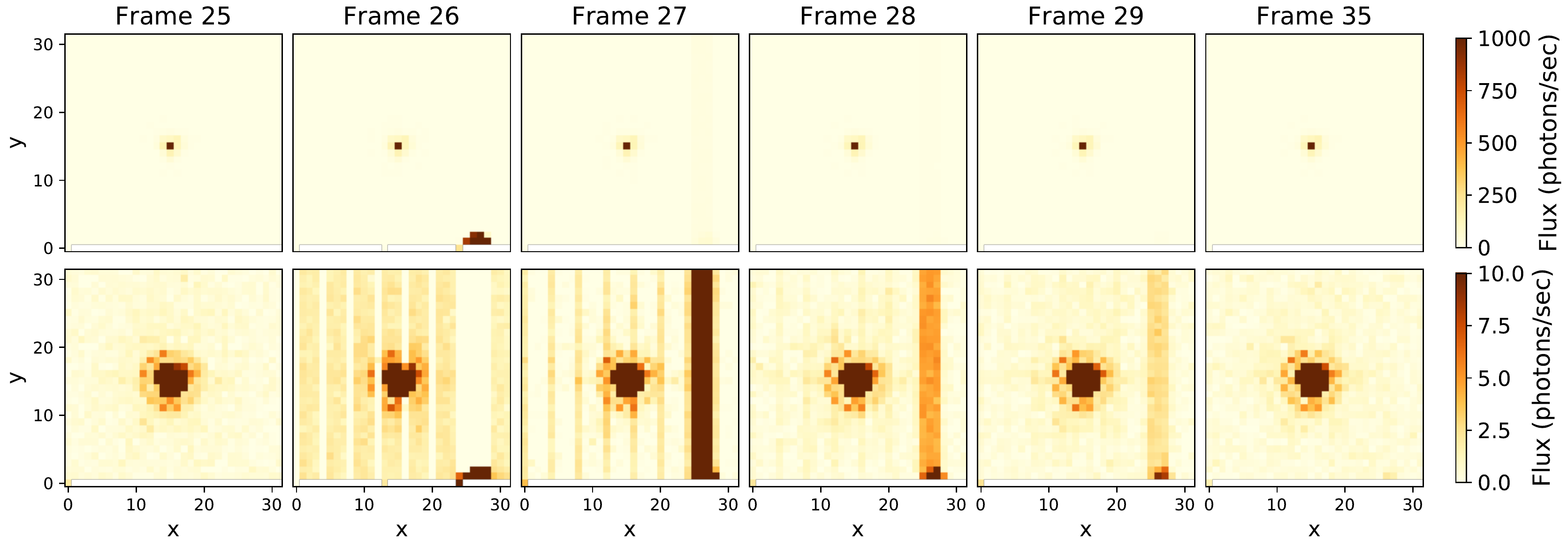}
\caption{\Spitzer\ data frames showing strong cosmic ray hit on the \Spitzer\ IRAC detector which led to changes in the background flux during AOR5. Each column shows the same Basic Calibrated Data (BCD, provided by the \Spitzer\ Science Center) frame but at a different contrast. The star is located in the center of every frame, whereas the cosmic ray hit can be seen in the lower right of the second frame. All frames in this plot have been taken from the same BCD cube which typically consist out of 64 images with $32 \times 32$ pixel.}
\label{fig:frames}
\end{figure*}

\begin{figure*}[!h]
\centering
\includegraphics[width=1.0\textwidth]{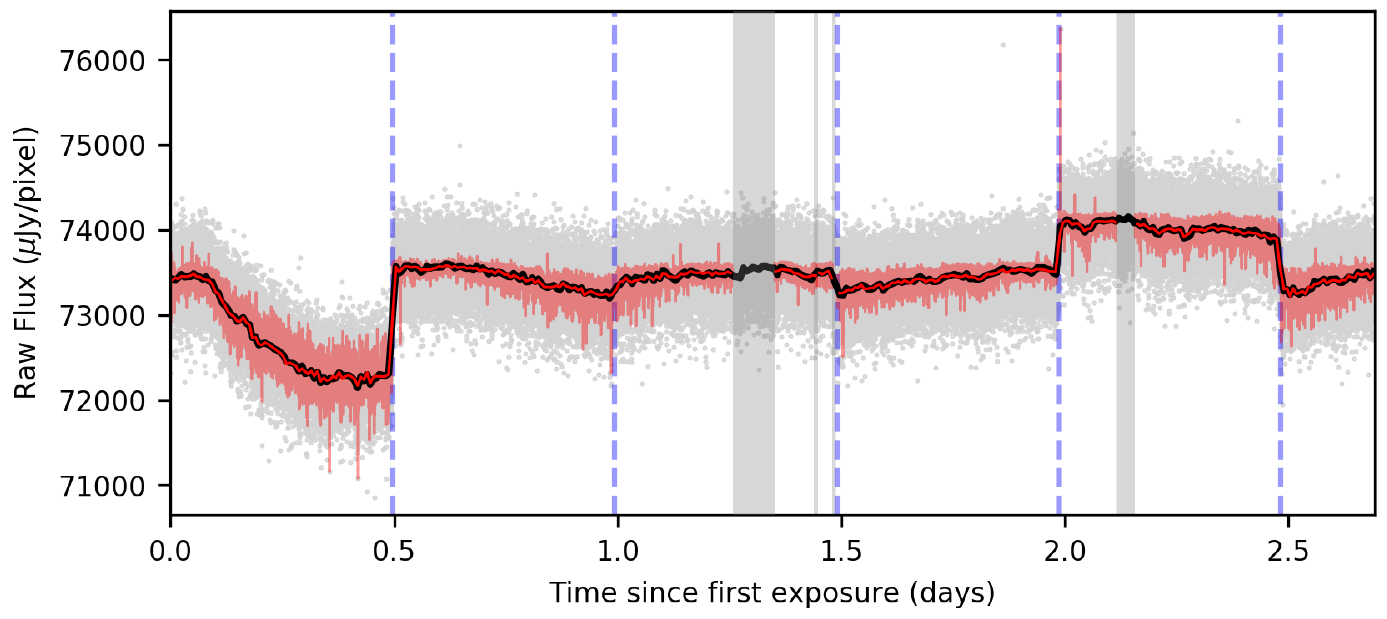}
\caption{The raw, unbinned light curve is shown with grey dots with the best fitting model in light red. The vertical, dashed blue lines indicate the start of a new Astronomical Observation Request (AOR). The data in the vertical, grey shaded regions were removed for the fit due to systematic effects. The solid black (red) line shows the raw light curve (best fitting model) binned down to 10 minutes. The planetary signatures (transit, phase curve variation or eclipse) are too weak to be seen in the raw data. E.g.: the transit depth of K2-141\,b is $\sim$425 ppm which leads to a dip of just $\sim$30 $\mu$Jy/pixel.}
\label{fig:raw_Spitzer}
\end{figure*}

\begin{figure*}[!h]
\centering
\includegraphics[width=1.0\textwidth]{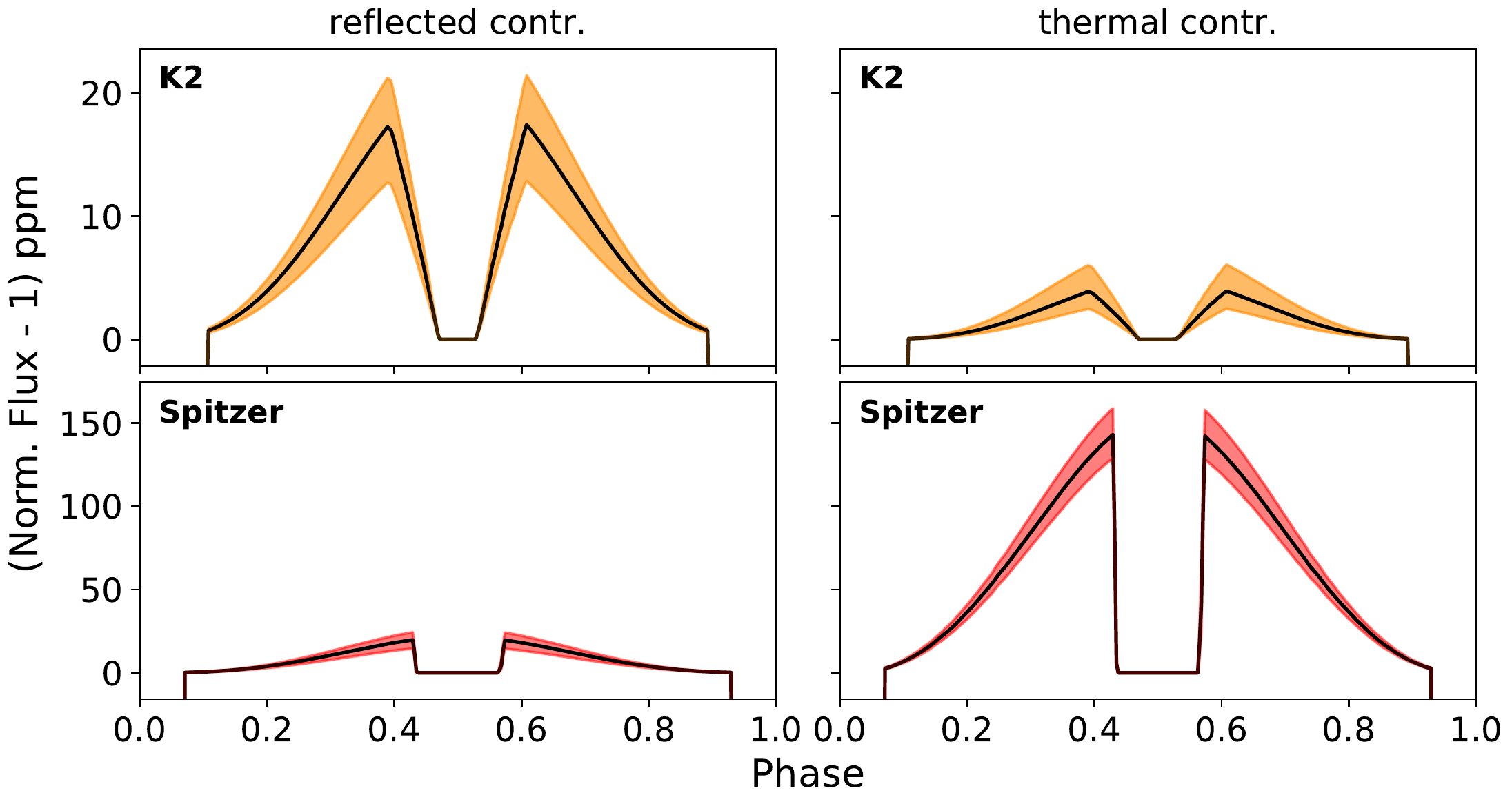}
\caption{Reflected and thermal contributions to the total flux in the \Kepler\ and \Spitzer\ bandpasses using our best fitting model (toy model without heat redistribution). The shaded areas show the $1\sigma$ uncertainties on the best fitting phase curve shape. The K2 phase curves shown here take the longer exposure time into account (30 minutes for K2 Campaign 12) which leads to a less steep ingress and egrees at the eclipse.}
\label{fig:contributions}
\end{figure*}

\clearpage
\subsection{\Spitzer\ pointings}
\label{sec:pointing}
\begin{figure*}[!h]
\centering
\includegraphics[width=0.47\textwidth]{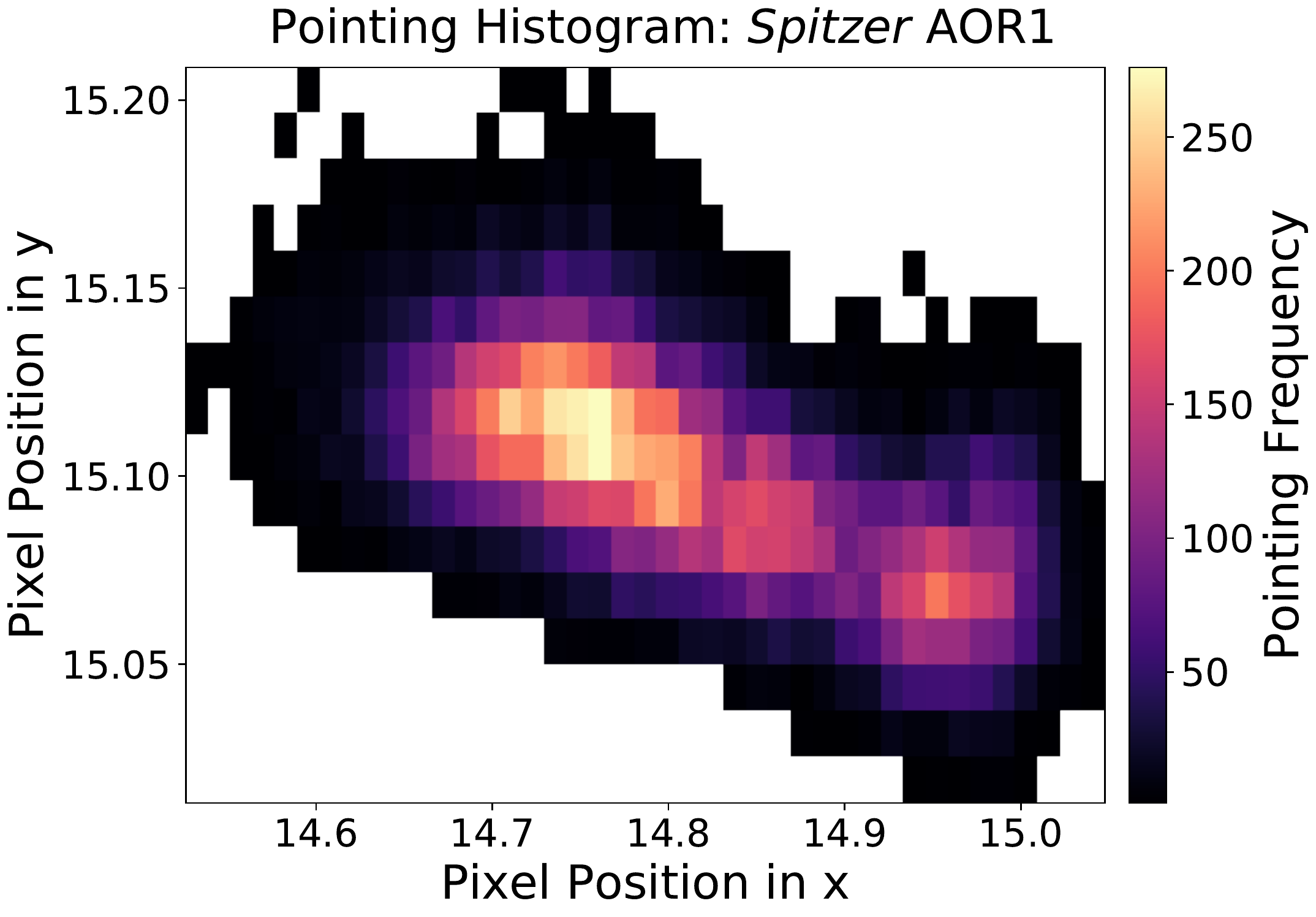}
\hspace{0.5cm}
\vspace{0.5cm}
\includegraphics[width=0.47\textwidth]{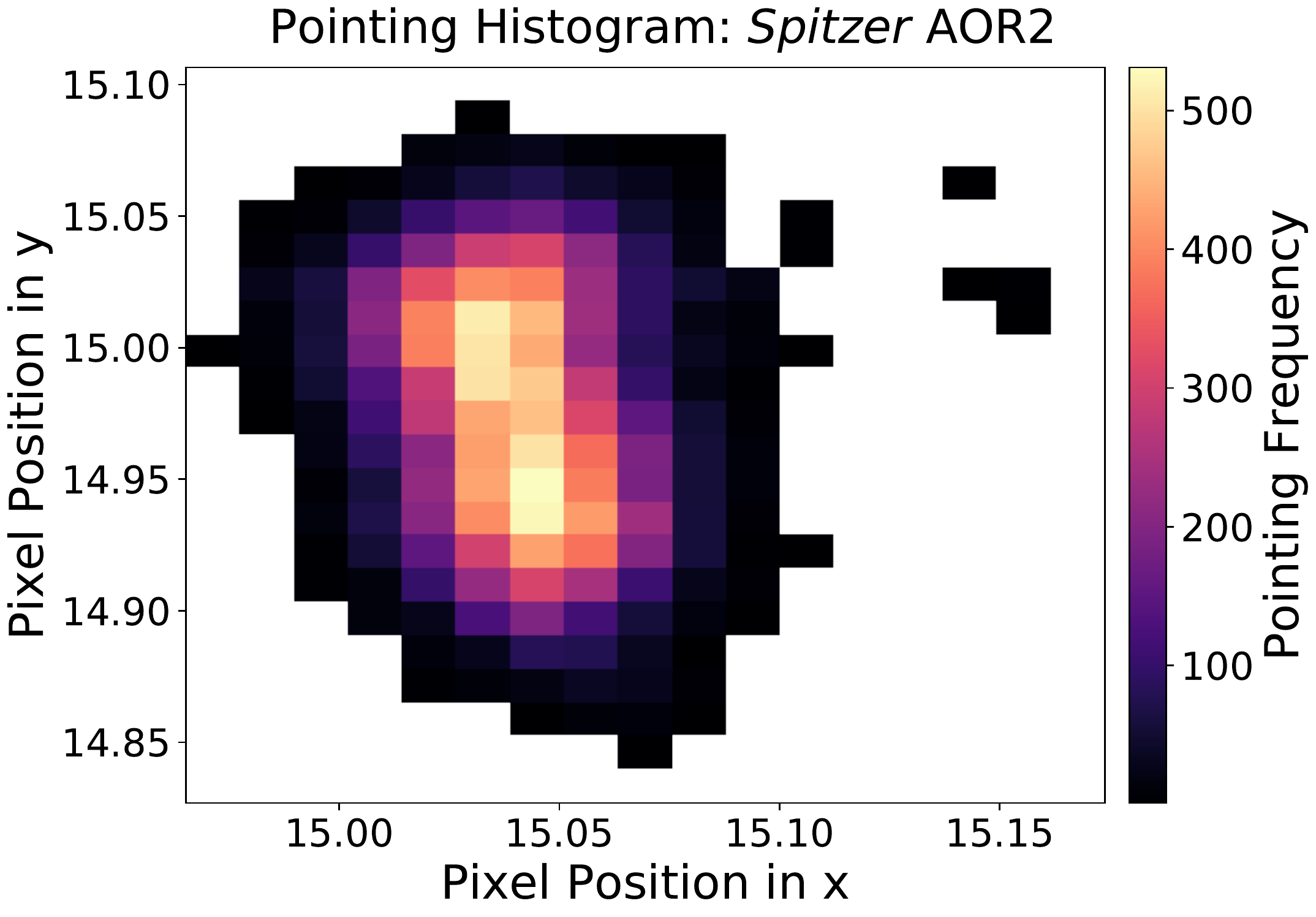}

\includegraphics[width=0.47\textwidth]{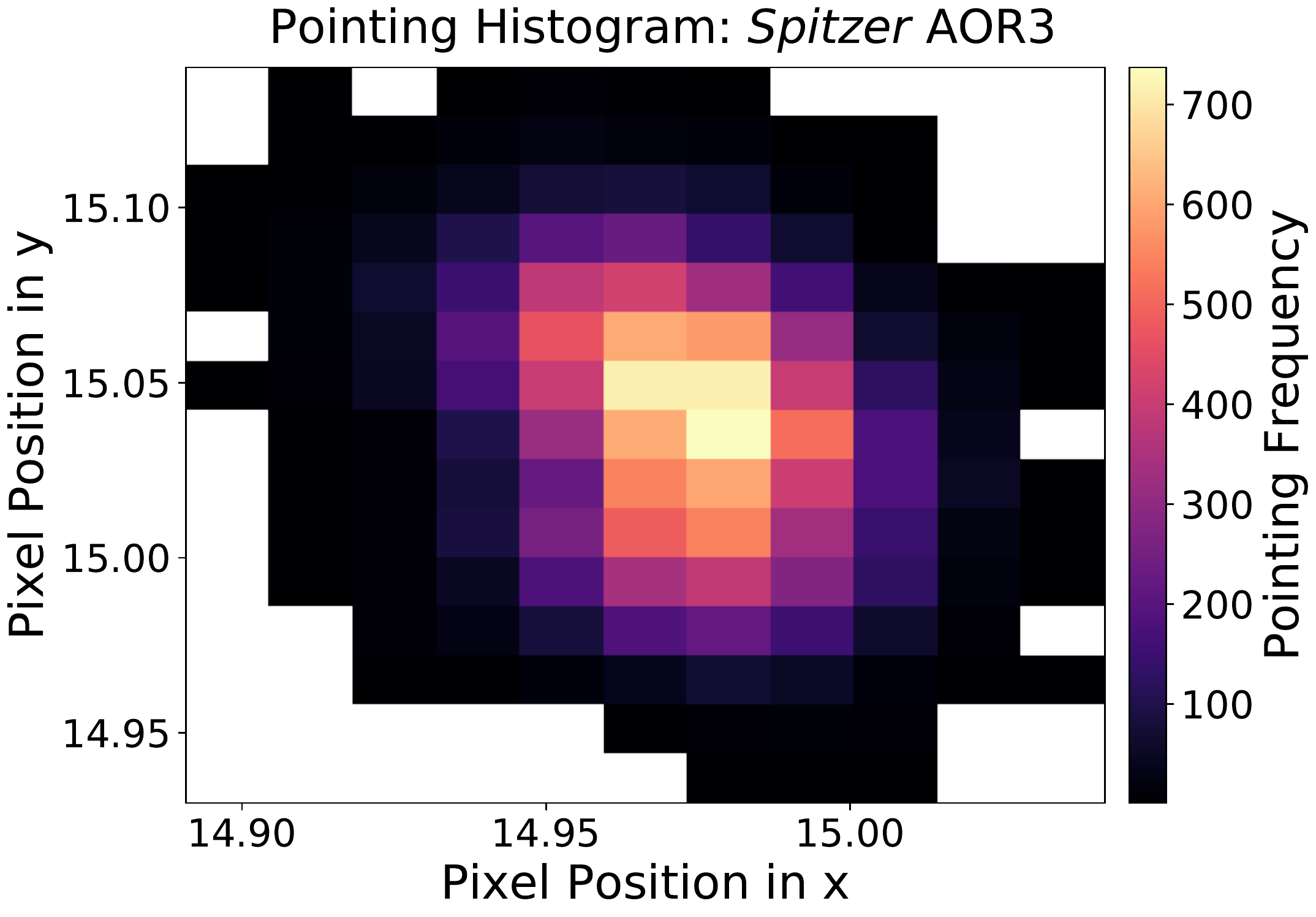}
\hspace{0.5cm}
\vspace{0.5cm}
\includegraphics[width=0.47\textwidth]{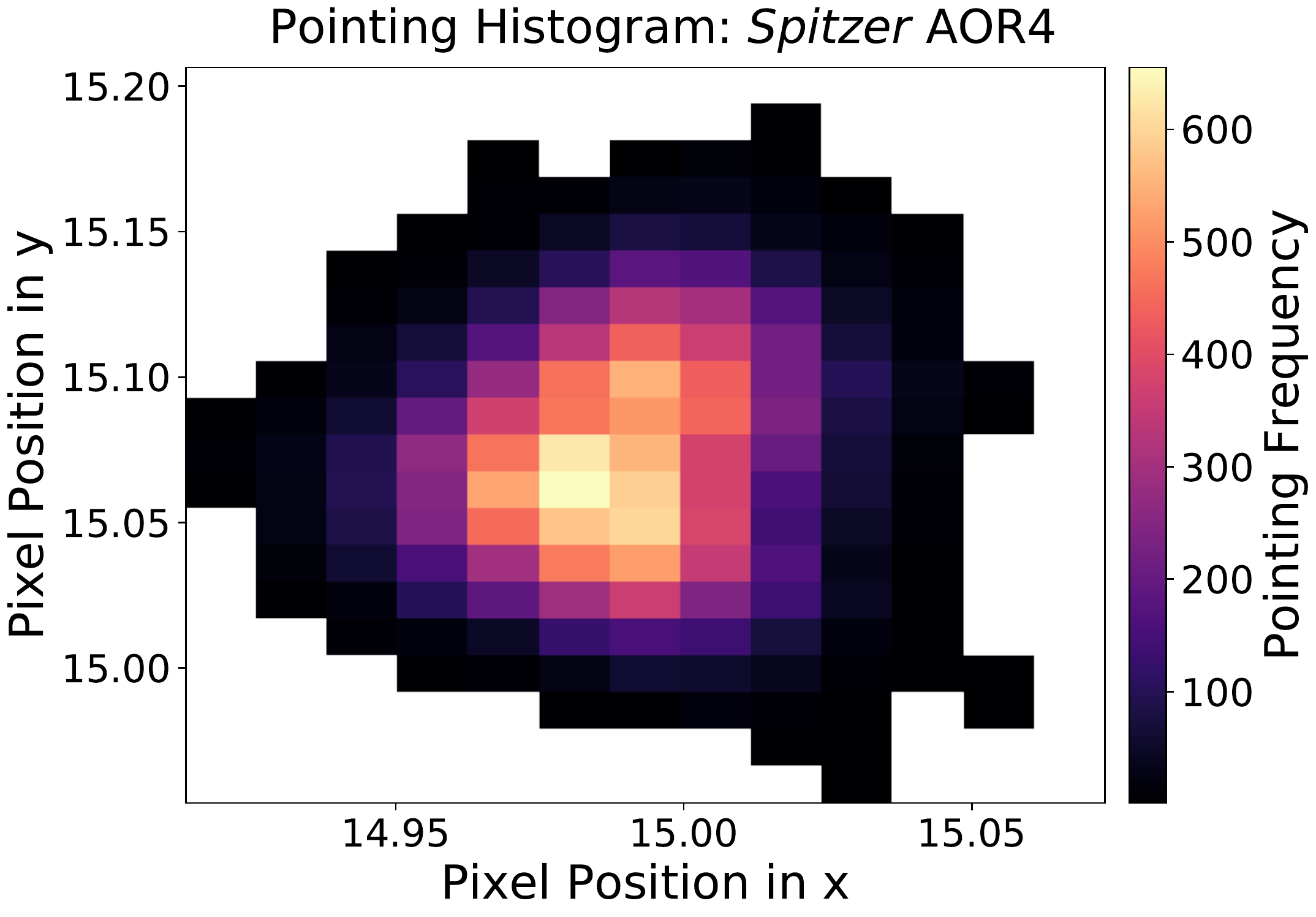}

\includegraphics[width=0.47\textwidth]{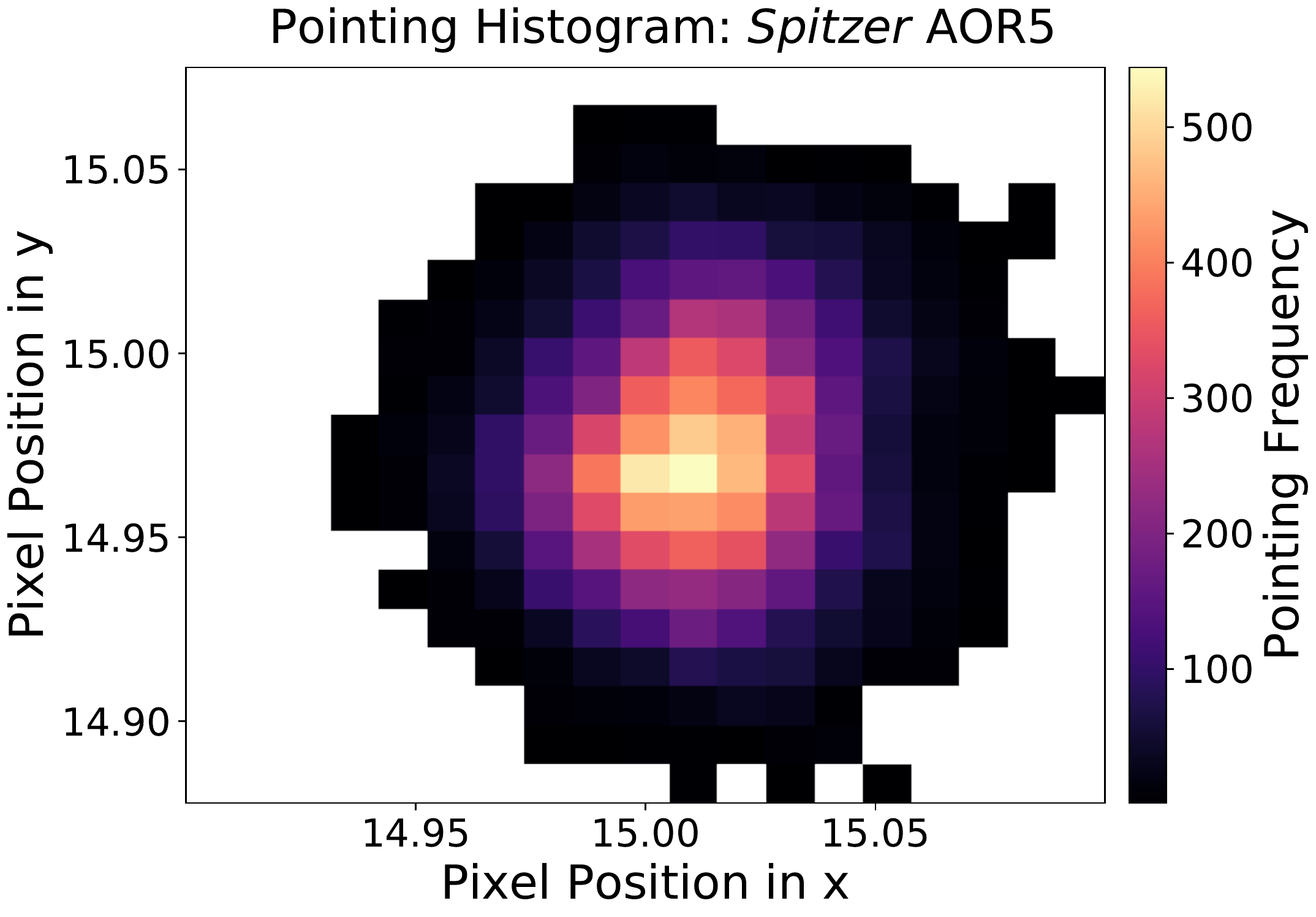}
\hspace{0.5cm}
\includegraphics[width=0.47\textwidth]{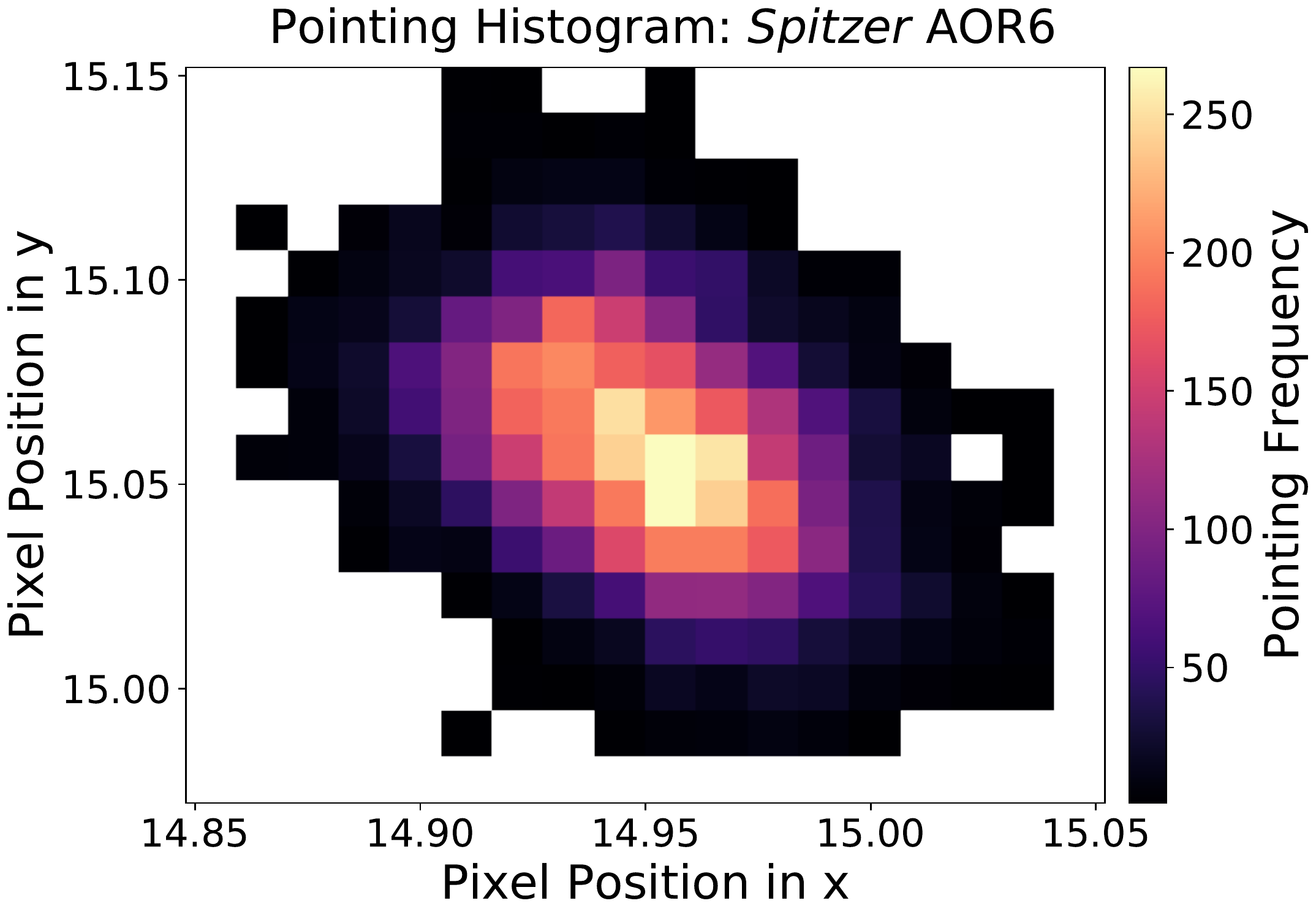}
\caption{Pointing diagrams for all six \Spitzer\ AORs. The color map indicates the frequency of exposures for which the centroid of the star hit a certain position.}
\end{figure*}

\clearpage
\subsection{\Spitzer\ BLISS maps}
\label{sec:bliss}
\begin{figure*}[!h]
\centering
\includegraphics[width=0.47\textwidth]{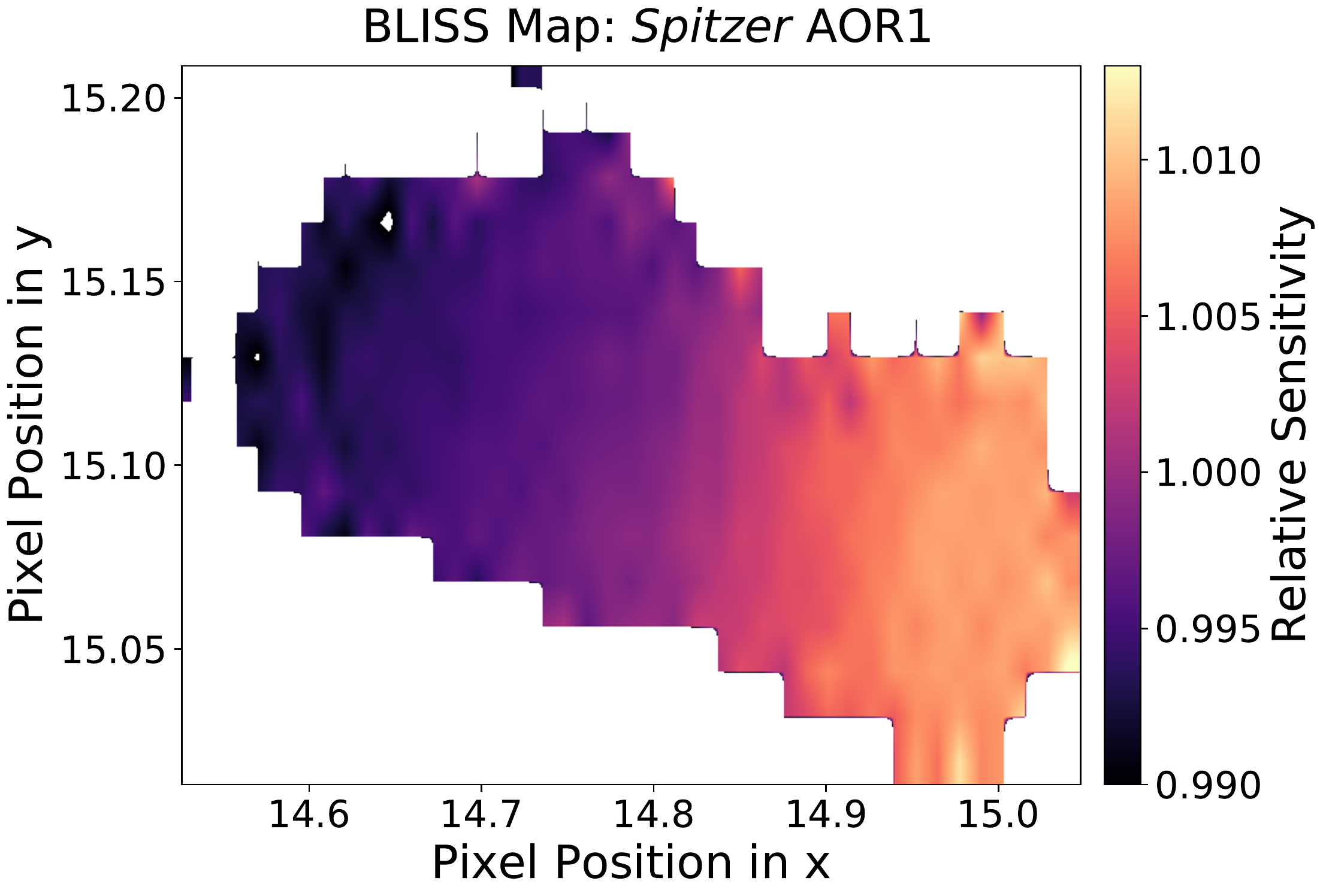}
\hspace{0.5cm}
\vspace{0.5cm}
\includegraphics[width=0.47\textwidth]{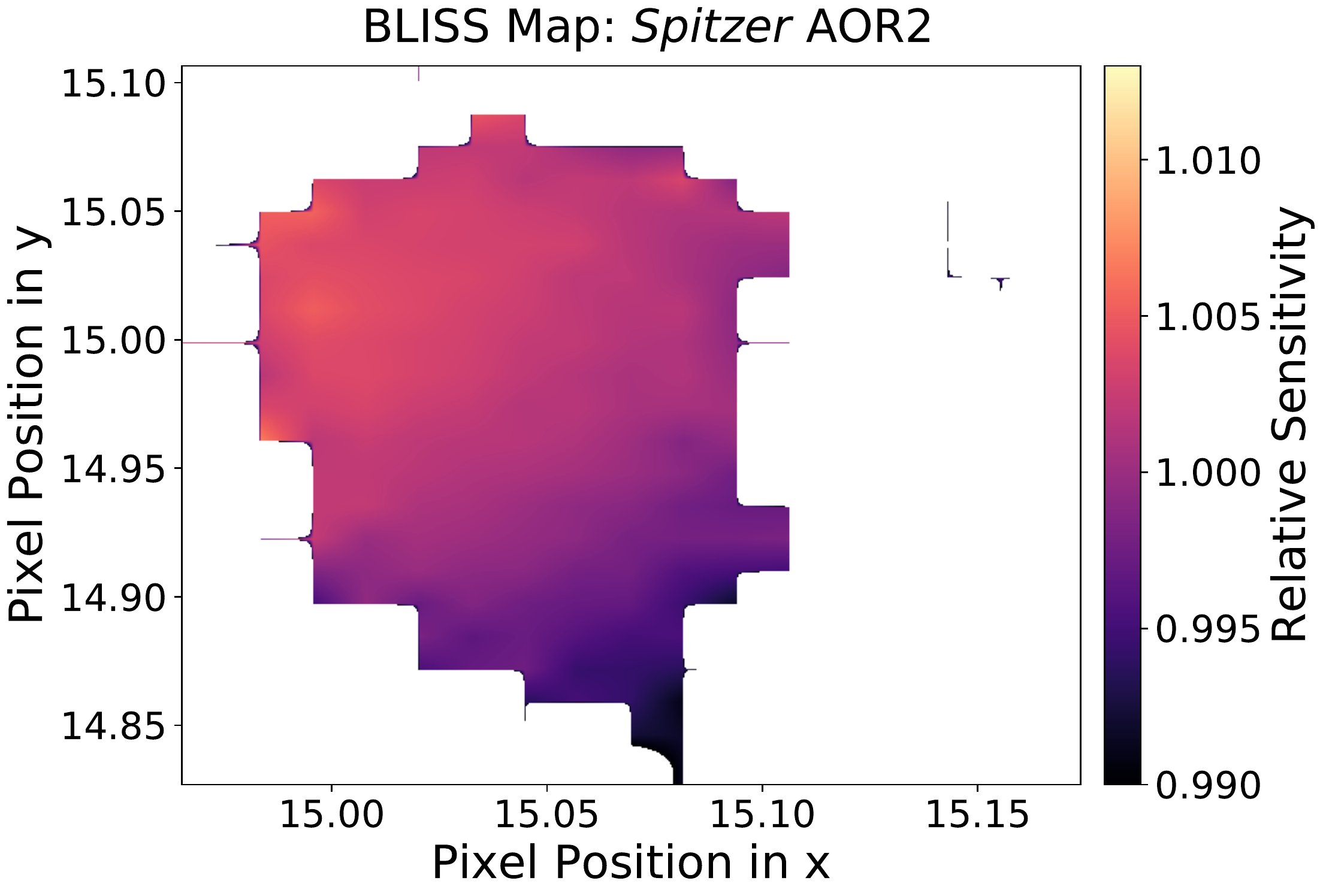}

\includegraphics[width=0.47\textwidth]{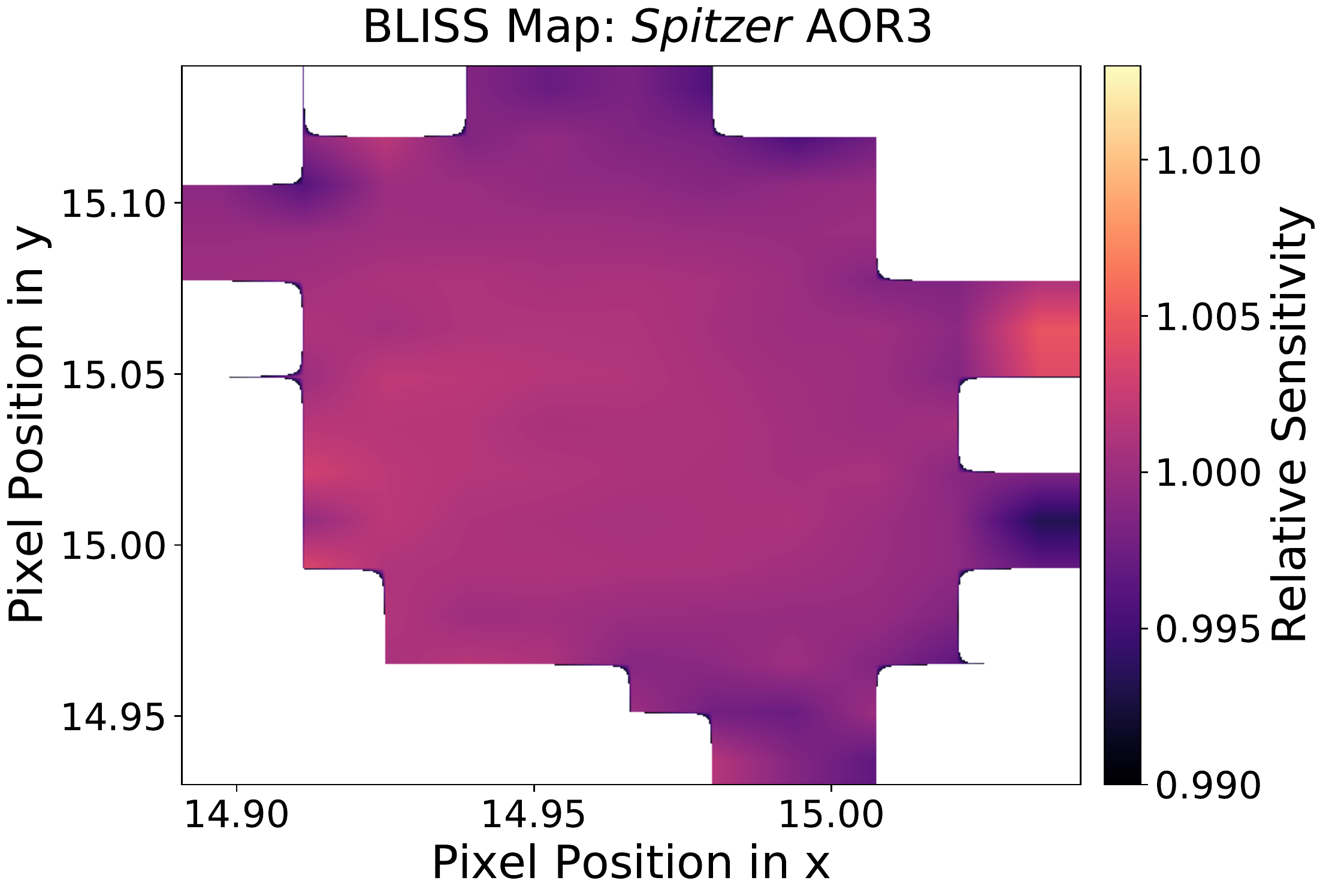}
\hspace{0.5cm}
\vspace{0.5cm}
\includegraphics[width=0.47\textwidth]{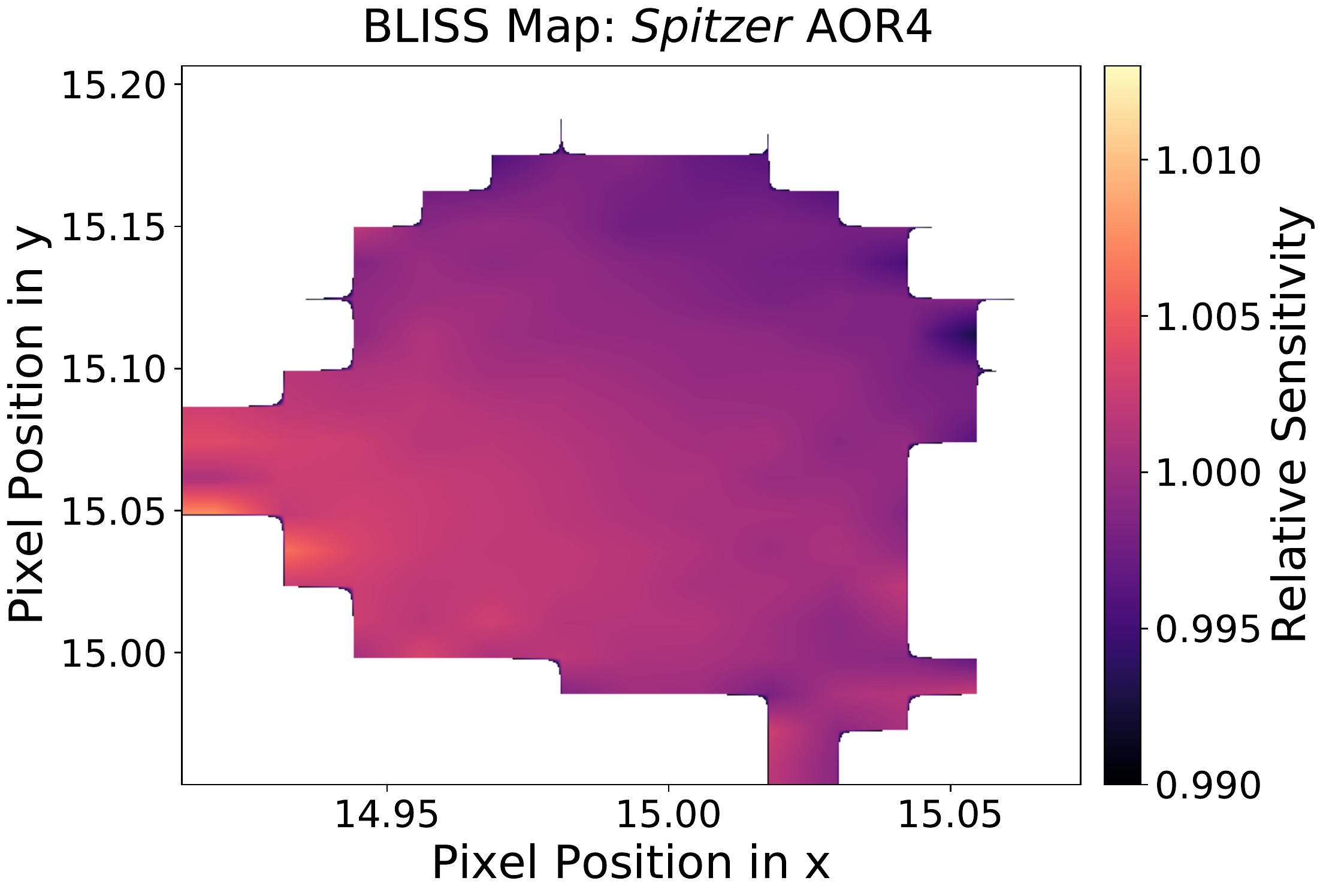}

\includegraphics[width=0.47\textwidth]{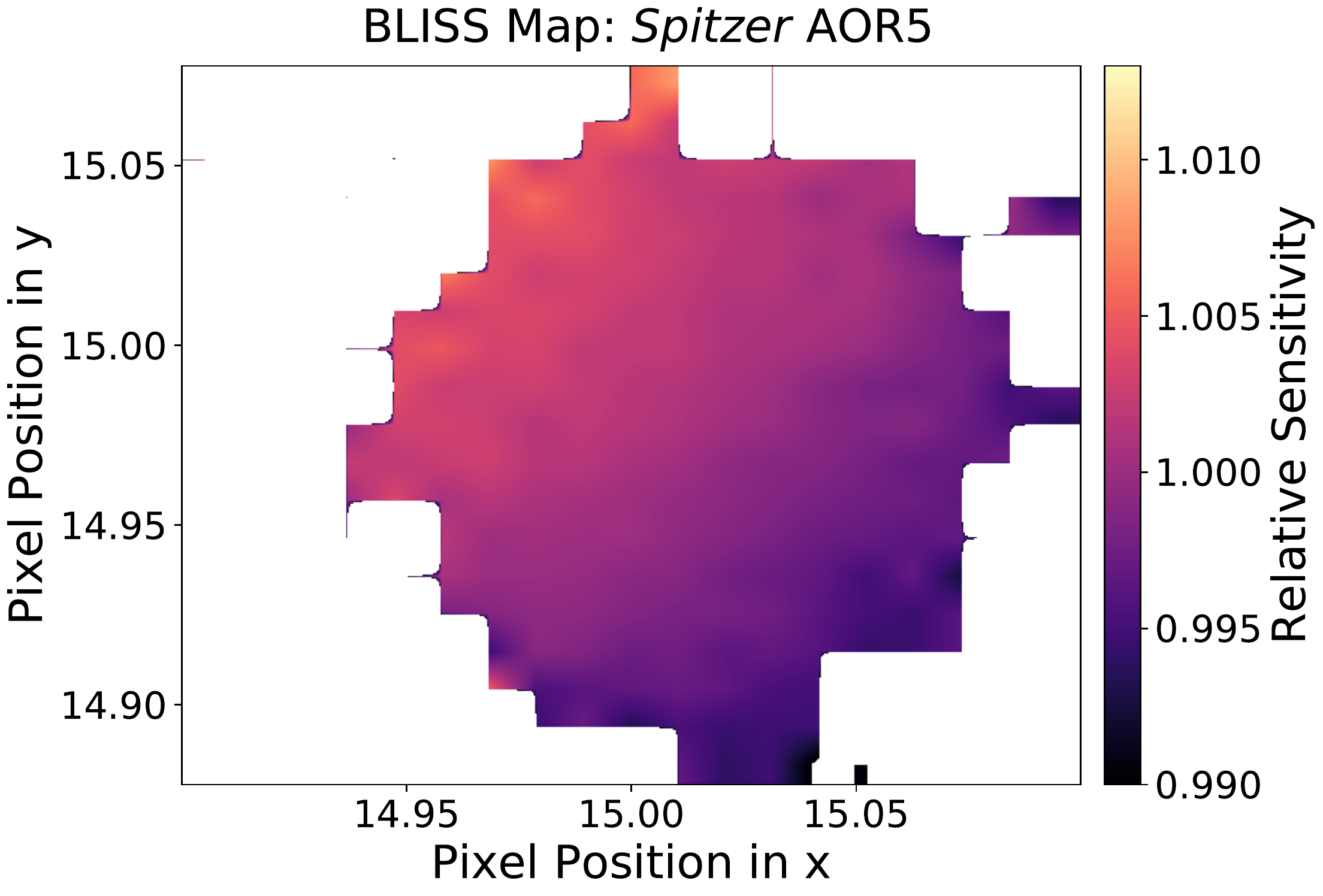}
\hspace{0.5cm}
\includegraphics[width=0.47\textwidth]{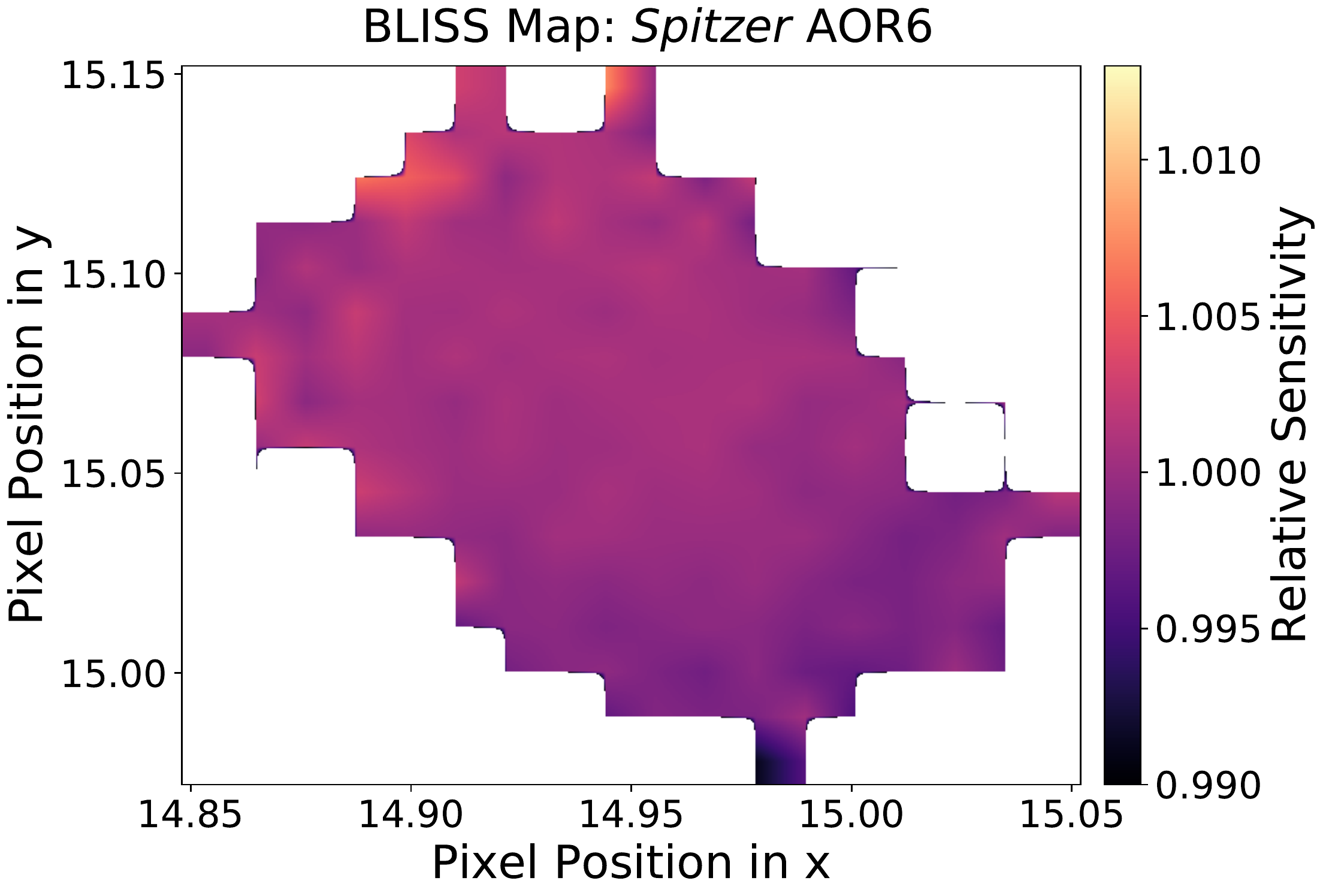}
\caption{BLISS sensitivity maps for all six \Spitzer\ AORs.}
\end{figure*}

\clearpage
\subsection{\Spitzer\ Allan deviation plots}
\label{sec:all_allan_spitzer}
\begin{figure*}[!h]
\centering
\includegraphics[width=0.47\textwidth]{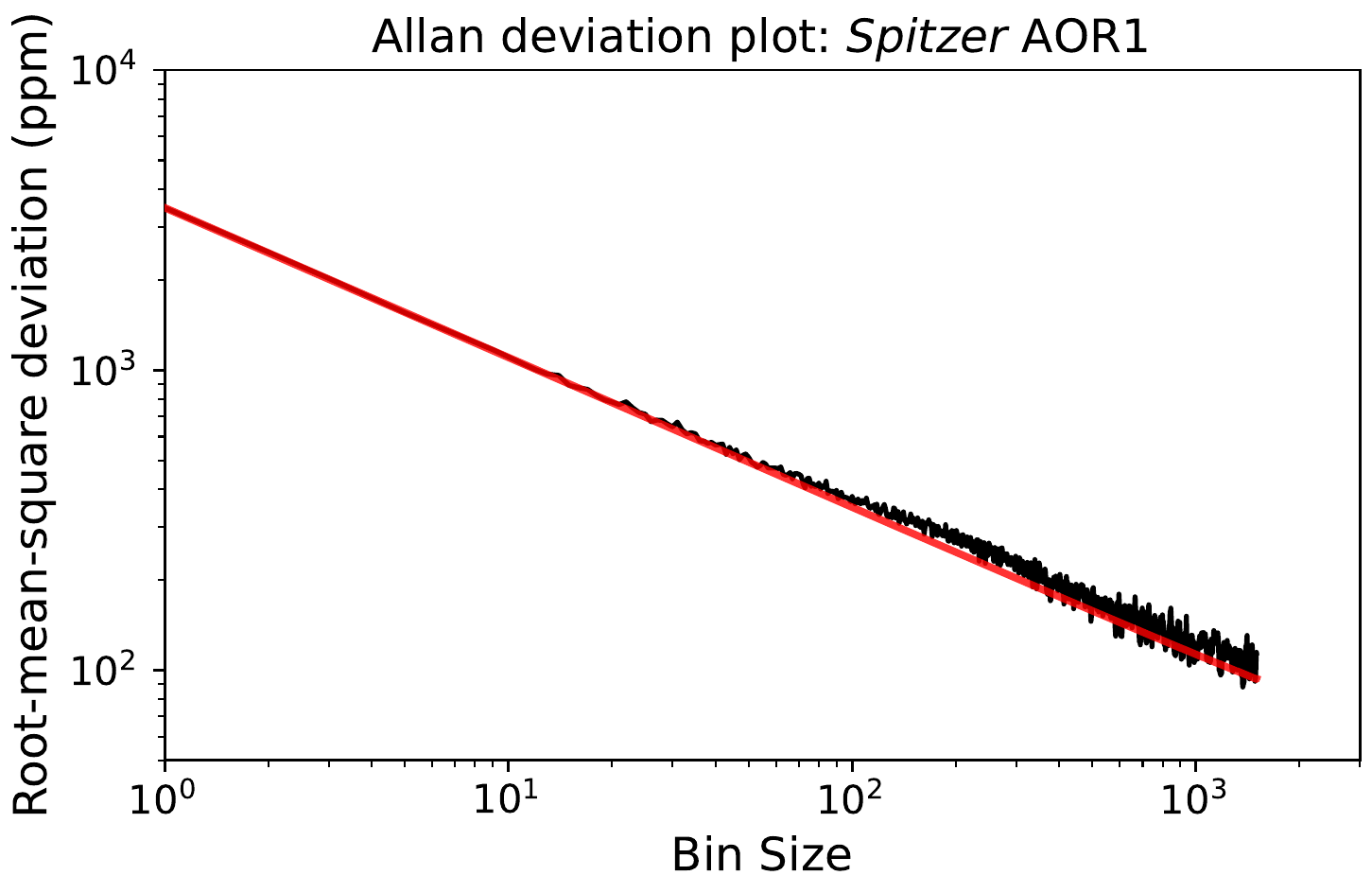}
\hspace{0.5cm}
\vspace{0.5cm}
\includegraphics[width=0.47\textwidth]{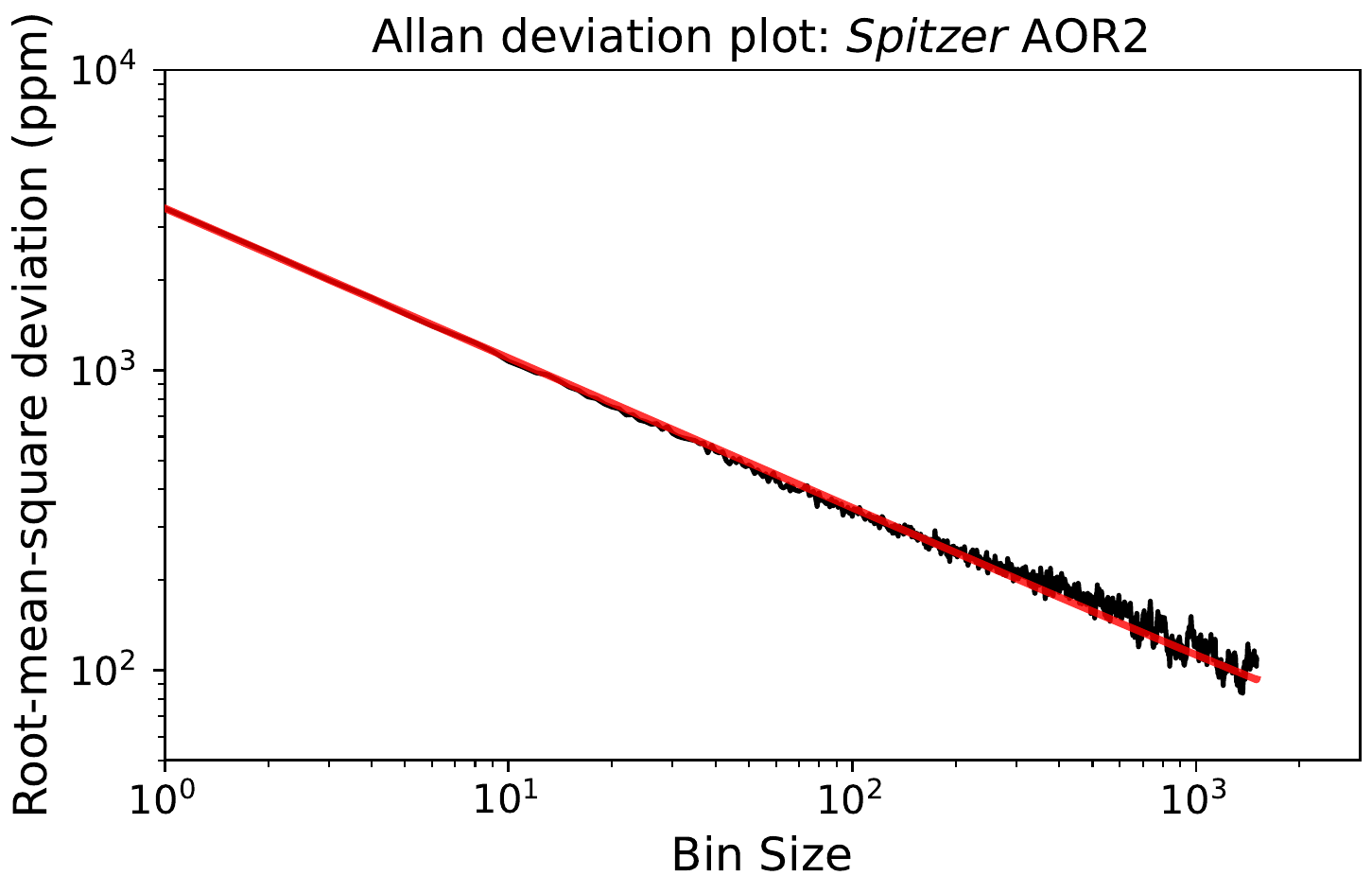}

\includegraphics[width=0.47\textwidth]{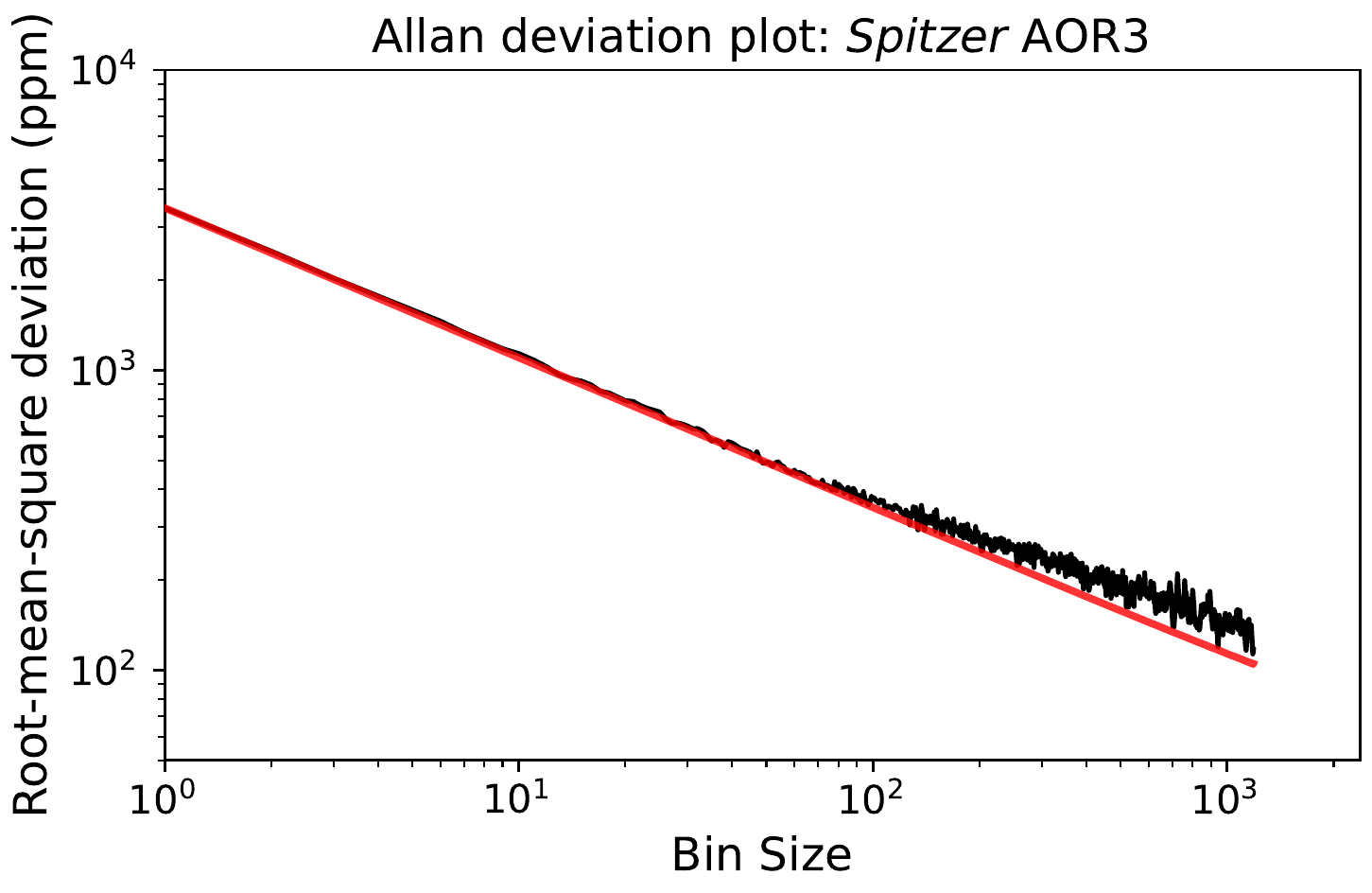}
\hspace{0.5cm}
\vspace{0.5cm}
\includegraphics[width=0.47\textwidth]{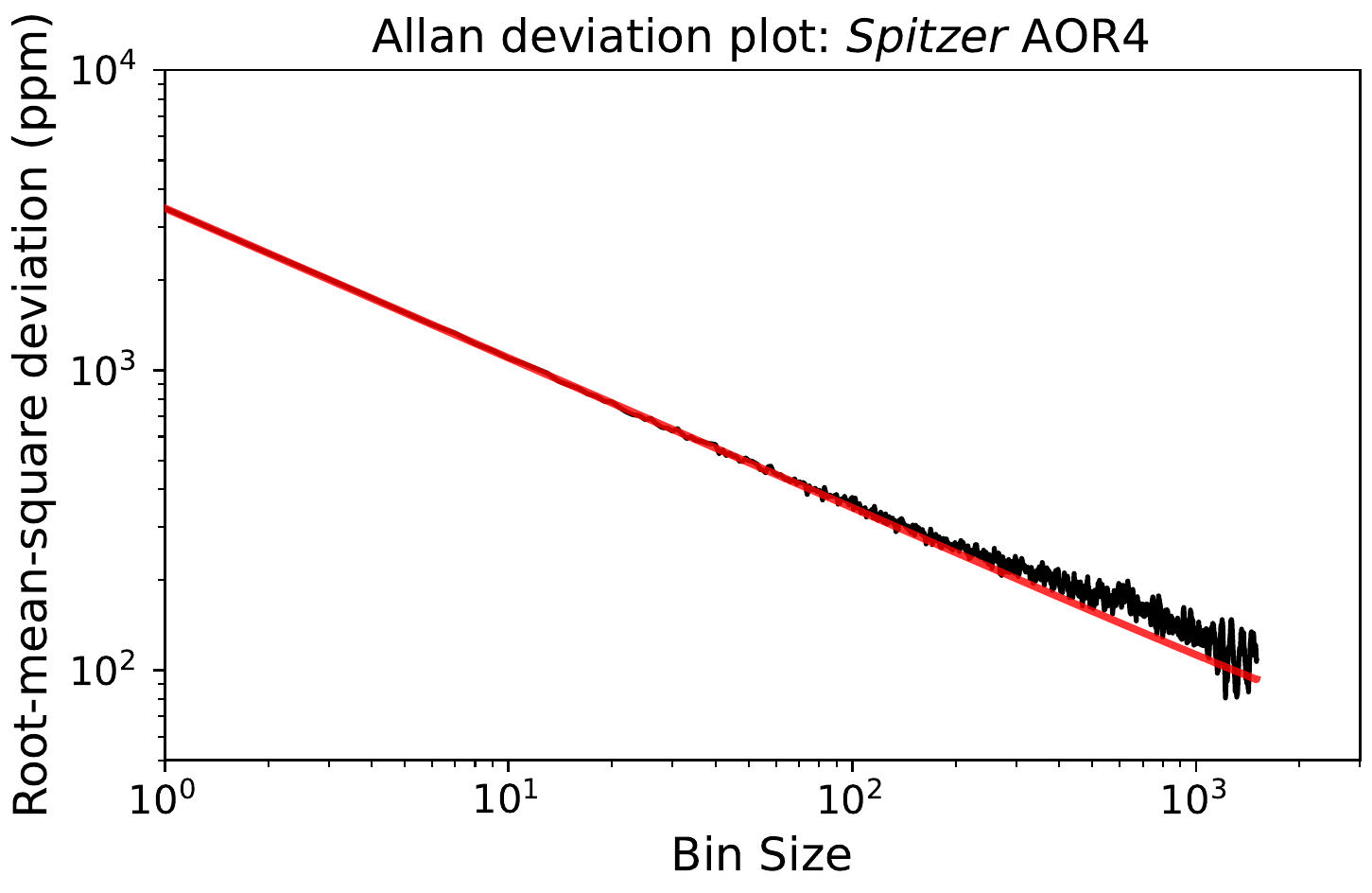}

\includegraphics[width=0.47\textwidth]{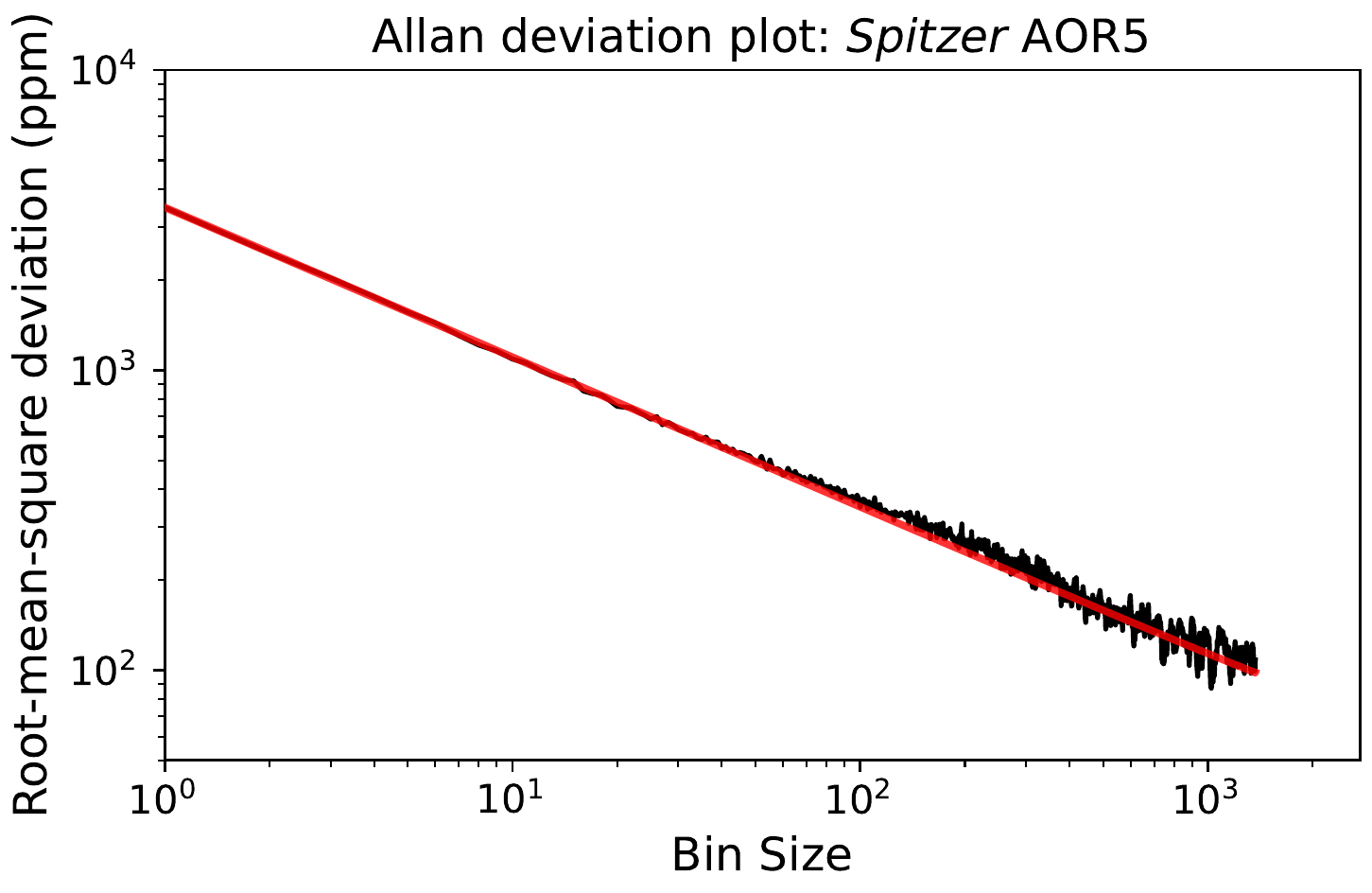}
\hspace{0.5cm}
\includegraphics[width=0.47\textwidth]{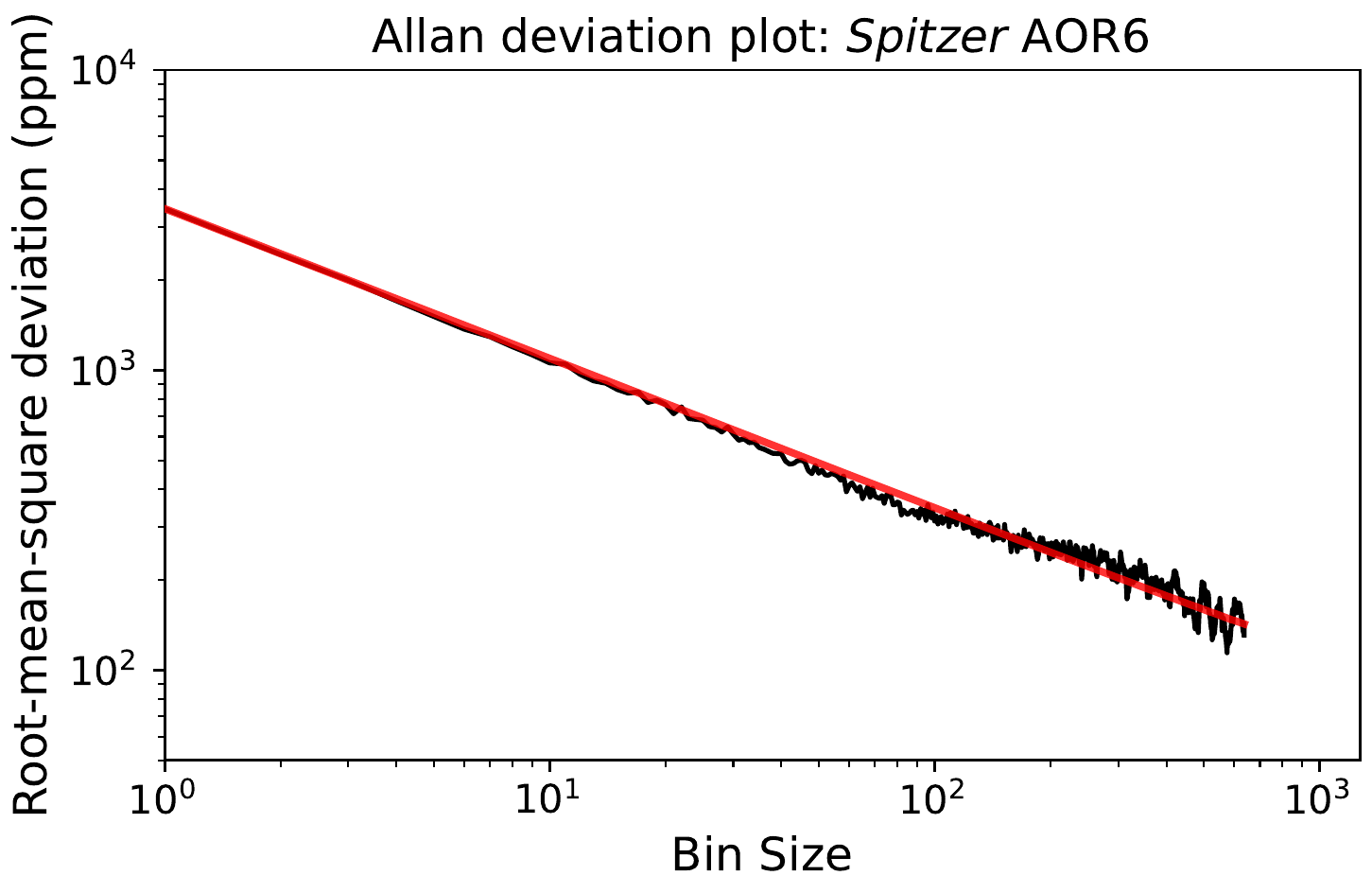}
\caption{Allan deviation plots for all six \Spitzer\ AORs. The residuals (black curve) are calculated by taking the difference of the full dataset (\Spitzer\ and \Kepler) and the Toy model without redistribution fit. A bin size of one depicts no binning at all. The red line shows the expected root-mean-square (rms) for Gaussian noise following the inverse square root law.}
\end{figure*}

\clearpage
\subsection{\Spitzer\ fit: Sinusoidal ($\phi$ = 0)} \label{fig:corner1}
\begin{figure*}[!h]
\centering
\includegraphics[trim=1.6cm 1.6cm 0cm 0cm, clip, width=1\textwidth]{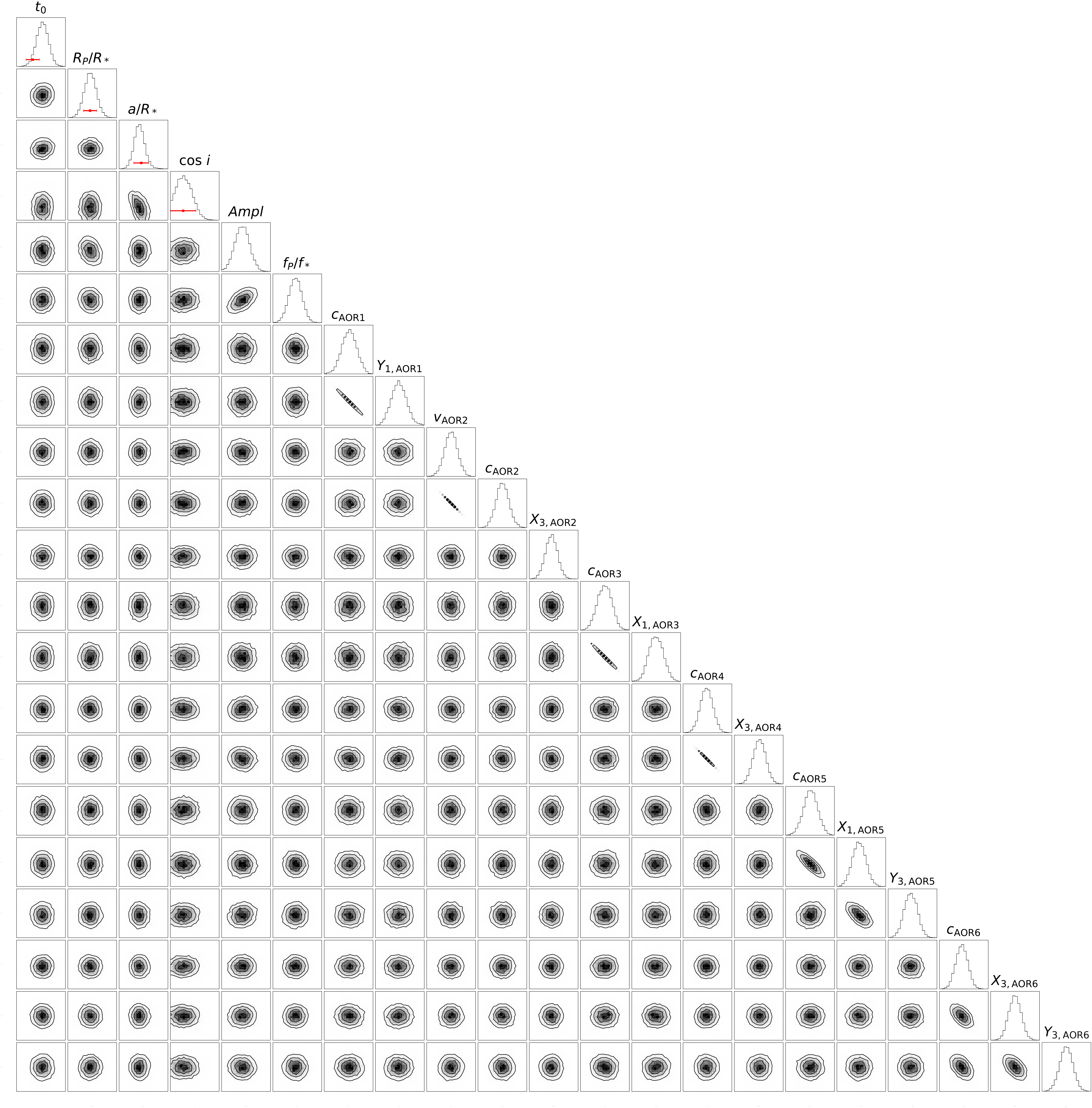}
\caption{MCMC corner plot for the sinusoidal model fit without a hotspot offset ($\phi$ = 0) to the \Spitzer\ data. The red bars for $t_0$, $R_p/R_*$, \ar\ and $\cos i$ show the Gaussian priors which were used in this fit. The prior values and the best fit values are listed in Table \ref{tab:Spitzer values}. The resulting values for the systematic parameters are in Table \ref{tab:systematic_values}. A list with all fit parameters can be found in Table \ref{tab:Spitzer_models}.}
\label{}
\end{figure*}

\clearpage
\subsection{\Spitzer\ fit: Sinusoidal ($\phi$ free)}
\begin{figure*}[!h]
\centering
\includegraphics[trim=1.6cm 1.6cm 0cm 0cm, clip, width=1\textwidth]{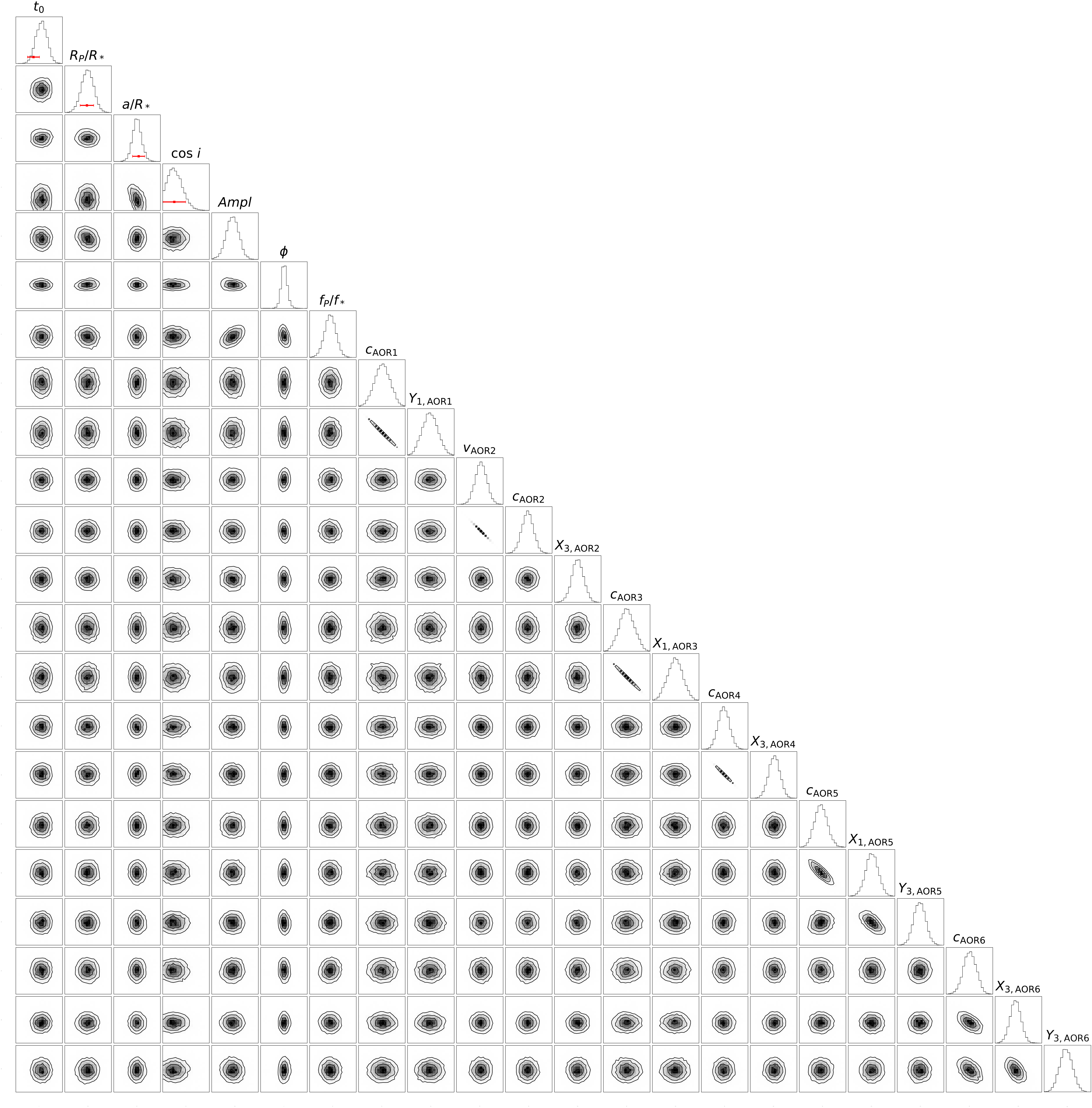}
\caption{MCMC corner plot for the sinusoidal model fit with a hotspot offset ($\phi$ free) to the \Spitzer\ data. The red bars for $t_0$, $R_p/R_*$, \ar\ and $\cos i$ show the Gaussian priors which were used in this fit. The prior values and the best fit values are listed in Table \ref{tab:Spitzer values}. The resulting values for the systematic parameters are in Table \ref{tab:systematic_values}. A list with all fit parameters can be found in Table \ref{tab:Spitzer_models}.}
\label{}
\end{figure*}

\clearpage
\subsection{\Spitzer\ fit: Two Temperature Model} \label{app:S2T}
\begin{figure*}[!h]
\centering
\includegraphics[trim=1.6cm 1.6cm 0cm 0cm, clip, width=1\textwidth]{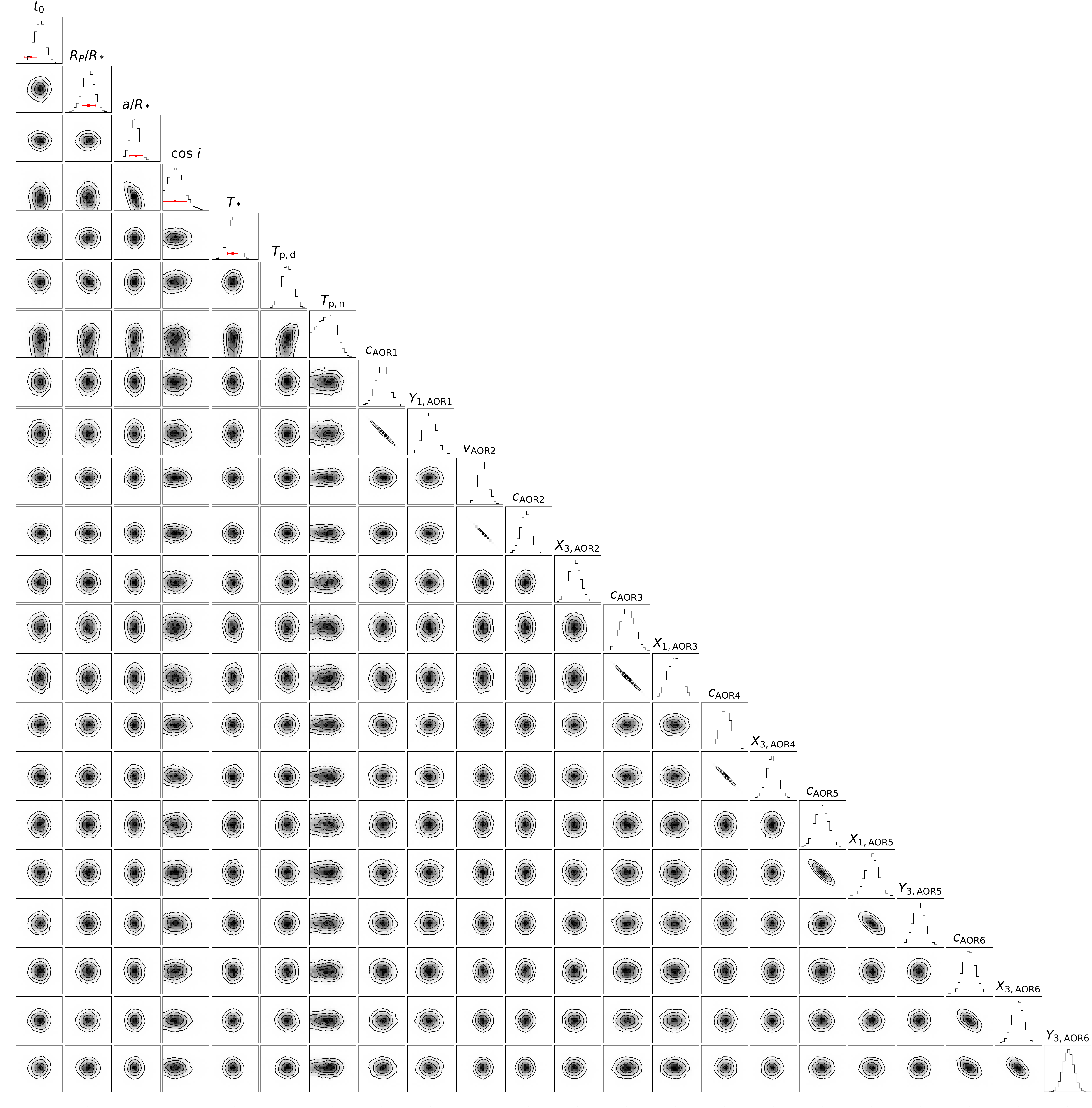}
\caption{MCMC corner plot for the two temperature model fit to the \Spitzer\ data. The red bars for $t_0$, $R_p/R_*$, \ar\, $\cos i$ and $T_*$ show the Gaussian priors which were used in this fit. The prior values and the best fit values are listed in Table \ref{tab:Spitzer values}. The resulting values for the systematic parameters are in Table \ref{tab:systematic_values}. A list with all fit parameters can be found in Table \ref{tab:Spitzer_models}.}
\label{}
\end{figure*}

\clearpage
\subsection{Joint (K2 and \Spitzer) fit: Toy Model ($F$ = 0)}
\begin{figure*}[!h]
\centering
\includegraphics[trim=1.6cm 1.6cm 0cm 0cm, clip, width=1\textwidth]{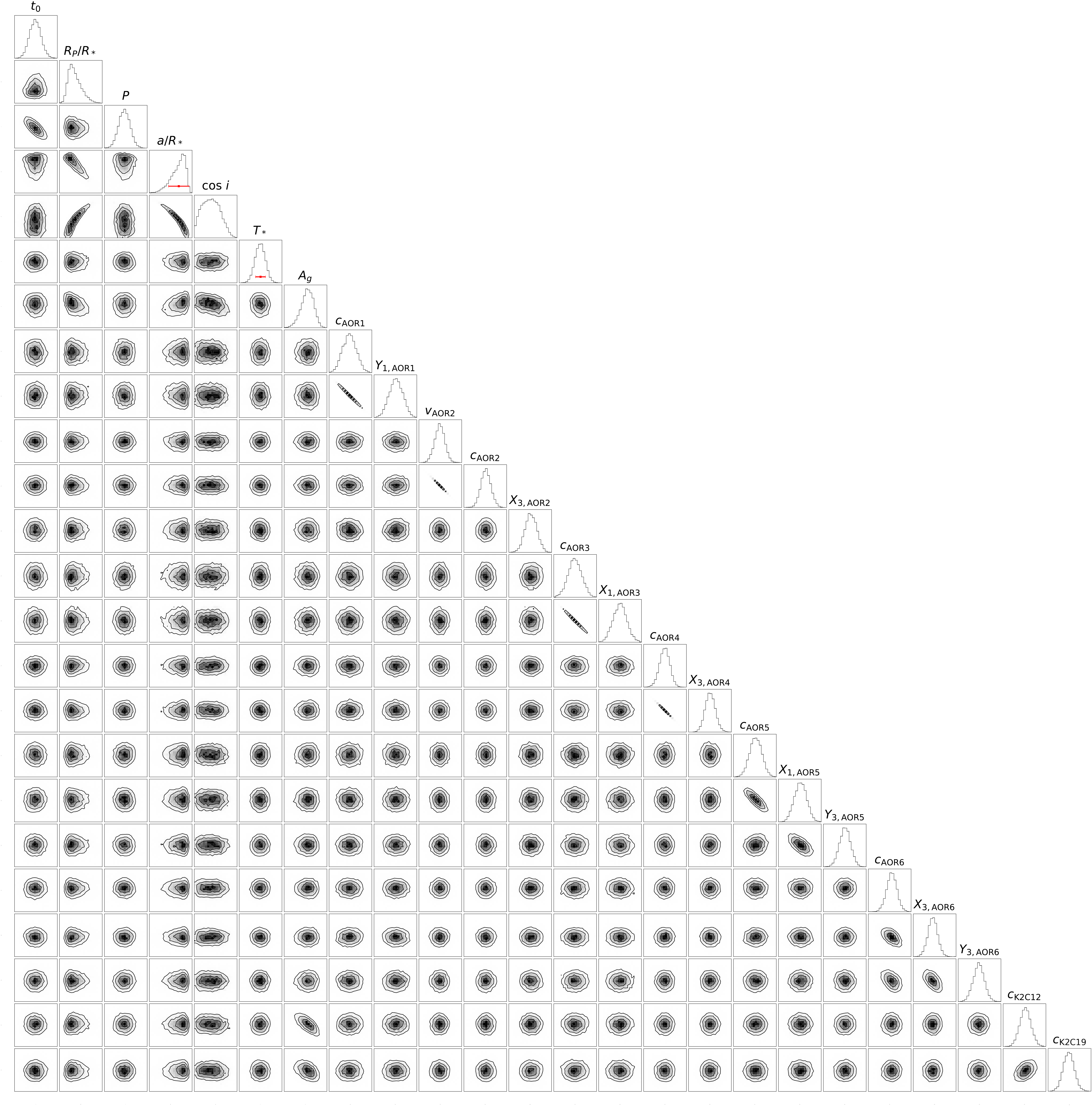}
\caption{MCMC corner plot for the toy model fit without heat redistribution ($F$ = 0) to the joint dataset, i.e. \Spitzer\ and K2. The red bars for \ar\ and $T_*$ show the Gaussian priors which were used in this fit. The prior values and the best fit values are listed in Table \ref{tab:Joint values}. The resulting values for the systematic parameters are in Table \ref{tab:systematic_values}. A list with all fit parameters can be found in Table \ref{tab:joint_models}.}
\label{}
\end{figure*}

\clearpage
\subsection{Joint (K2 and \Spitzer) fit: Toy Model ($F$ free)}
\begin{figure*}[!h]
\centering
\includegraphics[trim=1.6cm 1.6cm 0cm 0cm, clip, width=1\textwidth]{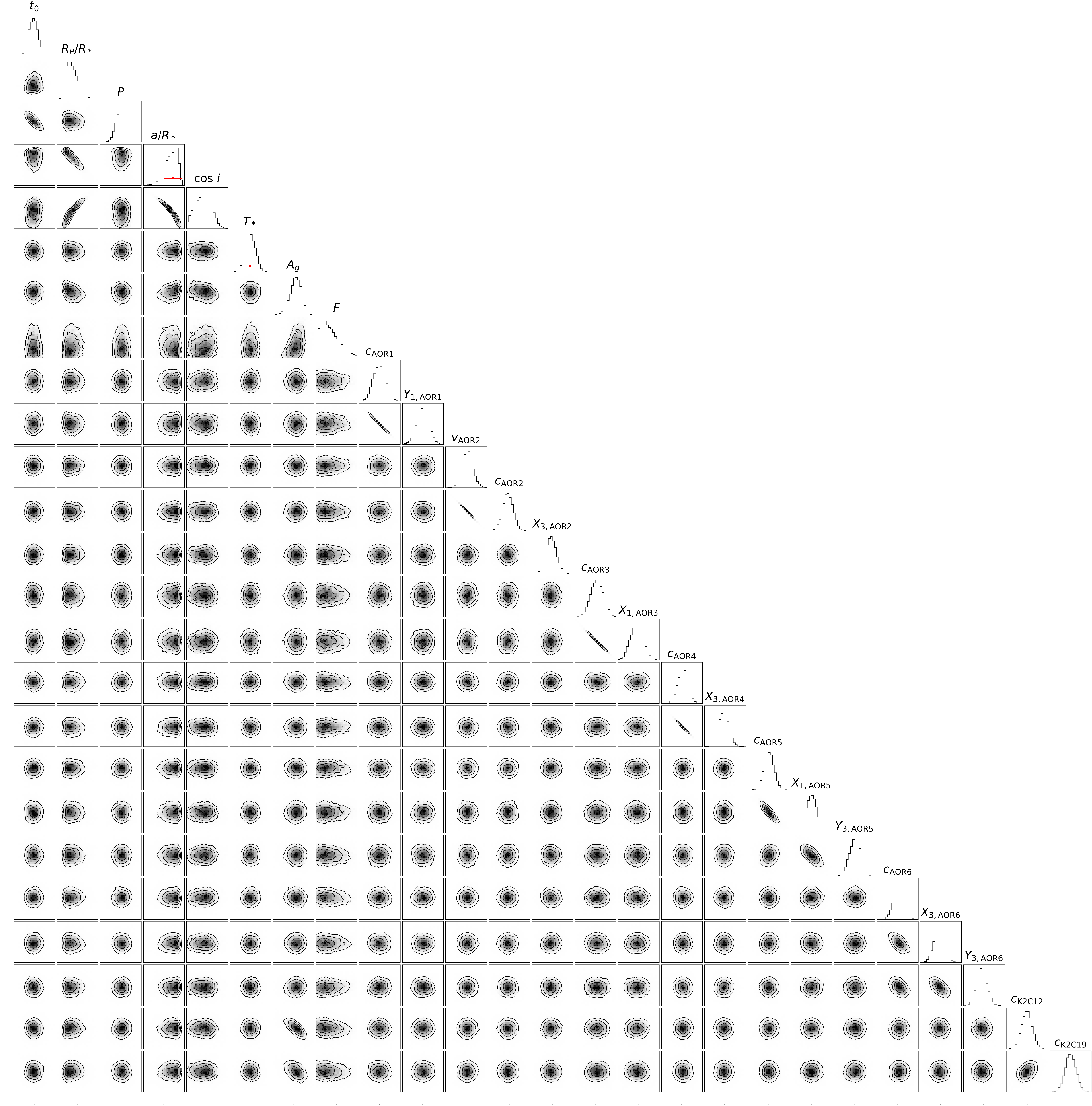}
\caption{MCMC corner plot for the toy model fit with heat redistribution ($F$ free) to the joint dataset, i.e. \Spitzer\ and K2. The red bars for \ar\ and $T_*$ show the Gaussian priors which were used in this fit. The prior values and the best fit values are listed in Table \ref{tab:Joint values}. The resulting values for the systematic parameters are in Table \ref{tab:systematic_values}. A list with all fit parameters can be found in Table \ref{tab:joint_models}.}
\label{}
\end{figure*}

\clearpage
\subsection{Joint (K2 and \Spitzer) fit: Two Temperature Model}\label{fig:corner6}
\begin{figure*}[!h]
\centering
\includegraphics[trim=1.6cm 1.6cm 0cm 0cm, clip, width=1\textwidth]{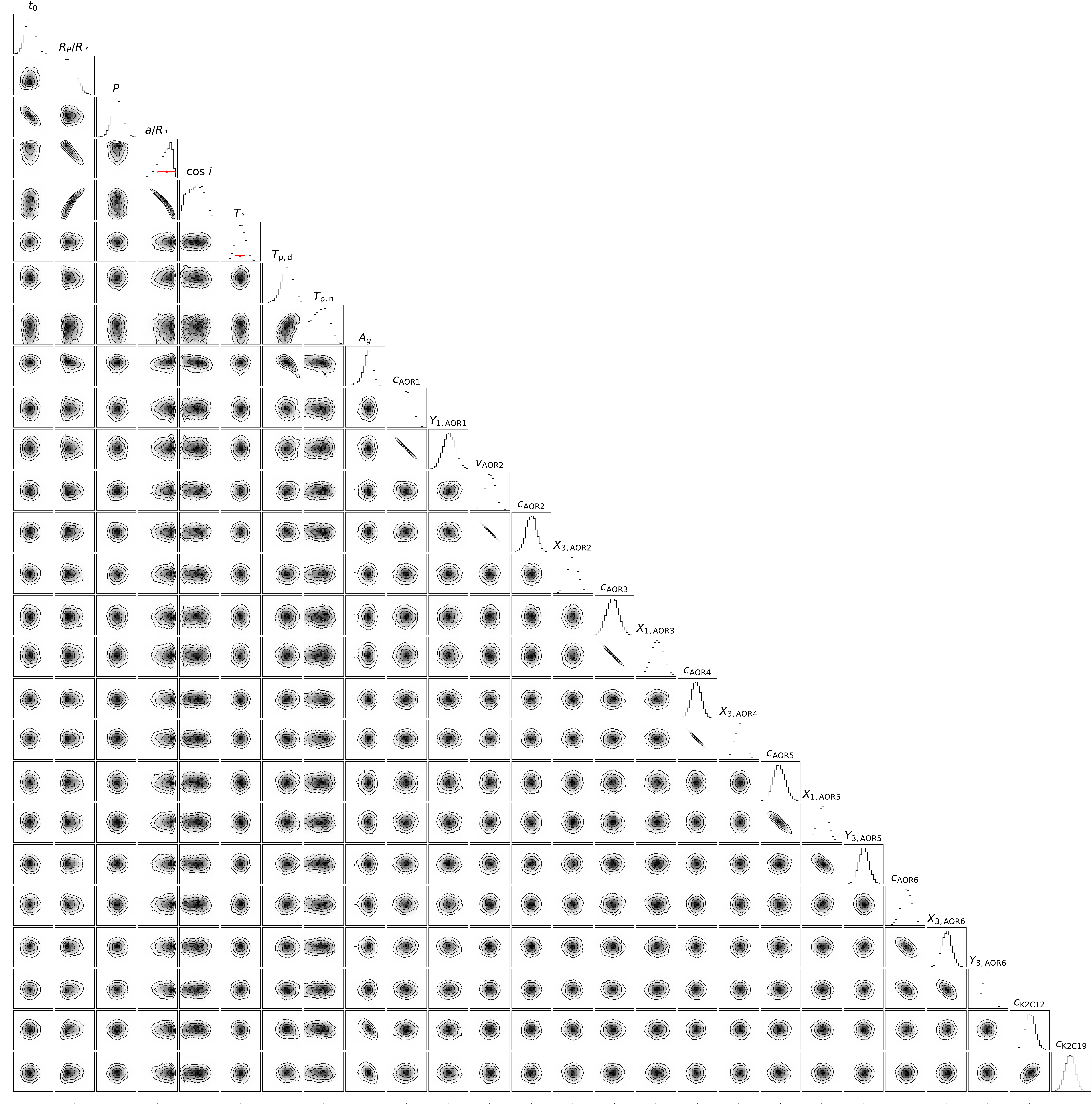}
\caption{MCMC corner plot for the two temperature model fit to the joint dataset, i.e. \Spitzer\ and K2. The red bars for \ar\ and $T_*$ show the Gaussian priors which were used in this fit. The prior values and the best fit values are listed in Table \ref{tab:Joint values}. The resulting values for the systematic parameters are in Table \ref{tab:systematic_values}. A list with all fit parameters can be found in Table \ref{tab:joint_models}.}
\label{}
\end{figure*}

\end{document}